\documentclass[3p,fleqn,times]{elsarticle}

\usepackage{amsmath}
\usepackage{pgfplots}
\usepackage{xcolor}
\usepackage{graphicx}
\usepackage{soul}
\usepackage{subcaption}
\usepackage{booktabs}
\usepackage{multirow}
\usepackage{mathabx}
\usepackage{mathrsfs}

\newcommand\numberthis{\addtocounter{equation}{1}\tag{\theequation}}

\makeatletter
\def\ps@pprintTitle{%
 \let\@oddhead\@empty
 \let\@evenhead\@empty
 \def\@oddfoot{}%
 \let\@evenfoot\@oddfoot}
\makeatother
\usepackage{fancyhdr}

\fancypagestyle{pprintTitle}{%
\lhead{}\chead{}\rhead{}
\lfoot{\textcopyright ~British Crown Owned Copyright 2023/AWE}\cfoot{\thepage}\rfoot{}

}

\title{A cell-centred Eulerian volume-of-fluid method for compressible multi-material flows}

\begin{document}
\begin{frontmatter}

\author[awe]{Timothy R. Law}
\ead{tim.law@awe.co.uk}
\author[awe]{Philip T. Barton}
\ead{phil.barton@awe.co.uk}

\affiliation[awe]{
    addressline={AWE Aldermaston},
    city={Reading},
    postcode={RG7 4PR},
    country={United Kingdon}
}

\begin{abstract}

    \noindent We present a practical cell-centred volume-of-fluid method developed within a
    pure Eulerian setting for the simulation of compressible solid-fluid
    problems.  The method builds on a previously published diffuse-interface
    Godunov-type scheme through the addition of a specialised mixed-cell update
    that is capable of maintaining sharp interfaces indefinitely. The mixed-cell
    update is local and may be viewed as an interface-sharpening extension to
    the underlying diffuse-interface scheme along the lines of other techniques
    such as Tangent of Hyperbola INterface Capturing (THINC), and hence the
    method can be straightforwardly extended to include other coupled physics.
    We validate the method on a range of challenging test problems including a
    collapsing metal shell, cylinder impacts and the three-dimensional simulation of a buried
    explosive charge.  Finally we demonstrate the robustness of the method, and
    its use in a multi-physics context, by modelling the BRL 105mm unconfined
    shaped charge with reactive high-explosive burn and rate-sensitive
    plasticity.

\textcopyright ~British Crown Owned Copyright 2023/AWE

\end{abstract}

\begin{keyword}
    Eulerian solid dynamics\sep%
    Solid-fluid coupling\sep%
    Volume-of-fluid\sep%
    Diffuse-interface\sep%
    Multi-physics
\end{keyword}

\end{frontmatter}

\pagestyle{pprintTitle}
\thispagestyle{pprintTitle}


\section{Introduction}
\label{sec:introduction}
\noindent Accurate and robust computational methods for the treatment of dynamic
problems involving multiple solid and fluid materials are vital across many
scientific and engineering disciplines. Applications include auto-mobile
crashworthiness, sheet metal forming, inertial confinement fusion, asteroid
impacts and supernova core-collapse to name but a few. Such problems frequently
involve large deformations which cannot be treated by purely Lagrangian methods
due to mesh tangling, and so evolving the interfaces between materials
explicitly becomes a necessary part of any viable numerical approach.

Many approaches have been developed to tackle this problem. Arbitrary
Lagrangian-Eulerian (ALE) methods are designed to avoid mesh tangling by
decoupling the motion of the computational mesh from the motion of the material.
ALE codes are often implemented using a Lagrange-plus-remap approach where the
mesh motion is formulated as an operator-split advection step appended to a pure
Lagrangian update, but direct ALE is also possible~\cite{benson92,barlow16}.
Fixed grid Eulerian methods on the other hand have many attractive
characteristics, chief amongst them being the complete avoidance of all
mesh-tangling related issues which can hamper robustness.  Historically, many
multi-material Eulerian methods have been developed based on solving hypoelastic
systems on staggered grids, with explicit artificial viscosity terms to resolve
shocks (see \cite{benson92} for a review of such methods). In more recent years
numerous Godunov-type methods, originally developed for compressible fluid flow,
have been extended to solid dynamics using hyperelastic models
\cite{plohr88,miller01,miller02,lomov05,titarev08,barton09,barton10,hill10,lopezortega14,barton19}.
Godunov methods use the solution of Riemann problems to define numerical fluxes
and introduce the required artificial viscosity implicitly.  A distinct
advantage of these pure Eulerian cell-centred methods is their amenability to
implementation within adaptive mesh refinement (AMR) frameworks, which is often
essential to render complex multi-physics problems computationally tractable.
The foremost challenge of integrating these methods into numerical schemes
suitable for simulating multi-material problems is the treatment of material
interfaces, which form internal boundaries.

Methods for treating material interfaces can be broadly divided into two
categories: \textit{interface-tracking} and \textit{interface-capturing}
(analogously to methods for handling shocks). Interface-tracking methods use an
indicator function that denotes the location of the interface, such as the zero
contour of a level-set~\cite{osher88}. Volume-of-fluid methods are another
notable example, where the fraction of each material present in each cell is
tracked and used to reconstruct an approximation of the
interface~\cite{hirt81,pilliod04}.  Cut cells may be discretised directly as
polyhedra within a finite-volume or finite-element
framework~\cite{miller02,barton11}, but this is complex and requires special
treatments in order to avoid impractically small allowable timesteps.
Alternatively, some approximate treatment may be used to evolve the solution
around the interface, based on the true jump conditions.  Examples include
ghost-fluid methods, pioneered by Fedkiw~\textit{et~al.}~\cite{fedkiw99}, where
the state in each material is extrapolated across the interface (either by
solution of the Riemann problem or some other method) allowing the use of an
unmodified single-material update, which, by virtue of the manipulated states
near the interface, captures the required interface boundary conditions
\cite{hill10,barton13,schoch13,lopezortega14,michael18}. A variety of closure
models have been developed for performing Lagrangian updates in cut cells,
typically based on asserting some kind of equilibrium between
materials~\cite{shashkov08,barlow14,klima20}.

Interface-capturing methods differ in that the location of the interface is
implicit in the solution. The update applied is the same irrespective of the
number of materials present in each cell. The downside of this approach is that
the interface tends to become smeared over time unless some sharpening
procedure is used, hence these methods may also be referred to as
diffuse-interface methods in contrast to the sharp (physical) interface usually
offered by interface-tracking. A recent review can be found
in~\cite{maltsev22}. Many diffuse-interface methods have been developed for
multi-fluid systems~\cite{allaire02,saurel18}, but relatively few for solid-fluid
interactions~\cite{favrie09,favrie12,barton19}, principally because obtaining a
consistent representation of the thermodynamic state in the mixture region is
challenging due to the great variation in material properties.

This paper is devoted to the development of a Godunov-type sharp-interface
method that may be viewed as a hybrid of interface-capturing and
interface-tracking methods. Our method is based on the diffuse-interface scheme
presented by Barton~\cite{barton19}, which is itself a development of the
five-equation multi-fluid model presented by
Allaire~\textit{et~al.}~\cite{allaire02} to include material strength.  We
design a specialised update procedure inspired by the multi-fluid methods of
Miller~and~Puckett~\cite{millerpuckett96}, and
Cutforth~\textit{et~al.}~\cite{cutforth21} for cells that contain more than one
material, which preserves the sharp character of interfaces by means of
volume-of-fluid advection. The solution within these cells is derived from the
Riemann problem posed in terms of the underlying diffuse-interface model, and a
pressure relaxation step is included to force the solution to converge to this
model. Our mixed-cell update may be viewed as an alternative to other interface
sharpening methods such as Tangent of Hyperbola INterface Capturing
(THINC)~\cite{xiao05,shyue14} (previously applied to this problem by
Barton~\cite{barton19}) and other anti-diffusion
schemes~\cite{shukla10,so12,kokh10}. The method is practical (not requiring any
complicated geometric considerations), robust (capable of treating multiple
solids and fluids undergoing extreme deformation), and extensible (amenable to
the addition of coupled physics).

The remainder of the paper is organised as follows. In Section~\ref{sec:theory}
we lay out the theoretical underpinnings of the method, including the evolution
equations and the thermodynamic and kinematic framework. In
Section~\ref{sec:numerical-method} we describe the numerical method in detail.
In Section~\ref{sec:results} we validate the method through application to a
strenuous series of problems culminating in the simulation of an explosive
charge buried in clay. Finally in Section~\ref{sec:application} we show the
potential of our method by using it to model the BRL 105mm unconfined shaped
charge, an extremely challenging multi-physics problem featuring reactive
high-explosive burn and rate-sensitive elasto-plastic solid dynamics. We
conclude the work in Section~\ref{sec:conclusions}.

\section{Governing theory}
\label{sec:theory}
\subsection{Evolution equations---derivation}

\noindent The sharp-interface method is presented as an extension to the
diffuse-interface method of Barton~\cite{barton19}. Under this model, materials
are allowed to mix at their interfaces. Each material $l$ is described by its
mass density $\rho^l$, volume-fraction $\phi^l$ and specific internal energy
$\mathscr{E}^l$. The model assumes mechanical equilibrium: materials in a
mixture share a common velocity $\mathbf{u}$ and mixture stress tensor
$\boldsymbol\sigma$. The basic conservation equations for mass, momentum and
energy are then as follows

\begin{eqnarray}
    \frac{\partial (\phi^l \rho^l)}{\partial t} + \nabla\cdot(\phi^l \rho^l \mathbf{u}) &=& 0\label{eq:continuity}\\
    \frac{\partial (\rho \mathbf{u})}{\partial t} + \nabla\cdot(\rho \mathbf{u}\otimes\mathbf{u}) &=& \nabla\cdot\boldsymbol\sigma\label{eq:momentum}\\
    \frac{\partial (\rho E)}{\partial t} + \nabla\cdot(\rho E \mathbf{u}) &=& \nabla\cdot(\boldsymbol\sigma \mathbf{u})\label{eq:energy}
\end{eqnarray}

\noindent where $\rho$, $\rho
E=\rho\mathscr{E}+\rho(\mathbf{u}\cdot\mathbf{u})/2$ and $\mathscr{E}$ are the
mixture density, total energy density and specific internal energy respectively,
which along with the volume-fraction are subject to the thermodynamically
consistent mixture rules

\begin{eqnarray}
    1 &=& \sum_{l} \phi^l \\
    \rho &=& \sum_{l} \phi^l \rho^l \\
    \rho \mathscr{E} &=& \sum_{l} \phi^l \rho^l \mathscr{E}^l \label{eq:mixture-energy}
\end{eqnarray}

In contrast to the model from Barton~\cite{barton19} which uses simple advection
equations for volume-fraction, here we use the volume-fraction evolution
equation from~\cite{millerpuckett96,lomov05,cutforth21} which is derived
directly from the continuity equation \eqref{eq:continuity} and therefore takes
account of the relative compressibility of each material due to the divergence
of the velocity field. First, expanding product derivatives gives

\begin{equation}
    \label{eq:vof-deriv1}
    \frac{\partial \phi^l}{\partial t} + \nabla\cdot(\phi^l \mathbf{u}) =
    -\frac{\phi^l}{\rho^l} \frac{\partial \rho^l}{\partial t} - \frac{\phi^l}{\rho^l} \mathbf{u}\cdot\nabla\rho^l
\end{equation}

\noindent If we assume that any compression that takes place is isentropic, and
that isotropic stress is maintained during the advection process, then the
pressure change in each material will be equal to the pressure change of the
mixture. Then we can use the definition of the isentropic bulk modulus $K_S =
\rho \left. \partial p / \partial \rho \right|_S$ to replace references to the
partial densities in \eqref{eq:vof-deriv1} with fractions of the mixture
density.

\begin{equation}
    \label{eq:vof-deriv2}
    \frac{K_S^l}{\rho^l} \partial \rho^l = \partial p^l = \partial p = \frac{\overline{K}_S}{\rho} \partial \rho
\end{equation}

\noindent where $\overline{K}_S = (\sum_l \phi^l / K_S^l)^{-1}$ is the mixture
bulk modulus. Substituting \eqref{eq:vof-deriv2} into \eqref{eq:vof-deriv1}
gives

\begin{equation}
    \label{eq:vof-deriv3}
    \frac{\partial \phi^l}{\partial t} + \nabla\cdot(\phi^l \mathbf{u}) =
    -\frac{\phi^l \overline{K}_S}{\rho K_S^l} \left[ \frac{\partial
    \rho}{\partial t} + \mathbf{u}\cdot\nabla\rho \right]
\end{equation}

\noindent Finally summing \eqref{eq:continuity} over materials to give the
continuity equation of the mixture and substituting $\partial \rho / \partial t
+ \mathbf{u}\cdot\nabla\rho = - \rho \nabla\cdot\mathbf{u}$ into
\eqref{eq:vof-deriv3}, we obtain the volume-fraction evolution
equation\footnote{We note that while the choice of right-hand side weighting in
the volume-fraction equation can be motivated physically as described,
phenomenological choices can also work well. We have tested substituting the
Gr\"uneisen parameter for the isentropic bulk modulus, to make the
volume-fraction evolution appear consistent with the internal energy evolution
described later, and found no significant difference in the performance of the
method.}

\begin{equation}
    \label{eq:vof}
    \frac{\partial \phi^l}{\partial t} + \nabla\cdot(\phi^l \mathbf{u}) =
    \frac{\phi^l \overline{K}_S}{K_S^l} \nabla\cdot\mathbf{u}
\end{equation}

The pressure for each material is assumed to be defined by an equation of state
of the Mie-Gr\"uneisen form

\begin{equation}
    \label{eq:pressure}
    p^l(\rho^l, \mathscr{E}^l, \mathrm{dev}(\mathbf{H}^l_e)) = p^l_\mathrm{ref}(\rho^l, \mathrm{dev}(\mathbf{H}^l_e)) + \rho^l \Gamma^l(\rho^l) (\mathscr{E}^l - \mathscr{E}^l_\mathrm{ref}(\rho^l, \mathrm{dev}(\mathbf{H}^l_e)))
\end{equation}

\noindent where $\Gamma^l$ is the Gr\"uneisen parameter and
$\mathrm{dev}(\mathbf{H}_e)$ is the deviatoric part\footnote{The matrix deviator
$\mathrm{dev}(M)$ is defined as $M- \frac{1}{3} \mathrm{tr}(M) \mathbf{I}$,
where $\mathrm{tr}(M)$ denotes the matrix trace.} of the $3\times{}3$ Hencky
elastic strain tensor.  The reference energy is assumed to admit an additive
decomposition, comprising a contribution due to cold compression or dilation,
$\mathscr{E}_c$, and a contribution due to isochoric shear strain,
$\mathscr{E}_s$

\begin{eqnarray}
    \mathscr{E}^l_\mathrm{ref}(\rho^l, \mathrm{dev}(\mathbf{H}^l_e)) &=& \mathscr{E}^l_c(\rho^l) + \mathscr{E}^l_s(\rho^l, \mathrm{dev}(\mathbf{H}^l_e)) \\
    p^l_\mathrm{ref}(\rho^l, \mathrm{dev}(\mathbf{H}^l_e)) &=& \rho^{2,l} \frac{\partial \mathscr{E}^l_\mathrm{ref}}{\partial \rho^l}
\end{eqnarray}

\noindent The shear energy is given by

\begin{equation}
    \label{eq:eos-energy-shear}
    \mathscr{E}^l_s(\rho^l, \mathrm{dev}(\mathbf{H}^l_e)) = \frac{G^l(\rho^l)}{\rho^l} \mathcal{J}^2(\mathrm{dev}(\mathbf{H}^l_e))
\end{equation}

\noindent where $G^l$ is the shear modulus and
$\mathcal{J}^2(\mathrm{dev}(\mathbf{H}_e)) =
\mathrm{tr}(\mathrm{dev}(\mathbf{H}_e) \mathrm{dev}(\mathbf{H}_e)^T)$ is the
second invariant of shear strain. In fluids the shear modulus is taken to be
zero, leading to zero shear energy contribution.

Following hyperelastic theory the Cauchy stress tensor is a product of
thermodynamic compatibility and is found through taking derivatives of internal
energy with respect to density and deviatoric strain (see \cite{barton19} for a
full derivation)

\begin{equation}
    \label{eq:stress_l}
    \boldsymbol\sigma^l(\rho^l, \mathscr{E}^l, \mathrm{dev}(\mathbf{H}^l_e)) =
        -p^l(\rho^l, \mathscr{E}^l, \mathrm{dev}(\mathbf{H}^l_e)) \mathbf{I} + 2 G^l (\rho^l) \mathrm{dev}(\mathbf{H}^l_e)
\end{equation}

\noindent Using the assumption that all components in a mixture are in
mechanical equilibrium and therefore have the same pressure, along with the
assumption that materials have a common deviatoric strain, the mixture stress
becomes

\begin{equation}
    \label{eq:stress}
    \boldsymbol\sigma(\rho, \mathscr{E}, \mathrm{dev}(\mathbf{H}_e)) =
        -p(\rho, \mathscr{E}, \mathrm{dev}(\mathbf{H}_e)) \mathbf{I} + 2 G (\rho) \mathrm{dev}(\mathbf{H}_e)
\end{equation}

\noindent where the mixture pressure $p$ and mixture shear modulus $G$ are given
by

\begin{eqnarray}
    \label{eq:mixture-pressure}
    p &=& \frac{\rho \mathscr{E} - \sum_l \left( \phi^l \rho^l \mathscr{E}^l_\mathrm{ref} - \phi^l p^l_\mathrm{ref} / \Gamma^l \right)}{\sum_l \phi^l / \Gamma^l}\\
    G &=& \frac{\sum_{l} \phi^l G^l / \Gamma^l}{\sum_l \phi^l / \Gamma^l}
\end{eqnarray}

\noindent The mixture Gr\"uneisen parameter $\overline{\Gamma}$ is defined as

\begin{equation}
    \overline{\Gamma} = \frac{1}{\sum_l \phi^l / \Gamma^l} \label{eq:mixture-gruneisen}
\end{equation}

In order to track elastic deformations in solid media, it is first noted that
the Hencky deviatoric strain can be defined in terms of the unimodular elastic
left stretch tensor $\overline{\mathbf{V}}_e$ according to

\begin{equation}
    \label{eq:hencky-strain}
    \text{dev}(\mathbf{H}_e) = \mathrm{ln}(\overline{\mathbf{V}}_e)
\end{equation}

\noindent An additional nine equations are then included to evolve
$\overline{\mathbf{V}}_e$

\begin{equation}
    \label{eq:stretch}
    \frac{\partial \overline{\mathbf{V}}_e}{\partial t} + \nabla\cdot(\overline{\mathbf{V}}_e \otimes \mathbf{u}) = \nabla \mathbf{u} \overline{\mathbf{V}}_e + \frac{2}{3}(\nabla\cdot{\mathbf{u}})\overline{\mathbf{V}}_e - \boldsymbol\Phi
\end{equation}

\noindent The source term $\boldsymbol\Phi$ represents relaxation due to
inelastic deformation, defined later.

It is sometimes necessary to evolve additional history variables to support the
closure models, For instance equivalent plastic strain or the reactant mass
fraction in a condensed-phase explosive. These additional history parameters can
be collectively denoted by the vectors $\boldsymbol\pi^l$ for each material, and
evolve according to

\begin{equation}
\frac{\partial \rho^{l}\phi^{l} \boldsymbol\pi^{l}}{\partial t} + \nabla\cdot( \rho^{l}\phi^{l} \boldsymbol\pi^{l} \otimes \mathbf{u}) = \rho^{l}\phi^{l} \dot{\boldsymbol\pi}^{l}\label{eq:history}
\end{equation}

\noindent For the interested reader, further examples of how multi-physics can
be introduced into the model in this way include the treatment of reactive
mixtures in \cite{wallis21}, continuum damage mechanics and fracture in
\cite{wallis21_wbn}, and phase transitions in~\cite{wilkinson21}.

Finally, around material interfaces we require an additional set of equations to
evolve the partial internal energies of each material.  In
\cite{millerpuckett96} an internal energy evolution equation was proposed which
added a $p\mathrm{d}v$ work term to the right-hand-side of the conservative
advection equation, inspired by the right-hand-side of the volume-fraction
equation \eqref{eq:vof}.  Here we propose an alternative formulation which is
equally self-consistent, but which uses the mixture rules to determine the
additional source terms, including those taking account of material strength.
By expanding the energy equation \eqref{eq:energy} into internal and kinetic
contributions we obtain

\begin{equation}
    \label{eq:internal-energy-0}
    \frac{\partial (\rho \mathscr{E})}{\partial t} + \nabla\cdot(\rho \mathscr{E} \mathbf{u}) - \boldsymbol\sigma : \nabla\mathbf{u} = 0
\end{equation}

\noindent Using the definition of stress \eqref{eq:stress} and substituting the
mixture rules for energy, pressure and the shear modulus gives

\begin{equation}
    \label{eq:internal-energy-2}
    \sum_l\left[ \frac{\partial (\phi^l \rho^l \mathscr{E}^l)}{\partial t} + \nabla\cdot(\phi^l \rho^l \mathscr{E}^l \mathbf{u}) - \frac{\phi^l \overline{\Gamma}}{\Gamma^l} (-p^l \mathbf{I} + 2 G^l \mathrm{dev}(\mathbf{H}_e)) : \nabla\mathbf{u}\right] = 0
\end{equation}

\noindent from which we assume the internal energy equations for individual
materials

\begin{equation}
    \label{eq:internal-energy}
\frac{\partial (\phi^l \rho^l \mathscr{E}^l)}{\partial t} + \nabla\cdot(\phi^l \rho^l \mathscr{E}^l \mathbf{u}) = \frac{\phi^l \overline{\Gamma}}{\Gamma^l} (-p^l \mathbf{I} + 2 G^l \mathrm{dev}(\mathbf{H}_e)) : \nabla\mathbf{u}
\end{equation}

\subsection{Evolution equations---summary}

\noindent The complete system of evolution equations for which we seek solutions
comprises \eqref{eq:continuity}-\eqref{eq:energy}, \eqref{eq:vof},
\eqref{eq:stretch}, and \eqref{eq:history}. In vector form, this system can be
written:

\begin{equation}
    \label{eq:conservation-form}
    \frac{\partial \mathbf{U}}{\partial t} + \sum_k \frac{\partial
    \mathbf{F}_k(\mathbf{U})}{\partial x_k} = \sum_k \mathbf{S}_k(\mathbf{U}) +
    \mathbf{S}_{\mathrm{split}}(\mathbf{U})
\end{equation}

\noindent where the vector of conserved variables $\mathbf{U}$ and the fluxes in
the $x_k$ direction $\mathbf{F}_k(\mathbf{U})$ are defined (in Cartesian
coordinates) as

\begin{equation}
    \label{eq:conserved-vars}
    \mathbf{U} = \begin{pmatrix} \rho^l \phi^l \\ \phi^l\\ \rho u_i\\ \rho E\\
    \overline{V}_{e,ij}\\ \rho^l \phi^l \pi^l_m
    \end{pmatrix},\qquad
    \mathbf{F}_k(\mathbf{U}) = \begin{pmatrix}
        \rho^l \phi^l u_k \\
        \phi^l u_k \\
        \rho u_i u_k - \sigma_{ik} \\
        \rho E u_k - \sum_i \sigma_{ik} u_i \\
        \overline{V}_{e,ij} u_k - \overline{V}_{e,kj} u_i \\
        \rho^l \phi^l \pi^l_m u_k \\
    \end{pmatrix}
\end{equation}

\noindent The source terms are separated reflecting the splitting strategy used
in the numerical implementation

\begin{equation}
    \mathbf{S}_k(\mathbf{U}) = \begin{pmatrix}
        0 \\
        \phi^l \overline{K}_S / K^l_S \cdot \partial u_k / \partial x_k \\
        0 \\
        0 \\
        \frac{2}{3} \overline{V}_{e,ij} \partial u_k/\partial x_k - u_i \partial \overline{V}_{e,kj}/\partial x_k \\
        0 \\
    \end{pmatrix}, \quad
    \mathbf{S}_{\mathrm{split}}(\mathbf{U}) = \begin{pmatrix}
        0 \\
        0 \\
        0 \\
        0 \\
        -\boldsymbol\Phi \\
        \rho^l \phi^l \dot{\pi}^l_m \\
    \end{pmatrix}
\end{equation}

\noindent To assist in determining solutions in mixed cells, the system is
complemented by the evolution equations for internal energies
\eqref{eq:internal-energy}.

\subsection{Quasi-linear primitive formulation and wave speeds}

\noindent Our choice of numerical method requires solution of the quasi-linear
primitive form of the evolution equations, which neglecting the inelastic and
history-rate source terms can be written

\begin{equation}
    \label{eq:primitive-form}
    \frac{\partial \mathbf{W}}{\partial t} + \sum_k \mathbf{A}_k(\mathbf{W}) \frac{\partial \mathbf{W}}{\partial x_k} = \mathbf{0}
\end{equation}

\noindent where $\mathbf{W}$ is the vector of primitive variables

\begin{equation}
    \label{eq:primitive-vars}
    \mathbf{W} = (\rho^l \phi^l, \phi^l, u_i, p,
    \overline{V}_{e,ij}, \pi^l_m
    )^T
\end{equation}

\noindent The matrix in the quasi-linear formulation, $\mathbf{A}_k$ is given by
(in the $x$-direction)

\begin{equation}
    \label{eq:primitive-jacobian}
    \mathbf{A}_x(\mathbf{W}) = \left(\begin{smallmatrix}
        u & 0 & \rho^l \phi^l & 0 & 0 & 0 & 0 & 0 & 0 & 0 & 0 & 0 & 0 & 0 & 0 & 0 \\
        0 & u & 0 & 0 & 0 & 0 & 0 & 0 & 0 & 0 & 0 & 0 & 0 & 0 & 0 & 0 \\
        B^l_1 & C^l_1 & u & 0 & 0 & 1/\rho & A^1_{11} & A^1_{21} & A^1_{31} & A^1_{12} & A^1_{22} & A^1_{32} & A^1_{13} & A^1_{23} & A^1_{33} & 0 \\
        B^l_2 & C^l_2 & 0 & u & 0 & 0 & A^2_{11} & A^2_{21} & A^2_{31} & A^2_{12} & A^2_{22} & A^2_{32} & A^2_{13} & A^2_{23} & A^2_{33} & 0 \\
        B^l_3 & C^l_3 & 0 & 0 & u & 0 & A^3_{11} & A^3_{21} & A^3_{31} & A^3_{12} & A^3_{22} & A^3_{32} & A^3_{13} & A^3_{23} & A^3_{33} & 0 \\
        0 & 0 & \rho a^2 & 0 & 0 & u & 0 & 0 & 0 & 0 & 0 & 0 & 0 & 0 & 0 & 0 \\
        0 & 0 & -\frac{2}{3} \overline{V}_{e,11} & 0 & 0 & 0 & u & 0 & 0 & 0 & 0 & 0 & 0 & 0 & 0 & 0 \\
        0 & 0 & \frac{1}{3} \overline{V}_{e,21} & -\overline{V}_{e,11} & 0 & 0 & 0 & u & 0 & 0 & 0 & 0 & 0 & 0 & 0 & 0 \\
        0 & 0 & \frac{1}{3} \overline{V}_{e,31} & 0 & -\overline{V}_{e,11} & 0 & 0 & 0 & u & 0 & 0 & 0 & 0 & 0 & 0 & 0 \\
        0 & 0 & -\frac{2}{3} \overline{V}_{e,12} & 0 & 0 & 0 & 0 & 0 & 0 & u & 0 & 0 & 0 & 0 & 0 & 0 \\
        0 & 0 & \frac{1}{3} \overline{V}_{e,22} & -\overline{V}_{e,12} & 0 & 0 & 0 & 0 & 0 & 0 & u & 0 & 0 & 0 & 0 & 0 \\
        0 & 0 & \frac{1}{3} \overline{V}_{e,32} & 0 & -\overline{V}_{e,12} & 0 & 0 & 0 & 0 & 0 & 0 & u & 0 & 0 & 0 & 0 \\
        0 & 0 & -\frac{2}{3} \overline{V}_{e,13} & 0 & 0 & 0 & 0 & 0 & 0 & 0 & 0 & 0 & u & 0 & 0 & 0 \\
        0 & 0 & \frac{1}{3} \overline{V}_{e,23} & -\overline{V}_{e,13} & 0 & 0 & 0 & 0 & 0 & 0 & 0 & 0 & 0 & u & 0 & 0 \\
        0 & 0 & \frac{1}{3} \overline{V}_{e,33} & 0 & -\overline{V}_{e,13} & 0 & 0 & 0 & 0 & 0 & 0 & 0 & 0 & 0 & u & 0 \\
        0 & 0 & 0 & 0 & 0 & 0 & 0 & 0 & 0 & 0 & 0 & 0 & 0 & 0 & 0 & u \\
    \end{smallmatrix}\right)
\end{equation}

\noindent where

\begin{align}
    A^i_{jk} &= -\frac{1}{\rho} \frac{\partial \mathrm{dev}(\boldsymbol\sigma)_{i1}}{\partial \overline{V}_{e,jk}} \\
    B^l_i &= -\frac{1}{\rho} \frac{\partial \mathrm{dev}(\boldsymbol\sigma)_{i1}}{\partial (\rho^l \phi^l)} \\
    C^l_i &= -\frac{1}{\rho} \frac{\partial \mathrm{dev}(\boldsymbol\sigma)_{i1}}{\partial \phi^l} \\
\end{align}

\noindent The derivative of the tensor logarithm is given by Jog~\cite{jog08},
and so

\begin{equation}
    A^i_{jk} = -2b^2 \delta_{ij} \delta_{1k} \overline{V}^{-1}_{e,i1}
\end{equation}

\noindent The bulk sound speed $a$ and the shear speed $b$ are defined next.

The wave speeds are required by the primitive variable formulation, the Riemann
solver, and in order to estimate the allowable time-step.  In \cite{barton19} it
was shown how the non-linear characteristic wave speeds derived from the
characteristic polynomial of $\mathbf{A}_k$ are a function of the eigenvalues of
the $3\times3$ {\it acoustic tensor}. In lieu of a costly evaluation of this
tensor the fastest longitudinal wave speed is estimated as

\begin{equation}
    \label{eq:long-speed}
    c^{2,l}(\rho^l, p^l, \mathrm{dev}(\mathbf{H}^l_e)) = a^{2,l}(\rho^l, p^l, \mathrm{dev}(\mathbf{H}^l_e)) + \frac{4}{3} b^{2,l}(\rho^l)
\end{equation}

\noindent The bulk sound speed is computed by differentiating
\eqref{eq:pressure} with respect to density at constant entropy, and using the
relation $\mathrm{d}\mathscr{E} = T \mathrm{d}S - p \mathrm{d}v$

\begin{align*}
    a^{2,l}(\rho^l, p^l, \mathrm{dev}(\mathbf{H}^l_e)) = \; &\frac{\partial
    p^l_\mathrm{ref}(\rho^l, \mathrm{dev}(\mathbf{H}^l_e))}{\partial \rho^l} +
    \frac{(\Gamma^l(\rho^l) + 1) p^l - p^l_\mathrm{ref}(\rho^l,
    \mathrm{dev}(\mathbf{H}^l_e))}{\rho^l} + \numberthis \label{eq:bulk-speed} \\
        &\;\frac{p^l - p^l_\mathrm{ref}(\rho^l,
        \mathrm{dev}(\mathbf{H}^l_e))}{\Gamma^l(\rho^l)} \frac{\partial
        \Gamma^l}{\partial \rho^l} - \rho^l \Gamma^l(\rho^l) \frac{\partial
        \mathscr{E}^l_\mathrm{ref}(\rho^l, \mathrm{dev}(\mathbf{H}^l_e))}{\partial \rho^l}
\end{align*}

\noindent The shear speed is given by

\begin{equation}
    \label{eq:shear-speed}
    b^{2,l}(\rho^l) = \frac{G^l(\rho^l)}{\rho^l}
\end{equation}

\noindent In cells containing more than one material,  the mixture wave speed is
calculated according to

\begin{equation}
    \rho c^2 = \frac{\sum_{l} \phi^l \rho^l c^{2,l} / \Gamma^l}{\sum _l \phi^l / \Gamma^l}
\end{equation}

\subsection{Closure models}
\label{sec:closure-models}

\noindent It is noted that the Mie-Gr\"uneisen framework embodies many specific
equations of state, depending on the functional form assumed for the reference
energy/pressure, Gr\"uneisen parameter and shear modulus. For example
(superscripts $l$ have been omitted for clarity):

\begin{itemize}

    \item The ideal gas law is given by $\mathscr{E}^l_\mathrm{ref} = p_\mathrm{ref} = 0$,
        $\Gamma = \gamma - 1$ and $G = 0$.

    \item The stiffened gas equation of state describing denser fluids is given
        by $\mathscr{E}_\mathrm{ref} = \mathscr{E}_\infty$, $p_\mathrm{ref} = -\gamma p_\infty$,
        $\Gamma = \gamma - 1$ and $G = 0$.

    \item The Jones-Wilkins-Lee (JWL) equation of state~\cite{kury65,lee68,jwl} describing explosive products (and
        sometimes reactants) is given by $\Gamma = \Gamma_0$, $G = 0$ and
        \begin{align*}
            \mathscr{E}_\mathrm{ref} &= \frac{A}{R_1 \rho_0} \exp\left({-R_1 \frac{\rho_0}{\rho}}\right) + \frac{B}{R_2 \rho_0} \exp\left({-R_2 \frac{\rho_0}{\rho}}\right) \\
            p_\mathrm{ref} &= A \exp\left({-R_1 \frac{\rho_0}{\rho}}\right) + B \exp\left({-R_2 \frac{\rho_0}{\rho}}\right)
        \end{align*}

    \item The equation of state from~\cite{dorovskii83} describing elastic solids is given by $\Gamma = \Gamma_0$ and
        \begin{align*}
            \mathscr{E}_\mathrm{ref} &= \frac{K_0}{2\rho_0\alpha^2}
            \left(\left(\frac{\rho}{\rho_0}\right)^\alpha - 1\right)^2 +
            \mathscr{E}_s \\
            p_\mathrm{ref} &= \rho^2 \frac{\partial \mathscr{E}_\mathrm{ref}}{\partial \rho} \\
            G &= G_0 \left( \frac{\rho}{\rho_0} \right)^{\beta + 1}
        \end{align*}

\end{itemize}

Plastic relaxations are incorporated via the source term $\boldsymbol\Phi$
following the method of convex potentials. The Von Mises flow rule is used,
leading to the following functional form

\begin{equation}
    \label{eq:plastic-source}
    \boldsymbol\Phi = \chi \sqrt{\frac{3}{2}} \frac{\mathrm{dev}(\boldsymbol\sigma)}{||\mathrm{dev}(\boldsymbol\sigma)||} \overline{\mathbf{V}}_e
\end{equation}

\noindent where $||\mathbf{M}|| = \mathcal{J}(\mathbf{M})$ is the Frobenius norm
and $\chi \geq 0$ is the plastic rate. The latter is a closure model which must
be provided for each material, and the following mixture rule is used

\begin{equation}
    \label{eq:chi-mixture}
    \chi = \frac{\sum_l \phi^l \chi^l / \Gamma^l}{\sum_l \phi^l / \Gamma^l}
\end{equation}

\noindent In this work we consider both ideal plasticity, and the rate-dependent
model from Johnson and Cook~\cite{johnson85}. For the former the plastic rate is
given by

\begin{equation}
    \label{eq:ideal-plasticity}
    \chi^l = \chi^l_0 H \left( \sqrt{\frac{3}{2}} ||\mathrm{dev}(\boldsymbol\sigma)|| - \sigma_Y \right)
\end{equation}

\noindent where $H(s)$ is the Heaviside function, $\sigma_Y$ is the constant
yield stress, and $\chi^l_0 \rightarrow \infty$. For the rate-dependent model,
we have

\begin{align}
    \chi^l &= \chi^l_0 \mathrm{exp} \left[ \frac{1}{c_3} \left( \sqrt{\frac{3}{2}} \frac{||\mathrm{dev}(\boldsymbol\sigma)||}{\sigma_Y(\varepsilon^l_p)} - 1 \right) \right] \\
    \sigma_Y(\varepsilon^l_p) &= (c_1 + c_2(\varepsilon^l_p)^n) \left[ 1- \left(\frac{T - T_0}{T_{\mathrm{melt}} - T_0}\right)^m \right]
\end{align}

\noindent where $\chi^l_0$ is the reference rate, $c_3$ controls the rate
dependency, $c_1$ is the yield stress, $c_2$ is the strain hardening factor, $n$
is the strain hardening exponent, $T_{\mathrm{melt}}$ is the melting temperature
of the material, $T_0$ is a reference temperature, and $m$ is the thermal
softening exponent. This model is a function of the equivalent plastic strain
$\varepsilon^l_p$, so an additional evolution equation is required to track this
quantity

\begin{equation}
    \label{eq:epst}
    \frac{\partial \phi^l \rho^l \varepsilon^l_p}{\partial t} + \nabla\cdot(\phi^l \rho^l \varepsilon^l_p \mathbf{u}) = \phi^l \rho^l \chi^l
\end{equation}

\noindent which is added as a history variable for the corresponding material.

Condensed-phase explosives are treated using the reactive burn model
from~\cite{wallis21}, where the reactants and products are represented as a
physical mixture. The reactant mass fraction, denoted $\lambda$, is added as a
history variable for the corresponding material with the evolution equation

\begin{equation}
    \frac{\partial \phi^l \rho^l \lambda^l}{\partial t} + \nabla\cdot(\phi^l \rho^l \lambda^l \mathbf{u}) = \phi^l \rho^l \dot{\lambda}^l
\end{equation}

\noindent Various forms for the reaction rate $\dot{\lambda}$ have been
proposed, notably the ignition and growth model~\cite{lee80}, viz.

\begin{align}
    -\dot{\lambda} =\, &I \lambda^b \left(\frac{\rho}{\rho_0} - 1 - a\right)^x H(F_{ig} - f) \, + \\
        &G_1 \lambda^c f^d p^y H(F_{G_1} - f) \, + \\
        &G_2 \lambda^e f^g p^z H(f - F_{G_2}) \\
    f =\, &1 - \lambda
\end{align}

\noindent where $H(s)$ denotes the Heaviside function.

\section{Numerical method}
\label{sec:numerical-method}
\noindent We leverage the AMReX block-structured adaptive mesh refinement
framework in our implementation~\cite{amrex}. AMReX has native support for
advanced architectures such as Graphics Processing Units (GPUs) which have the
potential to dramatically accelerate calculations. All of the 2D and 3D
simulations presented later in this work are performed on NVIDIA A100 GPUs.

Solutions are found using structured computational grids consisting of
hexahedral cells with cell centres denoted by the indices $i,j,k$. Each cell
$\mathcal{C}_{ijk}$ has the dimensions $\Delta x_{i}=x_{i+1/2}-x_{i-1/2}$,
$\Delta y_{j}=y_{j+1/2}-y_{j-1/2}$, $\Delta z_{k}=z_{k+1/2}-z_{k-1/2}$, where
$i\pm\frac{1}{2}$, $j\pm\frac{1}{2}$, $k\pm\frac{1}{2}$ denote cell boundary
quantities. Each cell therefore forms the control volume ${V}_{ijk}=\Delta
x_i\Delta y_j\Delta z_k$.  The method is implemented in a dimensionally split
fashion, which ensures coupling at cell corners is accounted for, and also
greatly simplifies the volume-of-fluid advection. A timestep thus consists of
$x$, $y$ and $z$ updates, followed by the split source term updates. Within each
cell the solution is advanced according to

\begin{align*}
    \mathbf{U}^{n,(1)}_{ijk} &= \mathbf{U}^n_{ijk} - \frac{\Delta t}{\Delta x_{i}} \left[ \mathbf{F}_{1,i+1/2} - \mathbf{F}_{1,i-1/2} \right] + \Delta t \mathbf{S}_1(\mathbf{U}^n_{ijk}) \\
    \mathbf{U}^{n,(2)}_{ijk} &= \mathbf{U}^{n,(1)}_{ijk} - \frac{\Delta t}{\Delta y_{j}} \left[ \mathbf{F}_{2,j+1/2} - \mathbf{F}_{2,j-1/2} \right] + \Delta t \mathbf{S}_2(\mathbf{U}^{n,(1)}_{ijk}) \\
    \mathbf{U}^{n,(3)}_{ijk} &= \mathbf{U}^{n,(2)}_{ijk} - \frac{\Delta t}{\Delta z_{k}} \left[ \mathbf{F}_{3,k+1/2} - \mathbf{F}_{3,k-1/2} \right] + \Delta t \mathbf{S}_3(\mathbf{U}^{n,(2)}_{ijk}) \\
    \mathbf{U}^{n+1}_{ijk} &= \mathbf{U}^{n,(3)}_{ijk} + \Delta t \mathbf{S}_{\mathrm{split}}(\mathbf{U}^{n,(3)}_{ijk}) \\
\end{align*}

\noindent where $\mathbf{F}_{k,i\pm1/2}$ are the cell-wall numerical
fluxes to be defined later. The volume-of-fluid method is incoporated into the
dimensionally split part of the update, described in detail below.

The basic procedure may be outlined as follows:

\begin{enumerate}

    \item For each coordinate direction:

        \begin{enumerate}

            \item Perform MUSCL reconstruction of the primitive variables and
                extrapolate to the half-time-step according to the MUSCL-Hancock
                method.

            \item Use the HLLD Riemann solver~\cite{lopezortega14,barton19} to
                compute interfacial states and numerical fluxes at all faces.

            \item Reconstruct material interfaces using the initial
                volume-fraction field and use the interfacial speeds from step
                1b to compute the volume of each material transported across
                each face.

            \item Local to material interfaces, apply the mixed-cell update
                formulae to evolve the state. Then relax each mixed-cell to
                pressure equilibrium and sum the partial internal energies to
                retrieve the total energy.

            \item Update state within single-material cells (i.e. all cells not
                updated in step 1d) using the HLLD fluxes from step 1b.

            \end{enumerate}

        \item Calculate the operator-split source terms. Reinstate the symmetry
            property for the unimodular elastic left stretch tensor.

\end{enumerate}

\noindent For clarity we describe the update in the $x$-direction, modifications
for the $y$- and $z$-directions are trivial.

\subsection{Reconstruction}

\noindent In order to obtain second-order accuracy in smooth single-material
regions, we use the MUSCL-Hancock method~\cite{vanleer97_iv}, wherein the input
data for the Riemann problems is computed by performing a limited MUSCL
reconstruction of the initial cell average states, and then extrapolating the
reconstructed states forward in time to the half-step $t+\Delta t/2$. These
states are then used to compute the numerical fluxes.  The method can be applied
to either the conserved or primitive variables, but testing has shown that
reconstruction of the primitive variables yields better results, with fewer
oscillations in the vicinity of material interfaces. This is in line with the
findings of Johnsen and Colonius~\cite{johnsen06}.

The reconstruction is carried out as follows. 
Given the piecewise constant vector of primitive variables $\mathbf{W}$ in each cell, the MUSCL method constructs a piecewise linear function from which the left and right boundary values can be written
\begin{eqnarray}
\mathbf{W}_{i-1/2,R} &=& \mathbf{W}_{i} - \frac{1}{2}\Phi(\boldsymbol\Delta_{R})\left(\mathbf{W}_{i+1}-\mathbf{W}_{i}\right),\\
\mathbf{W}_{i+1/2,L} &=& \mathbf{W}_{i} + \frac{1}{2}\Phi(\boldsymbol\Delta_{L})\left(\mathbf{W}_{i}-\mathbf{W}_{i-1}\right),
\end{eqnarray}
with
\begin{equation}
\boldsymbol\Delta_{L} = \frac{\mathbf{W}_{i+1}-\mathbf{W}_{i}}{\mathbf{W}_{i}-\mathbf{W}_{i-1}+\epsilon},\qquad
\boldsymbol\Delta_{R} = \frac{\mathbf{W}_i-\mathbf{W}_{i-1}}{\mathbf{W}_{i+1}-\mathbf{W}_{i}+\epsilon}
\end{equation}
where $\epsilon$ is a vanishingly small number used to avoid division by zero.
The function $\Phi(\boldsymbol\Delta)$  represents the slope limiter, which
prevents spurious oscillations and ensures that the Total Variation Diminishing
(TVD) property is maintained.
In the examples presented here the limiter of Van-Leer~\cite{vanleer97_iii} is used
\begin{equation}
\Phi(\boldsymbol\Delta) = \max\left(0,\frac{\boldsymbol\Delta+|\boldsymbol\Delta|}{1+|\boldsymbol\Delta|+\epsilon}\right)
\end{equation}
The extrapolation in time is carried out using
\begin{align}
    \mathbf{W}^{n+1/2}_{i-1/2,R} &= \mathbf{W}^{n}_{i-1/2,R} - \frac{\Delta t}{\Delta x} \mathbf{A}_x(\mathbf{W}^n_i)
    \left[\mathbf{W}^n_{i} - \mathbf{W}^{n}_{i-1/2,R}\right] \\
    \mathbf{W}^{n+1/2}_{i+1/2,L} &= \mathbf{W}^{n}_{i+1/2,L} - \frac{\Delta t}{\Delta x} \mathbf{A}_x(\mathbf{W}^n_i) 
    \left[\mathbf{W}^{n}_{i+1/2,L} - \mathbf{W}^n_i\right]
\end{align}

We note that in \cite{barton19} material interfaces were sharpened following
MUSCL reconstruction using the algebraic Tangent of Hyperbola INterface
Capturing (THINC) technique from Xiao~\textit{et~al.}~\cite{xiao05}, but that
this is unnecessary in our scheme as the interfaces are always confined to a
single cell by virtue of the specialised mixed-cell update. However it may still
be advantageous to apply the THINC technique to fields other than the material
volume-fractions. For example, Wallis~\textit{et~al.}~\cite{wallis21_wbn} use
THINC to sharpen the material damage field used for Continuum Damage Modelling
(CDM).

\subsection{Solution of Riemann problems}

\begin{figure}
    \centering
    \includegraphics{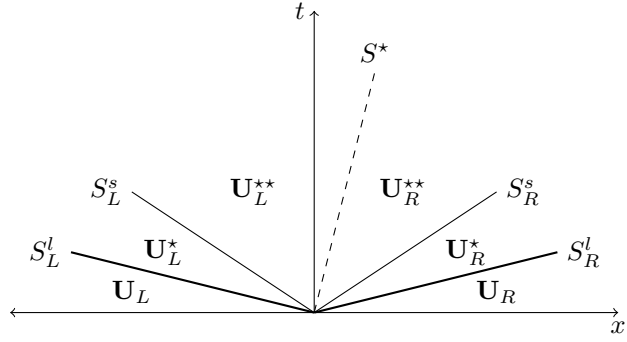}
    \caption{The HLLD solver assumes the solution to the Riemann problem
    contains six constant states interposed by five discontinuous waves: two
    fast longitudinal waves, two slow shear waves, and one central contact
    discontinuity.}
    \label{fig:hlld}
\end{figure}

\noindent The numerical fluxes and interfacial states are obtained using the
HLLD approximate Riemann solver, originally developed for magnetohydrodynamics
but repurposed for solid dynamics in \cite{lopezortega14,barton19}. The solution
consists of six constant states interposed by five waves (see
Figure~\ref{fig:hlld}). Given the reconstructed left and right states
$\mathbf{U}_L(\mathbf{W}^{n+\frac{1}{2}}_L)$ and
$\mathbf{U}_R(\mathbf{W}^{n+\frac{1}{2}}_R)$ as input, the fluxes are

\begin{equation}
    \label{eq:hlld-fluxes}
    \mathbf{F}_\mathrm{HLLD} = \begin{cases}
        \mathbf{F}(\mathbf{U}_L) \, &\text{if} \quad S^l_L \geq 0 \\
        \mathbf{F}(\mathbf{U}_L) + S^l_L(\mathbf{U}^\star_L - \mathbf{U}_L) \, &\text{if} \quad S^l_L < 0 \leq S^s_L \\
        \mathbf{F}(\mathbf{U}_L) + S^l_L(\mathbf{U}^\star_L - \mathbf{U}_L) + S^s_L(\mathbf{U}^{\star\star}_L - \mathbf{U}^\star_L) \, &\text{if} \quad S^s_L < 0 \leq S^\star \\
        \mathbf{F}(\mathbf{U}_R) + S^l_R(\mathbf{U}^\star_R - \mathbf{U}_R) + S^s_R(\mathbf{U}^{\star\star}_R - \mathbf{U}^\star_R) \, &\text{if} \quad S^\star < 0 \leq S^s_R \\
        \mathbf{F}(\mathbf{U}_R) + S^l_R(\mathbf{U}^\star_R - \mathbf{U}_R) \, &\text{if} \quad S^s_R < 0 \leq S^l_R \\
        \mathbf{F}(\mathbf{U}_R) \, &\text{if} \quad S^l_R < 0 \\
    \end{cases}
\end{equation}

\noindent where the left longitudinal, left shear, contact, right shear, and
right longitudinal wave speeds are given respectively by

\begin{align}
    S^l_L &= \mathrm{min}(u_L - c_L, u_R - c_R) \\
    S^s_L &= S^\star - b^\star_L \\
    S^\star &= \frac{\rho_L u_L (S^l_L - u_L) - \rho_R u_R (S^l_R - u_R) + \sigma_{L,11} - \sigma_{R,11}}{\rho_L(S^l_L - u_L) - \rho_R(S^l_R - u_R)} \\
    S^s_R &= S^\star + b^\star_R \\
    S^l_R &= \mathrm{max}(u_L + c_L, u_R + c_R)
\end{align}

\noindent Note that in fluids, $b=0$ and therefore $S^s_L = S^\star = S^s_R$ so
the solver reduces to the standard HLLC scheme.

The intermediate states $\mathbf{U}^\star_L$, $\mathbf{U}^{\star\star}_L$,
$\mathbf{U}^{\star\star}_R$ and $\mathbf{U}^\star_R$ are obtained through
manipulation of the jump conditions

\begin{align*}
    S^l_K (\mathbf{U}_K - \mathbf{U}_K^\star) &= \mathbf{F}(\mathbf{U}_K) - \mathbf{F}(\mathbf{U}_K^\star) \numberthis \label{eq:rankine-hugoniot1} \\
    S^s_K (\mathbf{U}_K^\star - \mathbf{U}_K^{\star\star}) &= \mathbf{F}(\mathbf{U}_K^\star) - \mathbf{F}(\mathbf{U}_K^{\star\star}) \numberthis \label{eq:rankine-hugoniot2}
\end{align*}

\noindent It is assumed that normal stress and normal velocity are
constant across shear waves, but may jump across longitudinal waves. Similarly,
the transverse stresses and velocities may jump across shear waves but are
assumed constant across longitudinal waves. Solving \eqref{eq:rankine-hugoniot1}
under these assumptions gives the states between the two pairs of longitudinal
and shear waves

\begin{equation}
    \label{eq:star-states}
    \mathbf{U}^\star_K = \chi_K \begin{pmatrix}
        \rho^l_K \phi^l_K \\
        \phi^l_K \\
        \rho^l_K S^\star \\
        \rho^l_K v_K \\
        \rho^l_K w_K \\
        \rho^l_K E_K + (S^\star - u_K) (\rho_K S^\star - \sigma_{K,11} / (S^l_K - u_K)) \\
        \overline{V}_{e,K,11} / \chi_K \\
        \overline{V}_{e,K,21} \\
        \overline{V}_{e,K,31} \\
        \overline{V}_{e,K,12} / \chi_K \\
        \overline{V}_{e,K,22} \\
        \overline{V}_{e,K,32} \\
        \overline{V}_{e,K,13} / \chi_K \\
        \overline{V}_{e,K,23} \\
        \overline{V}_{e,K,33} \\
        \rho^l_K \phi^l_K \pi^l_{K,m} \\
    \end{pmatrix}
\end{equation}

\noindent where

\begin{equation}
    \label{eq:chi}
    \chi_K = \frac{S^l_K - u_K}{S^l_K - S^\star}
\end{equation}

\noindent Between the two shear waves no-slip  boundary conditions---constant
transverse stresses and transverse velocities---are used to describe how the
solution varies across the contact wave, which reflects the underpinning
mechanical equilibrium of the model.  The intermediate states then become

\begin{equation}
    \label{eq:two-star-states}
    \mathbf{U}^{\star\star}_K = \mathbf{U}^\star_K + \frac{1}{S^\star - S^s_K} \begin{pmatrix}
        0 \\
        0 \\
        0 \\
        \sigma^{\star\star}_{K,21} - \sigma_{K,21} \\
        \sigma^{\star\star}_{K,31} - \sigma_{K,31} \\
        v^{\star\star}_K \sigma^{\star\star}_{K,21} - v_K \sigma_{K,21} + w^{\star\star}_K \sigma^{\star\star}_{K,31} - w_K \sigma_{K,31} \\
        0 \\
        \overline{V}^\star_{e,K,11} (v^{\star\star}_K - v_K) \\
        \overline{V}^\star_{e,K,11} (w^{\star\star}_K - w_K) \\
        0 \\
        \overline{V}^\star_{e,K,12} (v^{\star\star}_K - v_K) \\
        \overline{V}^\star_{e,K,12} (w^{\star\star}_K - w_K) \\
        0 \\
        \overline{V}^\star_{e,K,13} (v^{\star\star}_K - v_K) \\
        \overline{V}^\star_{e,K,13} (w^{\star\star}_K - w_K) \\
        0 \\
    \end{pmatrix}
\end{equation}

\noindent We also require the intermediate stresses which are given by

\begin{align*}
    \sigma_{11,K}^\star &= \sigma_{11,K} - \rho_K(S^l_K - u_K)(S^\star - u_K) \numberthis \label{eq:star-state-stress1} \\
    \sigma_{j1,K}^\star &= \sigma_{j1,K} \quad \mathrm{for} \; j=2,3 \numberthis \label{eq:star-state-stress2} \\
    \sigma_{11,K}^{\star\star} &= \sigma_{11,K}^\star \numberthis \label{eq:star-state-stress3} \\
    \sigma_{j1,K}^{\star\star} &= \frac{\alpha_L \alpha_R (u_{j,L} - u_{j,R}) + \alpha_L \sigma_{j1,R} - \alpha_R \sigma_{j1,L}}{\alpha_L - \alpha_R} \quad \mathrm{for} \; j=2,3 \numberthis \label{eq:star-state-stress4}
\end{align*}

\noindent where

\begin{equation*}
    \alpha_K = \rho^\star_K(S^\star - S^s_K)
\end{equation*}

The fluxes $\mathbf{F}_\mathrm{HLLD}$ are used directly to define the cell wall fluxes $\mathbf{F}_{k,i\pm1/2}$ to update
single-material cells following the standard MUSCL-Hancock method. In order to
perform the mixed-cell update the fluxes are not needed, but we do require the
states evaluated at the interface

\begin{equation}
    \label{eq:intermediate-states}
    \mathbf{U}^\mathrm{int} = \begin{cases}
        \mathbf{U}_L \, &\text{if} \quad S^l_L \geq 0 \\
        \mathbf{U}^\star_L \, &\text{if} \quad S^l_L < 0 \leq S^s_L \\
        \mathbf{U}^{\star\star}_L \, &\text{if} \quad S^s_L < 0 \leq S^\star \\
        \mathbf{U}^{\star\star}_R \, &\text{if} \quad S^\star < 0 \leq S^s_R \\
        \mathbf{U}^\star_R \, &\text{if} \quad S^s_R < 0 \leq S^l_R \\
        \mathbf{U}_R \, &\text{if} \quad S^l_R < 0 \\
    \end{cases}
\end{equation}

\noindent It is important to note that where the interfacial stresses are required
(e.g. in \eqref{eq:momentum-update}), these must be taken directly from
\eqref{eq:star-state-stress1}--\eqref{eq:star-state-stress4}. Evaluating them
from $\mathbf{U}^\mathrm{int}$ using the equation of state is invalid and will
cause the scheme to fail.

\subsection{Interface reconstruction}

\noindent Material interfaces are reconstructed using the method of
Youngs~\cite{youngs82,pilliod04}. In each cell, and for each material, the
gradient of the volume fraction field is approximated using finite differences
and an outward-facing normal vector is calculated

\begin{equation}
    \mathbf{n}^l = -\frac{\nabla \phi^l}{|\nabla \phi^l|}
\end{equation}

\noindent An oriented plane associated with this normal is then positioned such
that it cuts the cell producing the required volume fraction. This plane is
taken as a linear approximation of the interface within the cell. In cells with
two materials, the second material interface is simply the first with the sign
of the normal vector reversed. In cells with three or more materials, producing
an accurate reconstruction is more complicated and has been studied by many
authors~\cite{benson02,kucharik10,shashkov23}. This problem is outside the
scope of the current work, and we adopt a simple approach. The user of the code
specifies a global ordering of materials, and interfaces are computed between
adjacent pairs in this ordering using a cumulative sum of the individual volume
fraction fields. For instance, with four materials, we would compute the
interface of material one using $\phi^1$, material two using  $\phi^1+\phi^2$
and material three using $\phi^1+\phi^2+\phi^3$. The interface of material four
would be the same as material three with the normal sign reversed. 

\begin{figure}
    \centering
    \includegraphics{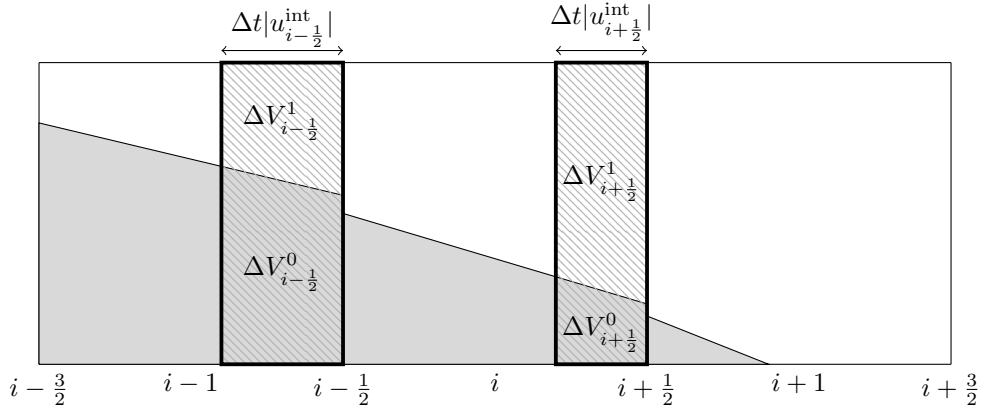}

    \caption{Illustration of the calculation of transported material volumes at
    each face. The filled regions show the reconstructed interface between two
    materials. The hatched regions denote the advected volumes, with widths given
    in terms of the interfacial speeds. The transported material volumes consist
    of the intersection of these regions with each material.}%

    \label{fig:youngs}
\end{figure}

Once we have the required material interfaces defined, we require the signed
transported volumes of each material across the left and right faces of the cell
(in the sweep direction), $\Delta V^l_{i-1/2}$ and $\Delta V^l_{i+1/2}$
respectively. These are computed as the intersection of the material polyhedra
underneath the interfaces and the cuboidal region traced out by the motion of
the continuum normal to and through the cell face by a distance $\Delta t \,
u^{\mathrm{int}}$, with sign equivalent to $u^\mathrm{int}$. In the
$x$-direction this cuboid has the signed volume $\Delta y \Delta z \Delta t \,
u^\mathrm{int}$, and the sum of the transported volumes $\sum_l \Delta V^l$ must
exactly equal this value at each face. A 2D illustration of this process is
given in Figure~\ref{fig:youngs}.

\subsection{Mixed-cell update}

\noindent The specialised mixed-cell update must be applied in all cells
containing more than one material, but additionally in cells that could become
mixed following the update. In practice this means all pure cells that are
adjacent to a mixed-cell in the direction of the sweep, or two adjacent pure
cells containing different materials (in the case where the material interface
coincides exactly with the adjoining face). Even in the most demanding
multi-material calculations this will typically be less than $10\%$ of the
total cell count, so the computational burden of this process is small.

Following interface reconstruction, the intermediate volume fractions
(neglecting contributions from the right-hand side of \eqref{eq:vof}) are
computed using volume-of-fluid advection

\begin{equation}
    \label{eq:vof-tilde-update}
    \tilde{\phi}^l = \phi^{l,n} - \frac{\Delta V^l_R - \Delta V^l_L}{V}
\end{equation}

\noindent where $V$ is the volume of the cell. In this section the subscripts
$L$ and $R$ are used as short-hand for $i-\frac{1}{2}$ and $i+\frac{1}{2}$
respectively to denote quantities located at the left and right faces of the
cell under consideration.

Following application of \eqref{eq:vof-tilde-update}, the sum of the volume
fractions $\tilde{\phi}^l$ in the cell may not equal one (whenever the velocity
divergence is non-zero). The difference ($1 - \sum_m \tilde{\phi}^m$) must be
redistributed amongst the materials in the cell in accordance with their
relative compressibility. To compute the right-hand side of \eqref{eq:vof} we
require the bulk modulus for each material following advection, denoted
$\tilde{K}_S^l$, which we obtain through a volume-weighted averaging
procedure as detailed in \cite{millerpuckett96,cutforth21}

\begin{equation}
    \tilde{K}^{l}_{S,i} = \frac{\Delta V^l_L \widehat{K}^l_{S,L} + \phi^l_i V_i K^l_{S,i} - \Delta V^l_R \widehat{K}^l_{S,R}}{\tilde{\phi}^l_i V_i}
\end{equation}

\noindent where the upwind bulk moduli at the left and right faces are given by

\begin{align*}
    \widehat{K}^{l}_{S,L} &= \begin{cases}
        K^l_{S,i-1} \, &\text{if} \quad u^\mathrm{int}_{i-\frac{1}{2}} > 0 \\
        K^l_{S,i} \, &\text{otherwise}
    \end{cases} \numberthis\label{eq:upwind-l} \\
    \widehat{K}^{l}_{S,R} &= \begin{cases}
        K^l_{S,i} \, &\text{if} \quad u^\mathrm{int}_{i+\frac{1}{2}} > 0 \\
        K^l_{S,i+1} \, &\text{otherwise}
    \end{cases} \numberthis\label{eq:upwind-r} 
\end{align*}

\noindent The post-advection bulk modulus of the mixture is then

\begin{equation}
    \widetilde{K}_S = \left( \sum_l \frac{\tilde{\phi}^{l}}{\tilde{K}^{l}_S}\right)^{-1}
\end{equation}

\noindent Note that the same approach is used to determine the post-advection
Gr\"uneisen parameters $\tilde{\Gamma}^l$ and $\widetilde{\Gamma}$ used in the
internal energy equations below. The final volume fractions are then given by

\begin{equation}
    \label{eq:vof-update}
    \phi^{l,n+1} = \tilde{\phi}^l + (1 - \sum_m \tilde{\phi}^m)
    \frac{\tilde{\phi}^l \widetilde{K}_S}{\tilde{K}_S^l}
\end{equation}

\noindent where the bracketed expression may be identified as the time integral
of the velocity divergence

\begin{equation}
    \label{eq:div-u}
    (1 - \sum_m \tilde{\phi}^m) = \Delta t \frac{u_R^\mathrm{int} - u_L^\mathrm{int}}{\Delta x}
\end{equation}

The mass, momentum and energy transported across the face of each mixed-cell are
computed using upwinding. This is important to ensure that the procedure is
self-consistent: when the cell is emptied of material $l$, the corresponding
contributions to mass, momentum and energy must be set to zero.  The relevant
upwind quantities (denoted with a wide hat $\widehat{q}$) are calculated at the
left and right faces of the cell as in \eqref{eq:upwind-l} and
\eqref{eq:upwind-r} respectively. We also require the transported density of the
mixture at each face, given by

\begin{equation}
    \widehat{\rho}_K = \frac{\sum_l \widehat{\rho}^l_K \Delta V^l_K}{\sum_l
    \Delta V^l_K}, \quad \mathrm{for} \: K=L, R
\end{equation}

The discrete update equations for the partial densities, momenta, elastic
stretches and history variables are as follows. The advective terms are
differenced in line with \eqref{eq:vof-tilde-update}. Note that the inelastic
and history-rate source terms are omitted as these are evaluated separately.

\begin{align*}
    \phi^{l,n+1} \rho^{l,n+1} = &\; \phi^{l,n} \rho^{l,n} -
        \frac{\Delta V^l_R \widehat{\rho}^{l}_R - \Delta V^l_L \widehat{\rho}^{l}_L}{V}
        \numberthis \label{eq:density-update} \\
    \rho^{n+1} u^{n+1}_i = &\; \rho^n u^n_i -
        \frac{\Delta V_R \widehat{\rho}_R \widehat{u}_{i,R} - \Delta V_L \widehat{\rho}_L \widehat{u}_{i,L}}{V} +
        \Delta t \frac{\sigma^\mathrm{int}_{i1,R} - \sigma^\mathrm{int}_{i1,L}}{\Delta x}
        \numberthis \label{eq:momentum-update} \\
    \overline{V}_{e,ij}^{n+1} = &\; \overline{V}_{e,ij}^n -
    \frac{\Delta V_R \widehat{\overline{V}}_{e,ij,R} -\Delta V_L
    \widehat{\overline{V}}_{e,ij,L}}{V} + \frac{2}{3} \overline{V}_{e,ij}^n (1 - \sum_l \tilde{\phi}^l) \; + \numberthis \label{eq:stretch-update} \\
    &\; \frac{\Delta t}{\Delta x} \left[
        (u^n_j - u^{\mathrm{int}}_{j,L}) \overline{V}^\mathrm{int}_{e,1j,L} +
        (u^{\mathrm{int}}_{j,R} - u^n_j) \overline{V}^\mathrm{int}_{e,1j,R} \right] \\
    \phi^{l,n+1} \rho^{l,n+1} \pi^{l,n+1}_m = &\; \phi^{l,n} \rho^{l,n}
    \pi^{l,n}_m -
        \frac{\Delta V^l_R \widehat{\rho}^{l}_R \widehat{\pi}^{l}_{m,R} - \Delta V^l_L
        \widehat{\rho}^{l}_L \widehat{\pi}^{l}_{m,L}}{V}
        \numberthis \label{eq:history-update}
\end{align*}

\noindent The momentum update \eqref{eq:momentum-update} is importantly not the
same as the one from~\cite{millerpuckett96}, which used the interfacial velocity
rather than the upwinded velocity in the advective terms. We found the original
to be highly unstable when a dense material is leaving a cell containing a light
material, as the dense material's contribution to the momentum is not correctly
set to zero. This resulted in large spikes in velocity which caused calculations
to fail. The new update does not exhibit these problems.

The discrete energy update based on \eqref{eq:internal-energy} is given by

\begin{align*}
    \phi^{l,n+1} \rho^{l,n+1} \mathscr{E}^{l,n+1} &= \, \phi^{l,n} \rho^{l,n} \mathscr{E}^{l,n}
    - \Delta_{\mathrm{adv}} -
    p^{l,\bigstar} \Delta_{\mathrm{vol}} +
    \Delta_{\mathrm{shear}} \numberthis \label{eq:energy-update} \\
    \Delta_{\mathrm{adv}} &= \frac{\Delta V^l_R \widehat{\rho}^{l}_R
    \widehat{\mathscr{E}}^{l}_R - \Delta V^l_L \widehat{\rho}^{l}_L \widehat{\mathscr{E}}^{l}_L}{V} \\
    \Delta_{\mathrm{vol}} &= (1 - \sum_m \tilde{\phi}^m) \frac{\tilde{\phi}^l
    \widetilde{\Gamma}}{\tilde{\Gamma}^l} \\
    \Delta_{\mathrm{shear}} &= 2 G^{l,\bigstar} \left[
        \sum_i \mathrm{dev}(\mathbf{H}_e)^\bigstar_{i1} \Delta t \frac{u^\mathrm{int}_{i,R} - u^\mathrm{int}_{i,L}}{\Delta x}
        \right] \frac{\tilde{\phi}^l \widetilde{\Gamma}}{\tilde{\Gamma}^l}
\end{align*}

\noindent Cutforth~\textit{et~al.} presented an energy update which reportedly
improved robustness significantly in complex simulations~\cite{cutforth21}.
Following their approach, the time-level of the pressure $p^{l,\bigstar}$ in the
discrete energy equation is taken to differ depending on whether the cell is in
compression or expansion:

\begin{equation}
    \label{eq:energy-update-stress}
    p^{l,\bigstar} = \begin{cases}
        p^{l,n} \, &\text{if} \quad u^\mathrm{int}_R - u^\mathrm{int}_L < 0 \\
        p^{l,n+1} \, &\text{otherwise}
    \end{cases}
\end{equation}

\noindent In order to ensure a consistent evaluation of the stress, the same
logic is applied when chosing the time-level of the shear modulus $G^{l,\bigstar}$
and the deviatoric strain $\mathrm{dev}(\mathbf{H}_e)^\bigstar$.

In the case of compression the update may be readily evaluated. In the implicit
case (expansion), we invoke the equation of state \eqref{eq:pressure} to
rearrange the right-hand side

\begin{equation}
    \label{eq:energy-update-implicit}
    \phi^{l,n+1} \rho^{l,n+1} \mathscr{E}^{l,n+1} = \frac
        {\phi^{l,n} \rho^{l,n} \mathscr{E}^{l,n} - \Delta_\mathrm{adv} + \Delta_{\mathrm{vol}} (\rho^{l,n+1} \Gamma^{l,n+1}
        \mathscr{E}^{l,n+1}_{\mathrm{ref}} - p^{l,n+1}_{\mathrm{ref}}) +
        \Delta_{\mathrm{shear}}}{1 + \Delta_{\mathrm{vol}} \Gamma^{l,n+1} /
        \phi^{l,n+1}}
\end{equation}

\noindent Then all of the variables on the right-hand side of
\eqref{eq:energy-update-implicit} are known: $\phi^{l,n+1}$ is given by
\eqref{eq:vof-update}, $\rho^{l,n+1}$ is given by \eqref{eq:density-update},
$\mathrm{dev}(\mathbf{H}_e)^{n+1}$ is given by \eqref{eq:stretch-update}, and
$\Gamma^{l,n+1}$, $\mathscr{E}^{l,n+1}_{\mathrm{ref}}$,
$p^{l,n+1}_{\mathrm{ref}}$ and $G^{l,n+1}$ are all assumed to depend only on
$\rho^{l,n+1}$ and $\mathrm{dev}(\mathbf{H}_e)^{n+1}$.

In the case of other equations of state where a closed-form rearrangement is not
possible (for instance reactive mixtures along the lines of~\cite{wallis21}, or
cavitating liquids along the lines of~\cite{wilkinson21}), a root-finding
procedure may be used to perform the implicit update.

\subsection{Pressure relaxation}

\noindent After the update mixed-cells will not in general be in pressure
equilibrium, which is a requirement of the base diffuse-interface method. A
pressure relaxation step is therefore included to iteratively adjust
volume-fractions and internal energies until equilibrium is attained. We first
outline the pressure relaxation scheme proposed by Miller and Puckett
in~\cite{millerpuckett96}, and then describe changes we have made to improve
robustness.

The following equations are solved to adjust the component pressures in
mixed-cells towards equilibrium in a controlled manner.

\begin{align}
    \overline{p} = p^l + \Delta p^l &= p^l - \frac{K^l_S}{\phi^l} \Delta \phi^l \\
    \sum_l \Delta \phi^l &= 0
\end{align}

\noindent These yield approximate expressions for the equilibrium pressure
$\overline{p}$ and the required change in volume-fraction for each material $\Delta
\phi^l$

\begin{align}
    \overline{p} &= \frac{\sum_l \phi^l p^l/K^l_S}{\sum_l \phi^l/K^l_S} \numberthis \label{eq:prelax-pbar} \\
    \Delta \phi^l &= \frac{\phi^l}{K^l_S} (p^l - \overline{p}) \numberthis \label{eq:prelax-vof}
\end{align}

\noindent Due to the linearisation implicit in \eqref{eq:prelax-vof}, Miller and
Puckett enforced a limit on the maximum relative volume change in a single
iteration of the relaxation, viz. $|\Delta \phi^l / \phi^l| \leq \delta$. When
the cell is under compression $\delta = 0.1$, and $\delta = 0.05$ in expansion.

On each iteration, \eqref{eq:prelax-pbar} is used to obtain an estimate of the
equilibrium pressure, which is then used in \eqref{eq:prelax-vof} to calculate
the amount by which each volume-fraction should change. If any of the values
exceed the threshold, all of them are rescaled (this ensures that they continue
to sum to zero)

\begin{equation}
    \Delta \phi^l \leftarrow \Delta \phi^l \min_l \frac{\delta}{|\Delta \phi^l / \phi^l|}
\end{equation}

\noindent Finally the volume-fractions are changed, and a $p\mathrm{d}v$
correction is made to each partial internal energy

\begin{align}
    \phi^l &\leftarrow \phi^l + \Delta \phi^l \\
    \rho^l \phi^l \mathscr{E}^l &\leftarrow \rho^l \phi^l \mathscr{E}^l -
    \overline{p} \Delta \phi^l
\end{align}

\noindent The process is repeated until each component pressure is within some
prescribed tolerance.

We found the above scheme to lack robustness, particularly within solids where
tiny changes in volume-fraction result in relatively large changes in pressure.
We observed oscillatory behaviour around the desired equilibrium point, where
the calculated volume-fraction changes would repeatedly overshoot the target
pressure, resulting in non-convergence. In order to address this we modify the
threshold adaptively.  Each time a sign-change is detected in $\Delta \phi^l$ we
reduce the threshold value $\delta$. The threshold is given by

\begin{equation}
    \label{eq:prelax-threshold}
    \delta = \delta_0 \mathrm{exp}(-\alpha N_\mathrm{sign})
\end{equation}

\noindent where $\delta_0$ is the initial threshold value, $\alpha$ is a free
parameter that controls the rate of fall-off, and $N_\mathrm{sign}$ is the
number of iterations in which a sign-change in $\Delta \phi^l$ has been
detected. We typically take $\delta_0 = 0.01$ and $\alpha = 0.1$.  Unlike Miller
and Puckett we do not distinguish between compression and expansion when
applying the threshold.

\subsection{Plastic update}

\noindent The plastic source term $\boldsymbol\Phi$ is evaluated as detailed by
Barton~\cite{barton19}. This update is independent of the volume-of-fluid method
and is performed identically in pure and mixed cells, but is summarised here for
completeness. First the following ordinary differential equation is solved using
the backwards Euler method to evolve the strain invariant for each material that
adheres to a plasticity law

\begin{equation}
    \label{eq:plastic-ode}
    \dot{\mathcal{J}}^{l} = -\sqrt{\frac{3}{2}} \chi^l, \quad \mathcal{J}^{l,\bullet} \in [0, \mathcal{J}^{l,\circ}]
\end{equation}

\noindent and the mixture strain invariant is obtained following

\begin{equation}
    \label{eq:J-mixture}
    \mathcal{J} = \frac{\sum_l \phi^l \mathcal{J}^l / \Gamma^l}{\sum_l \phi^l / \Gamma^l}
\end{equation}

\noindent where $\mathcal{J}^l \equiv 0$ in fluids. It is noted that since the
ODE only need be solved for materials where $\phi^l > 0$ the volume-of-fluid
method increases the efficiency of this update by virtue of there being fewer
mixed cells. Finally the mixture strain invariant at the new time level is used
to update the stretch tensor

\begin{equation}
    \overline{\mathbf{V}}^\bullet_e = \mathrm{exp} \left( \frac{\mathcal{J}^\bullet}{\mathcal{J}^\circ} \mathrm{dev}(\mathbf{H}^\circ_e) \right)
\end{equation}

\noindent In the course of this last step the symmetry and unimodularity of the
stretch tensor is reinstated, which may be lost during the hyperbolic update:
\begin{equation}
    \label{eq:stretch-symmetry}
    \overline{\mathbf{V}}_e \leftarrow \sqrt{\overline{\mathbf{V}}_e \overline{\mathbf{V}}_e^T}
\end{equation}

\section{Verification and validation}
\label{sec:results}
\subsection{Solid-solid Riemann problem}

\noindent The first test is a solid-solid Riemann problem taken from
\cite{barton09}, involving an interaction between aluminium and copper resulting
in a complex wave structure. The initial left and right states are as follows:

\begin{align*}
    \mathbf{u}_L &= \begin{pmatrix}
        2 \\
        0 \\
        0.1 \\
    \end{pmatrix} \mathrm{km\,s}^{-1}, \quad
    \mathbf{F}_{e,L} = \begin{pmatrix}
        1 & 0 & 0 \\
        -0.01 & 0.95 & 0.02 \\
        -0.015 & 0 & 0.9 \\
    \end{pmatrix}, \quad
    \mathscr{E}_{L} = \mathscr{E}_{\mathrm{ref},L} \\
    \mathbf{u}_R &= \begin{pmatrix}
        0 \\
        -0.03 \\
        -0.01 \\
    \end{pmatrix} \mathrm{km\,s}^{-1}, \quad
    \mathbf{F}_{e,R} = \begin{pmatrix}
        1 & 0 & 0 \\
        0.015 & 0.95 & 0 \\
        -0.01 & 0 & 0.9 \\
    \end{pmatrix}, \quad
    \mathscr{E}_{R} = \mathscr{E}_{\mathrm{ref},R}
\end{align*}

\noindent where $\mathbf{F}_e$ is the elastic deformation gradient tensor, from which the initial densities and elastic stretches are found via
\begin{align}
    \rho &= \frac{\rho_0}{\mathrm{det}|\mathbf{F}_e|} \\
    \overline{\mathbf{V}}_e &= \frac{\sqrt{\mathbf{F}_e \mathbf{F}^T_e}}{\sqrt[3]{\mathrm{det}|\mathbf{F}_e|}}
\end{align}

\noindent Both materials are assumed to be purely elastic and are governed by
the equation of state from~\cite{dorovskii83} with parameters given in
Table~\ref{tab:dorovskii-parameters}. The domain is $x \in [0, 1]$ with the
discontinuity initially located at $x=0.5$, with aluminium to the left and
copper to the right. The problem is run to time $0.5 \, \mu\mathrm{s}$ with CFL
number 0.1.

\begin{table*}
    \centering
    \begin{tabular}{lrrrrrr}
        \toprule
        Material & $\rho_0$ (g cm$^{-3}$) & $K_0$ (GPa) & $G_0$ (GPa) & $\Gamma_0$ & $\alpha$ & $\beta$ \\
        \midrule
        Aluminium 6061-T6 & 2.703 & 76.3 & 26.36 & 1.484 & 0.627 & 2.288 \\
        Copper & 8.930 & 136.45 & 39.38 & 2.0 & 1.0 & 3.0 \\
        Clay & 1.932 & 49.4 & 6.0 & 1.0 & 1.0 & - \\
        \bottomrule
    \end{tabular}

    \caption{Material parameters for the equation of state
    from~\cite{dorovskii83}, as given in Section~\ref{sec:closure-models}. Where
    $\beta$ is not specified a constant shear modulus is used: $G = G_0$.}%

    \label{tab:dorovskii-parameters}
\end{table*}

\begin{figure*}
    \centering
    \begin{subfigure}{0.5\textwidth}
        \centering
        \includegraphics[width=\textwidth,clip]{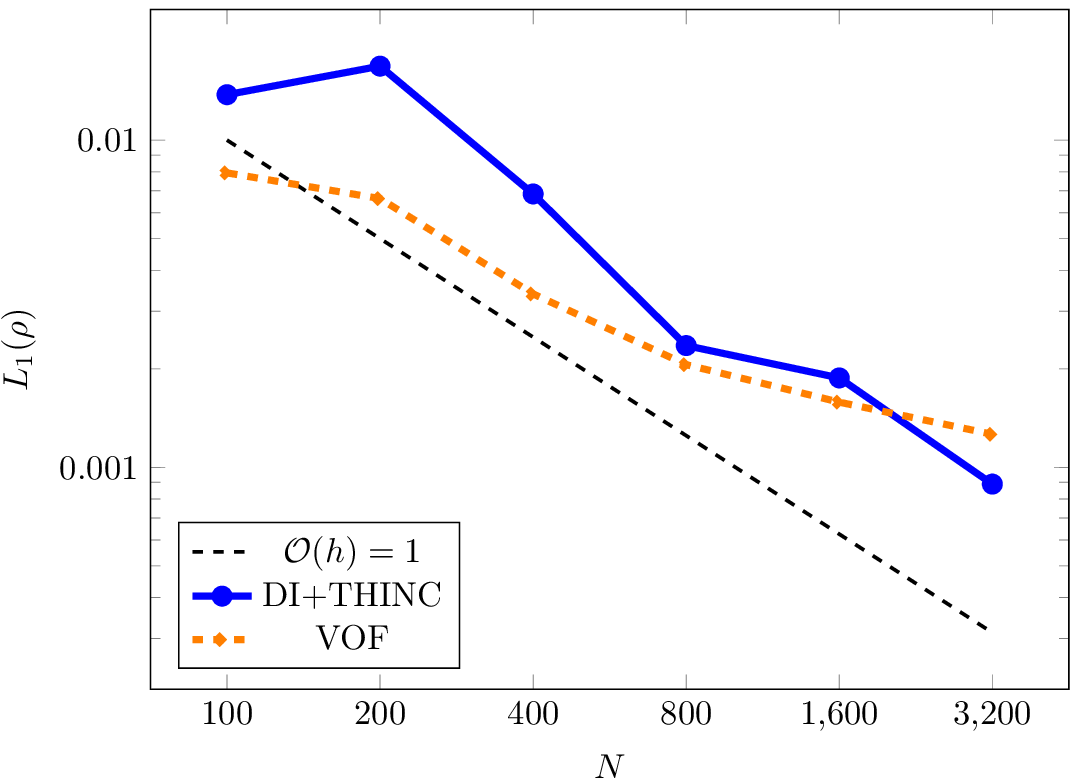}
        \caption{}%
        \label{fig:ssrp-l1-rho}
    \end{subfigure}%
    \begin{subfigure}{0.5\textwidth}
        \centering
        \includegraphics[width=\textwidth,clip]{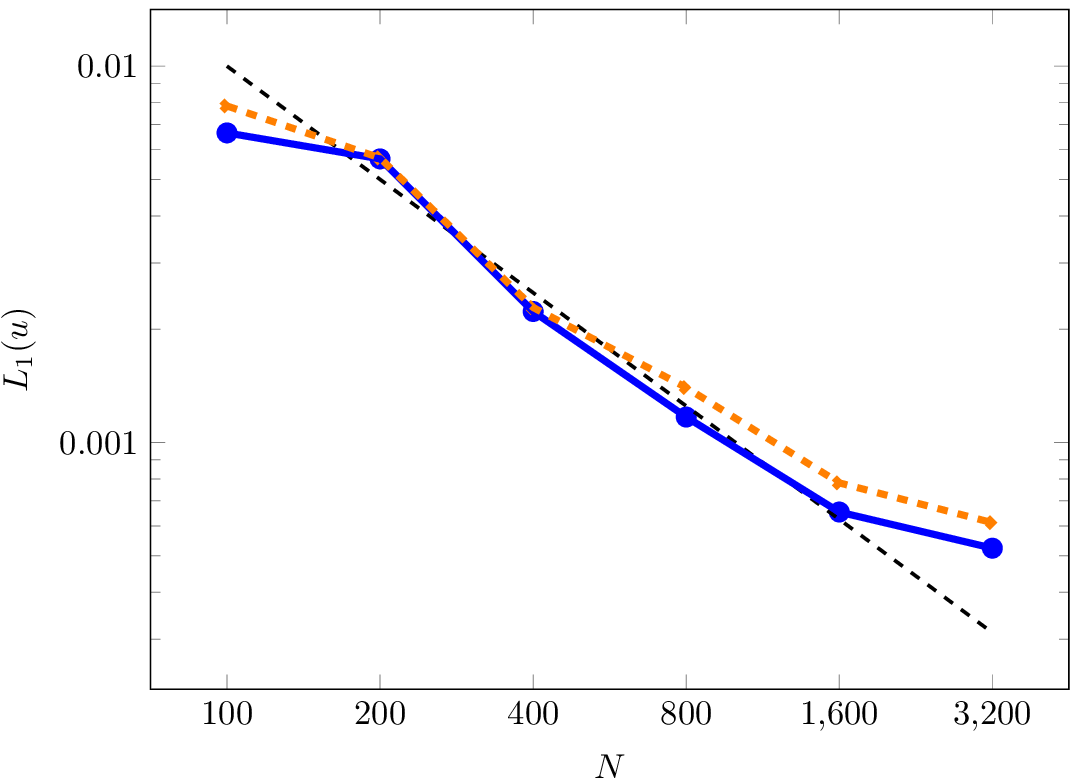}
        \caption{}%
        \label{fig:ssrp-l1-u}
    \end{subfigure}
    \begin{subfigure}{0.5\textwidth}
        \centering
        \includegraphics[width=\textwidth,clip]{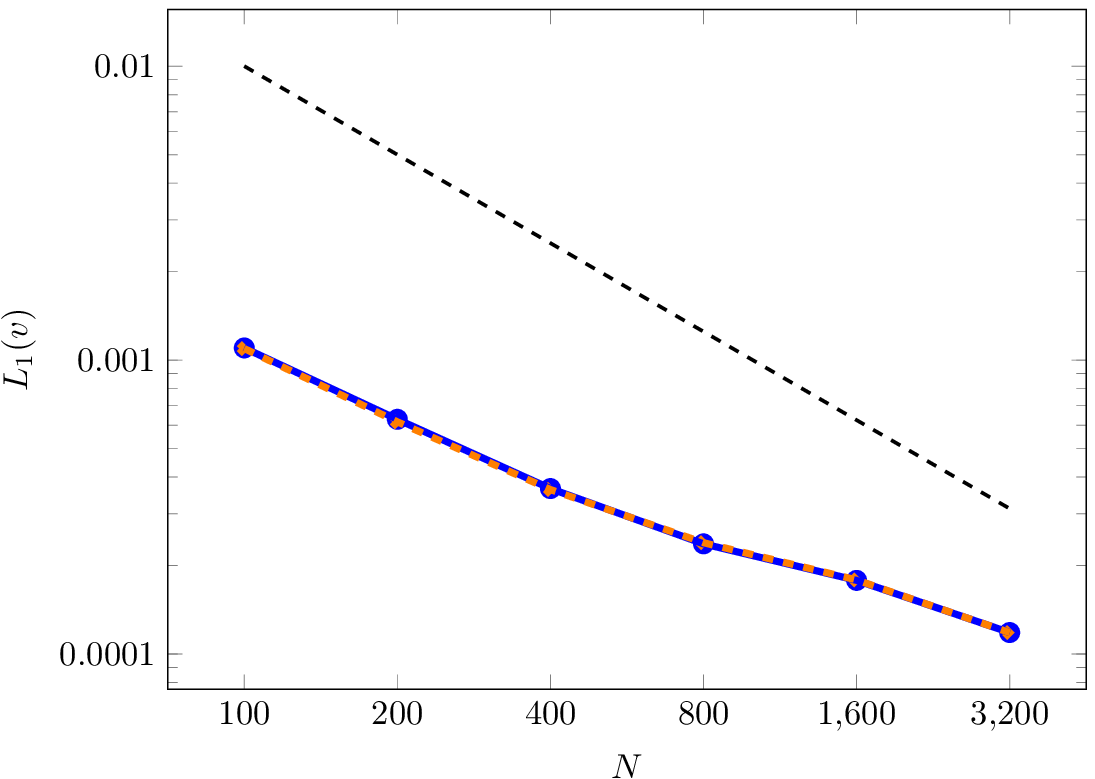}
        \caption{}%
        \label{fig:ssrp-l1-v}
    \end{subfigure}%
    \begin{subfigure}{0.5\textwidth}
        \centering
        \includegraphics[width=\textwidth,clip]{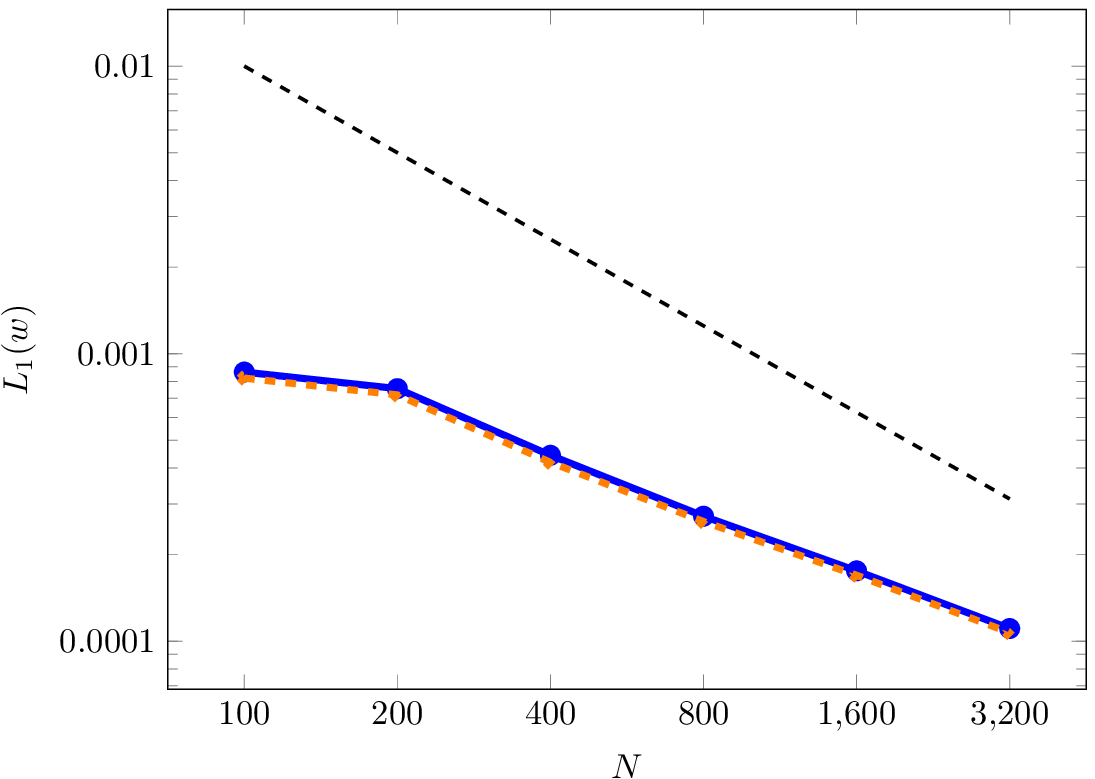}
        \caption{}%
        \label{fig:ssrp-l1-w}
    \end{subfigure}
    \caption{Solid-solid Riemann problem, $L_1$-norms versus exact solution.}%
    \label{fig:ssrp-l1}
\end{figure*}

Results are presented in Figure~\ref{fig:ssrp} for a mesh with 500 cells. The
volume-of-fluid method (labelled VOF) is compared against the base
diffuse-interface method with THINC interface sharpening~\cite{barton19}
(labelled DI+THINC) and the exact solution derived using the method
from~\cite{barton09}. Convergence is illustrated in Figure~\ref{fig:ssrp-l1},
which gives $L_1$-norms for density and velocity relative to the exact solution.

\begin{figure*}
    \centering
    \begin{subfigure}{0.5\textwidth}
        \centering
        \includegraphics[width=\textwidth,clip]{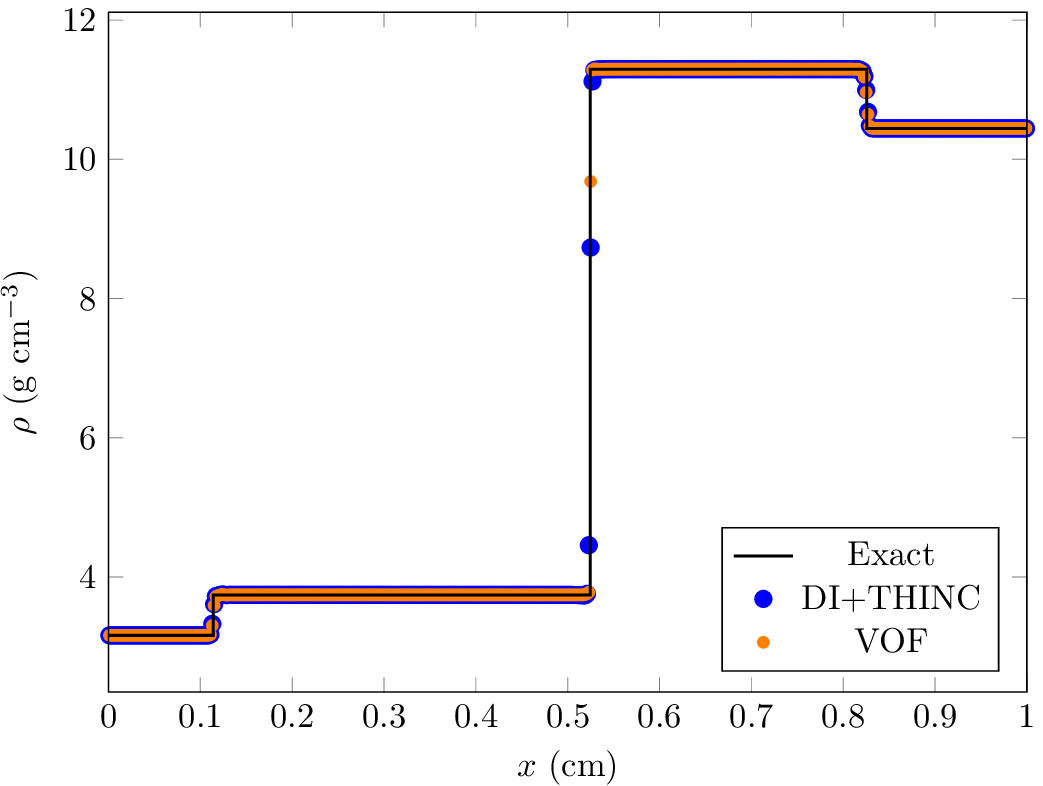}
        \caption{}%
        \label{fig:ssrp-rho}
    \end{subfigure}%
    \begin{subfigure}{0.5\textwidth}
        \centering
        \includegraphics[width=\textwidth,clip]{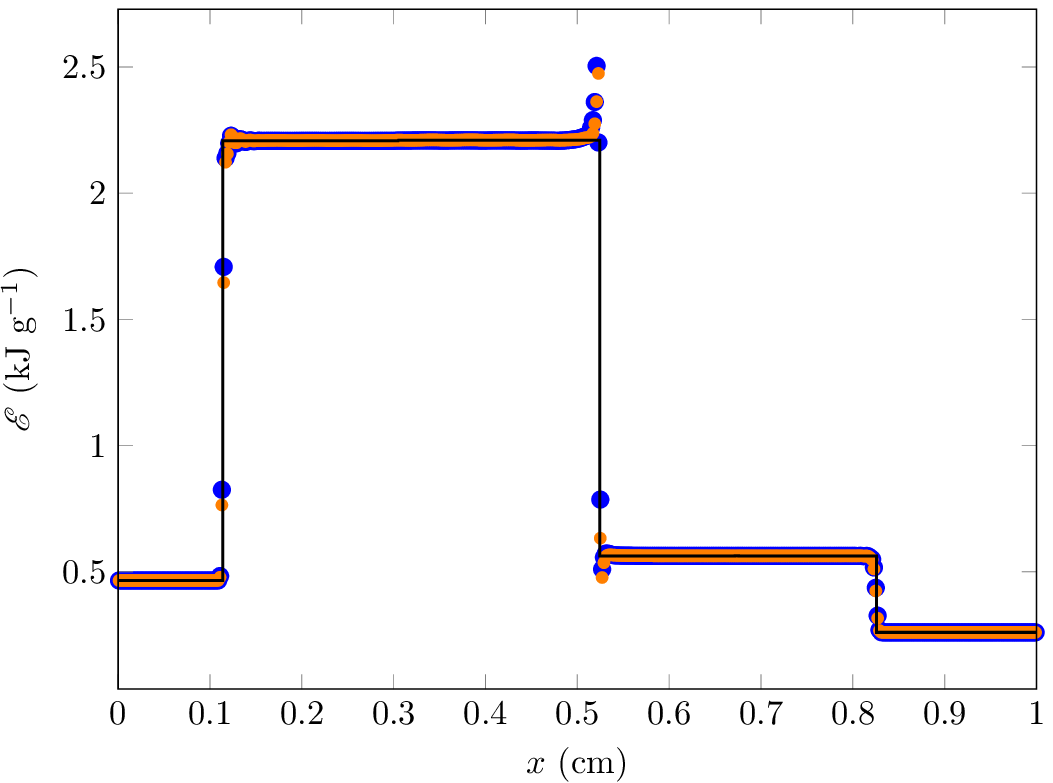}
        \caption{}%
        \label{fig:ssrp-ie}
    \end{subfigure}
    \begin{subfigure}{0.5\textwidth}
        \centering
        \includegraphics[width=\textwidth,clip]{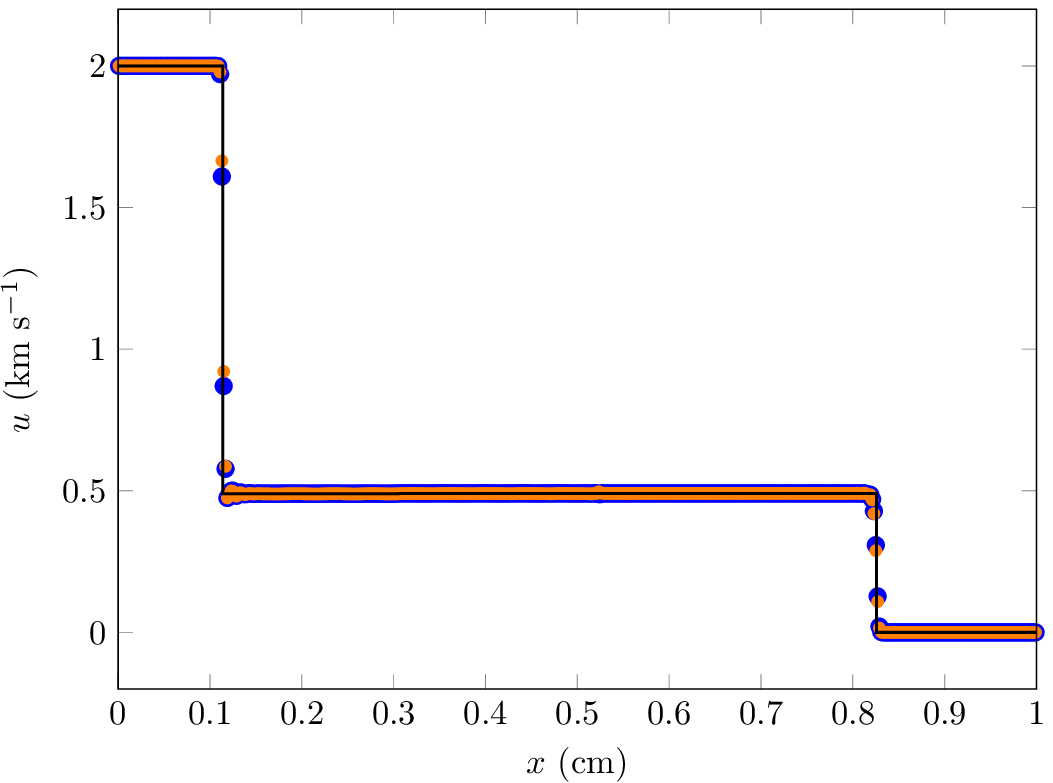}
        \caption{}%
        \label{fig:ssrp-u}
    \end{subfigure}%
    \begin{subfigure}{0.5\textwidth}
        \centering
        \includegraphics[width=\textwidth,clip]{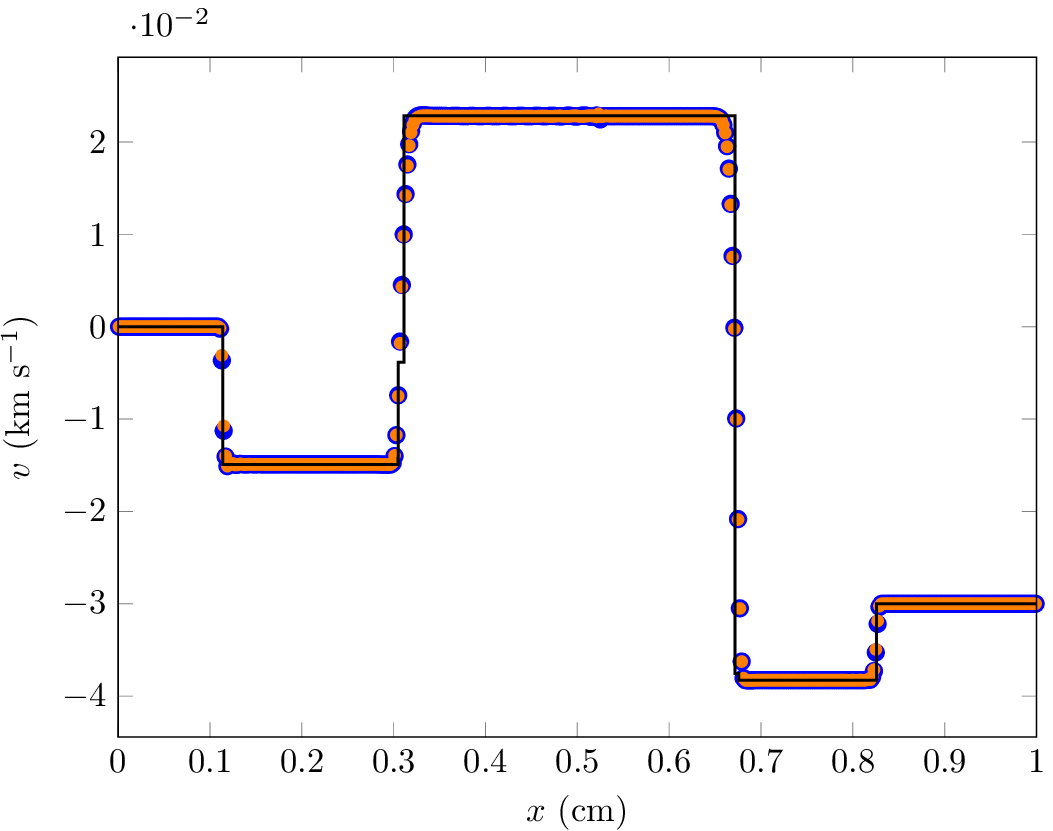}
        \caption{}%
        \label{fig:ssrp-v}
    \end{subfigure}
    \caption{Solid-solid Riemann problem.}%
\end{figure*}
\begin{figure*}\ContinuedFloat
    \begin{subfigure}{0.5\textwidth}
        \centering
        \includegraphics[width=\textwidth,clip]{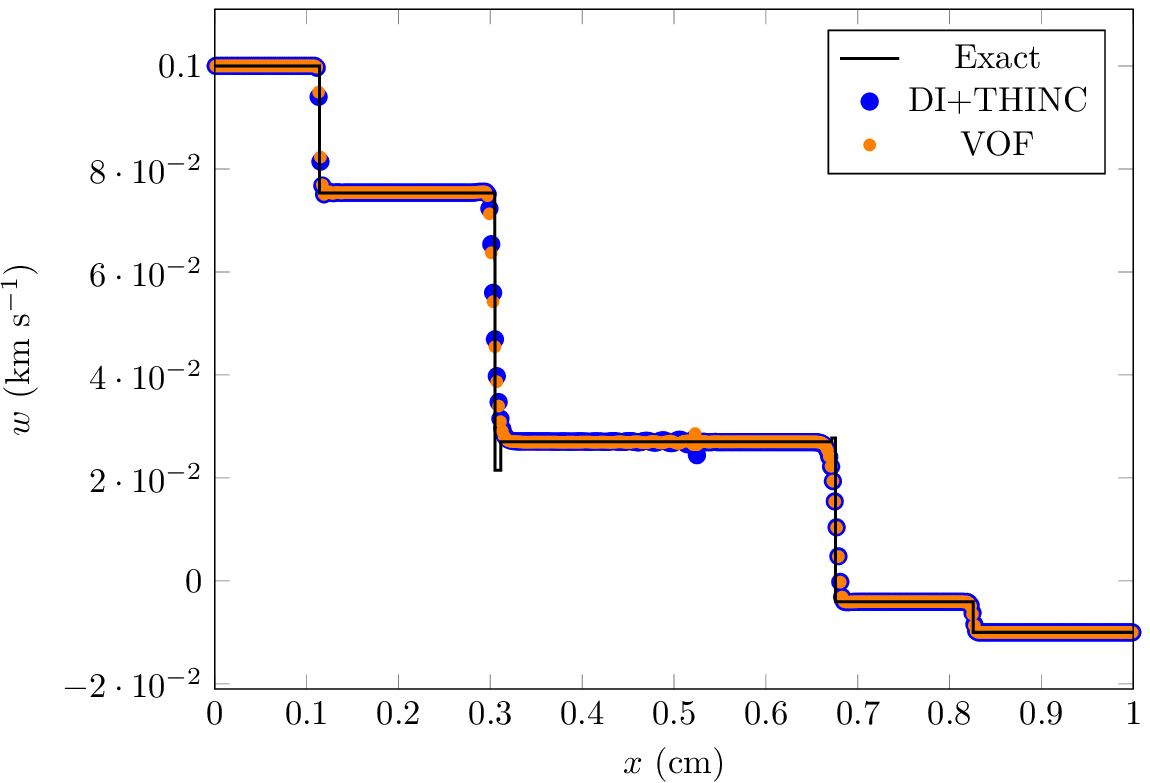}
        \caption{}%
        \label{fig:ssrp-w}
    \end{subfigure}%
    \begin{subfigure}{0.5\textwidth}
        \centering
        \includegraphics[width=\textwidth,clip]{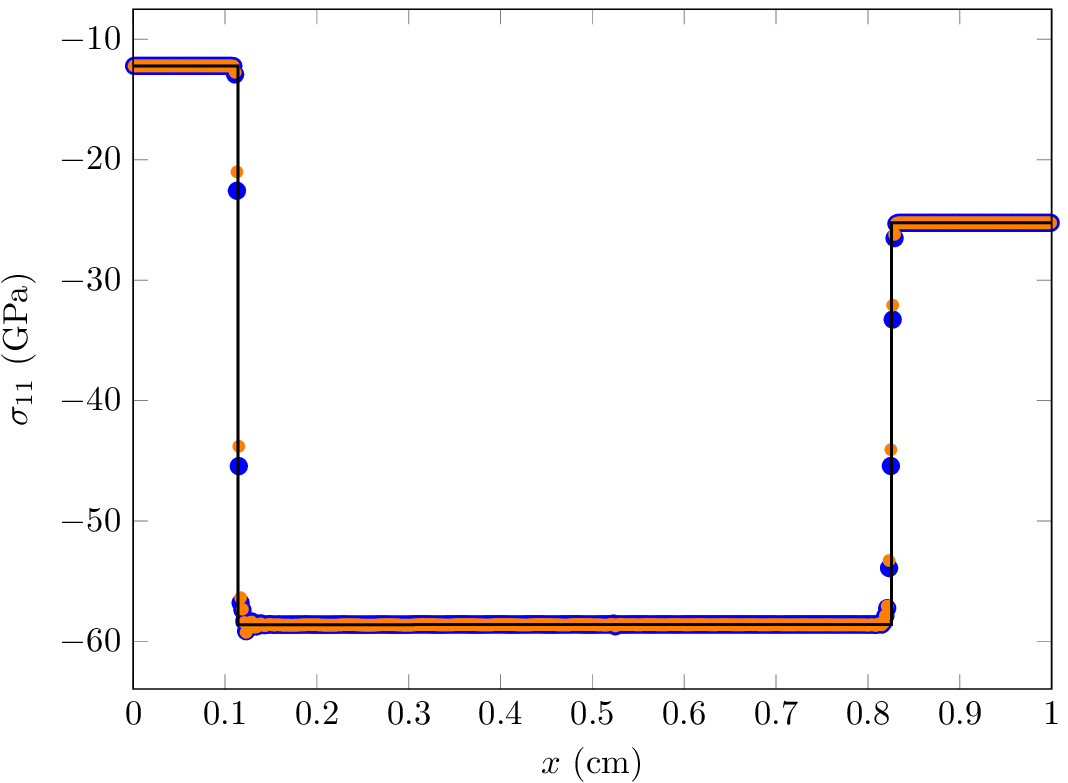}
        \caption{}%
        \label{fig:ssrp-sigma11}
    \end{subfigure}
    \begin{subfigure}{0.5\textwidth}
        \centering
        \includegraphics[width=\textwidth,clip]{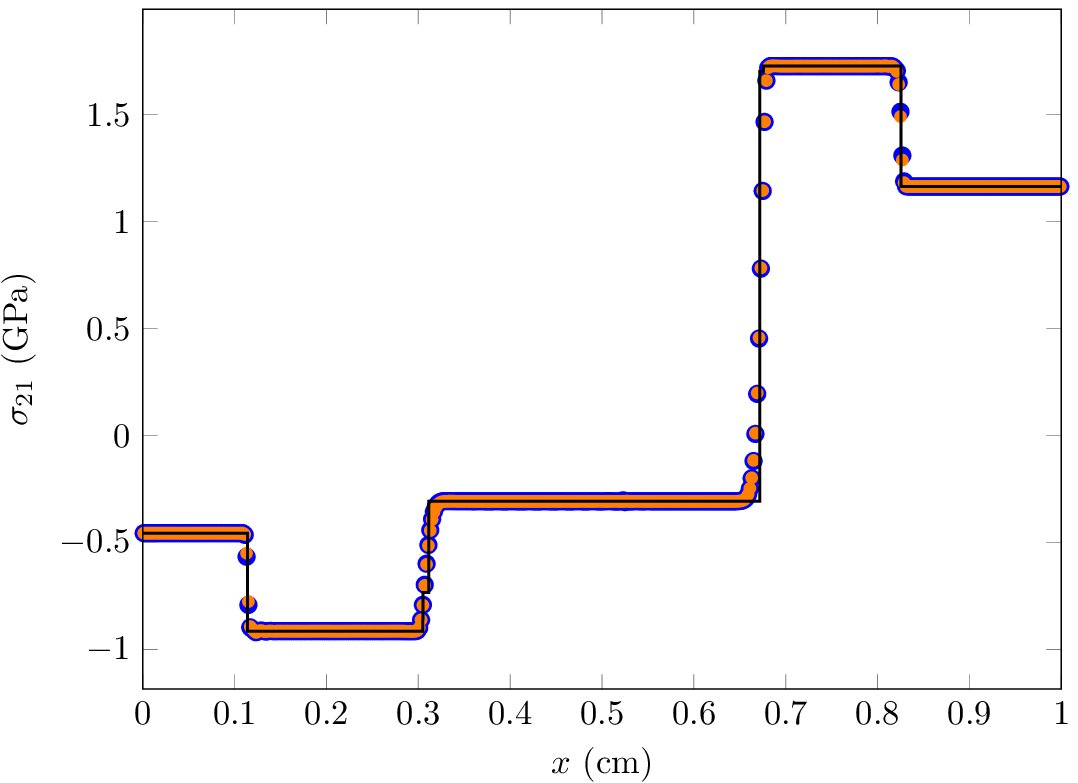}
        \caption{}%
        \label{fig:ssrp-sigma21}
    \end{subfigure}%
    \begin{subfigure}{0.5\textwidth}
        \centering
        \includegraphics[width=\textwidth,clip]{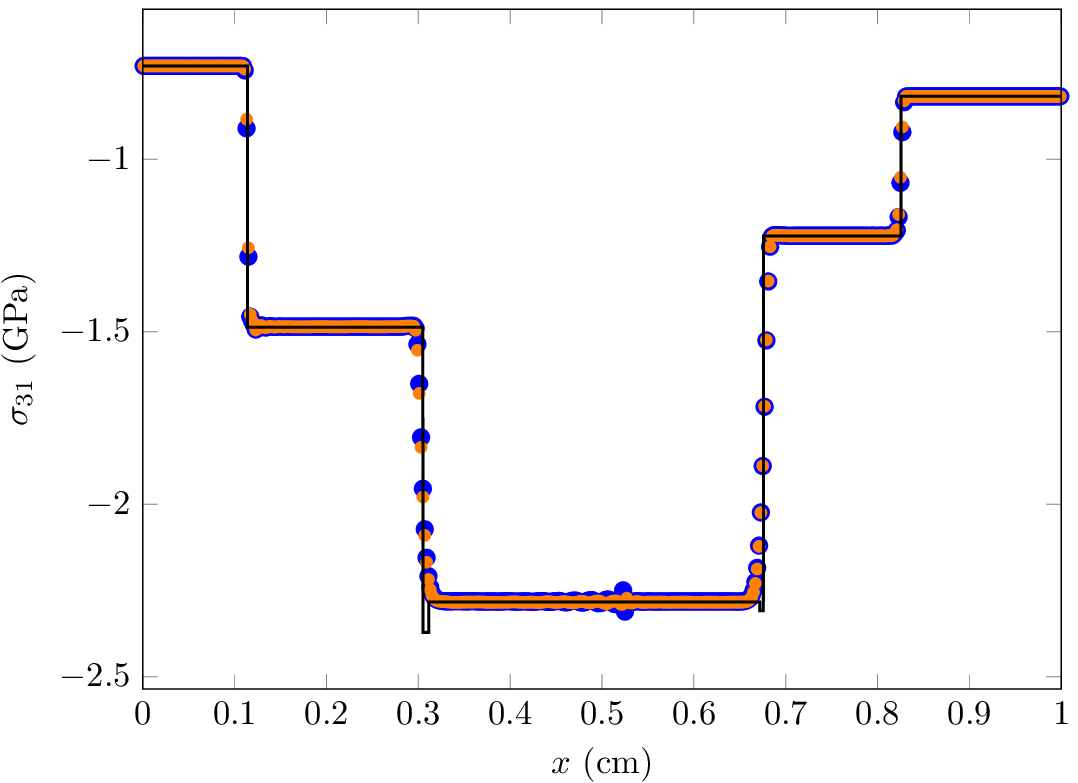}
        \caption{}%
        \label{fig:ssrp-sigma31}
    \end{subfigure}
    \caption{Solid-solid Riemann problem (cont.)}%
    \label{fig:ssrp}
\end{figure*}

As expected, away from the material interface the results are equivalent to the
diffuse-interface solution, however at the contact the ability of the
volume-of-fluid scheme to maintain a sharp interface is demonstrated. Barton
noted that THINC contributes to oscillations in the tangential stress and
velocity fields at the interface---these are notably reduced with the new
method.

\subsection{Solid-fluid Riemann problem}

\noindent The second test is a solid-fluid Riemann problem, again taken from
\cite{barton09}, featuring highly stressed moving aluminium in contact with
quiescent air. This test is used to verify that the volume-of-fluid method can
correctly treat an interface between two materials with radically different
properties. The initial left and right states are as follows:

\begin{align*}
    \mathbf{u}_L &= \begin{pmatrix}
        2 \\
        0 \\
        0.1 \\
    \end{pmatrix} \mathrm{km\,s}^{-1}, \quad
    \mathbf{F}_{e,L} = \begin{pmatrix}
        1 & 0 & 0 \\
        -0.01 & 0.95 & 0.02 \\
        -0.015 & 0 & 0.9 \\
    \end{pmatrix}, \quad
    \mathscr{E}_{L} = \mathscr{E}_{\mathrm{ref},L} \\
    \mathbf{u}_R &= \mathbf{0} \: \mathrm{km\,s}^{-1}, \quad
    \overline{\mathbf{V}}_{e,R} = \mathbf{I}, \quad
    \rho_R = 10^{-3} \: \mathrm{g\,cm}^{-3}, \quad
    p_R = 10^{-4} \: \mathrm{GPa}
\end{align*}

\noindent The aluminium is assumed to be purely elastic and is governed by the
equation of state from~\cite{dorovskii83} with parameters given in
Table~\ref{tab:dorovskii-parameters}. The air is governed by an ideal gas
equation of state with $\gamma = 1.4$.  The problem is run to time $0.6 \,
\mu\mathrm{s}$ with CFL number 0.1.

\begin{figure*}
    \centering
    \begin{subfigure}{0.5\textwidth}
        \centering
        \includegraphics[width=\textwidth,clip]{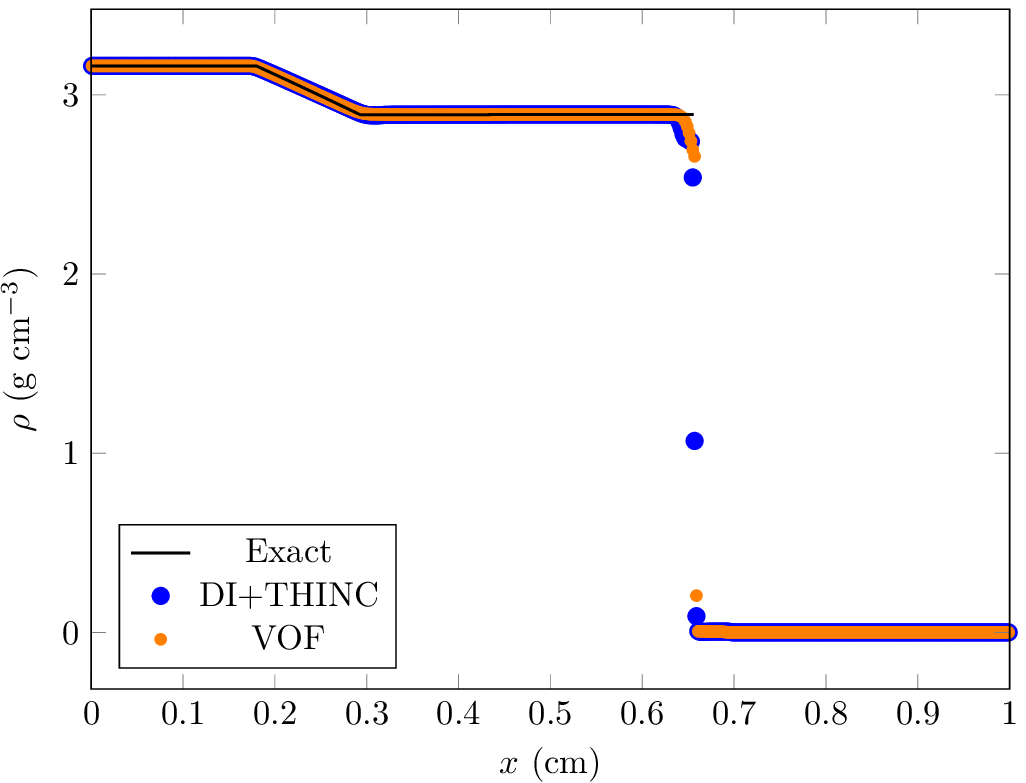}
        \caption{}%
        \label{fig:sgrp-rho}
    \end{subfigure}%
    \begin{subfigure}{0.5\textwidth}
        \centering
        \includegraphics[width=\textwidth,clip]{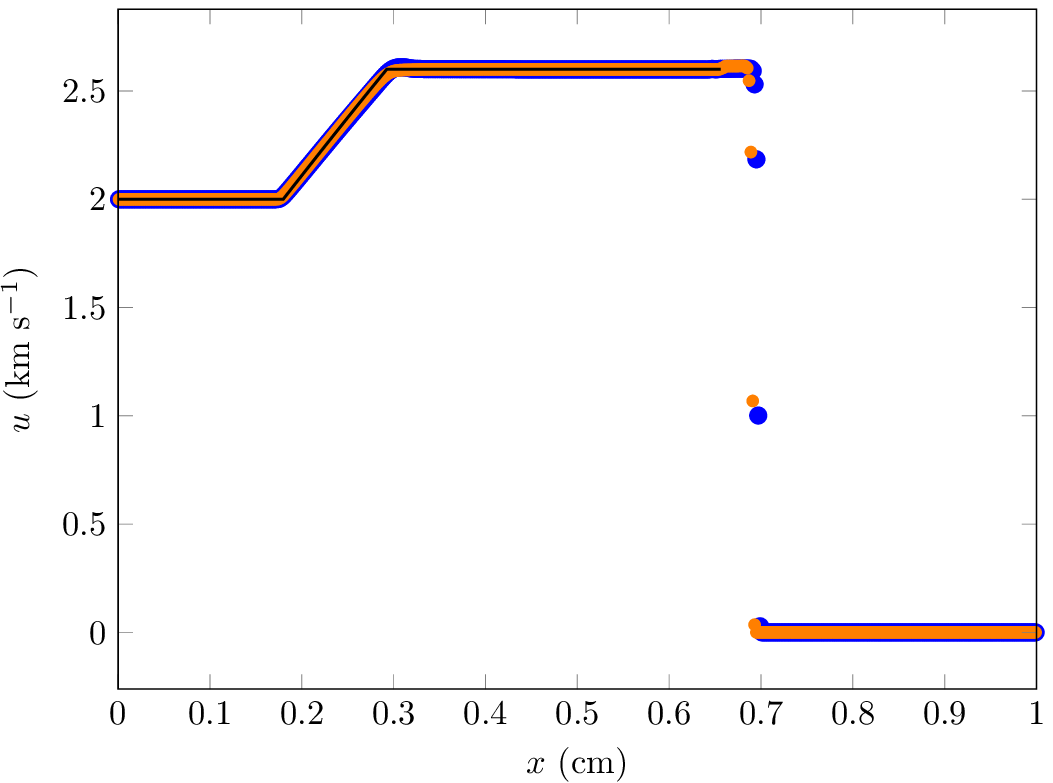}
        \caption{}%
        \label{fig:sgrp-u}
    \end{subfigure}
    \centering
    \begin{subfigure}{0.5\textwidth}
        \centering
        \includegraphics[width=\textwidth,clip]{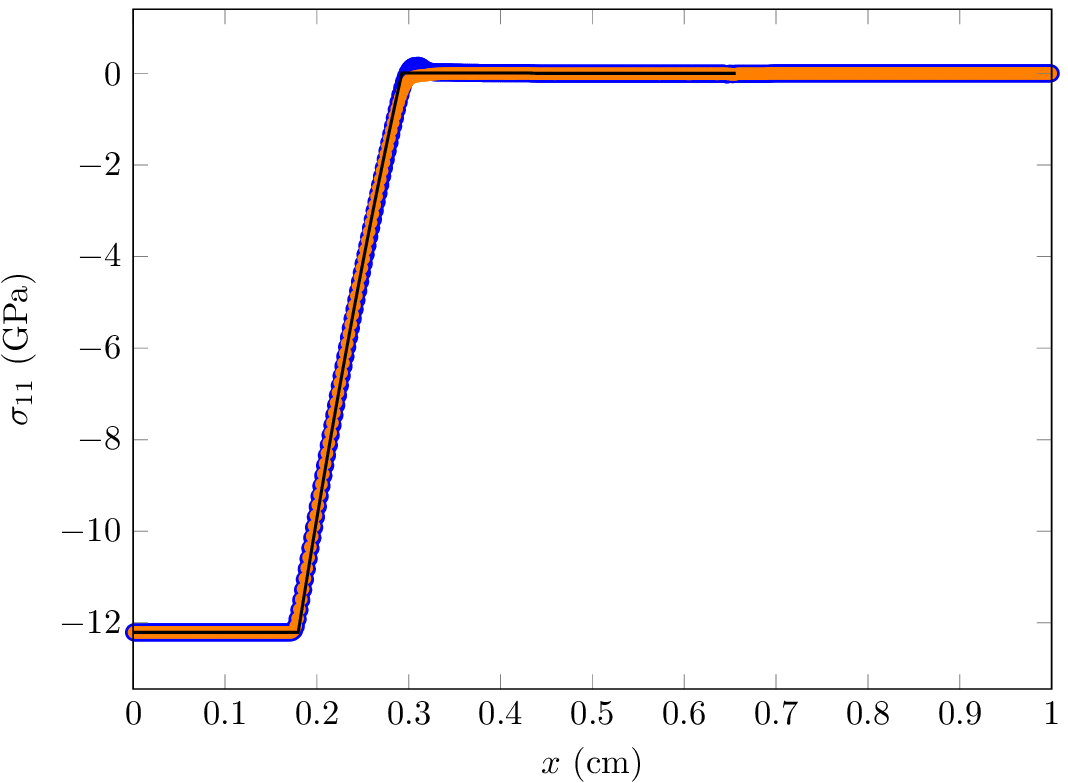}
        \caption{}%
        \label{fig:sgrp-sigma11}
    \end{subfigure}
    \caption{Solid-fluid Riemann problem.}%
    \label{fig:sgrp}
\end{figure*}

Results are shown in Figure~\ref{fig:sgrp}. The exact solution has been computed
for the case with vacuum instead of air, but this makes little difference to the
solution within the metal which is shown. The air supports a right-going shock
not present in the vacuum which can be seen in the normal velocity field.

\subsection{Shell collapse}

\noindent The shell collapse problem has been studied by many authors, an early
contributor being Verney~\cite{verney68}, and is a demanding test of energy
conservation and symmetry preservation. This can be seen in solutions obtained
using alternative Eulerian sharp interface treatments such as the ghost-fluid
method, where high resolution is typically required to obtain a reasonable
solution~\cite{barton13,lopezortega14}. In contrast our method exhibits smaller
conservation errors and is able to provide excellent results even on extremely
coarse meshes.

\begin{figure*}
    \centering
    \begin{subfigure}{0.25\textwidth}
        \centering
        \includegraphics[width=0.99\textwidth,clip]{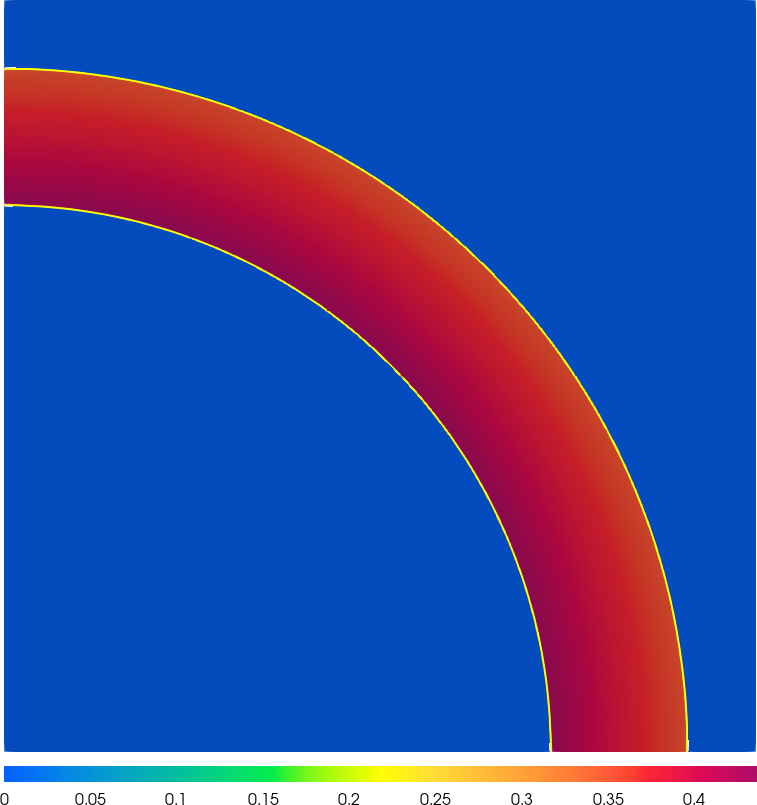}
        \caption{}%
        \label{fig:shell-collapse-0}
    \end{subfigure}%
    \begin{subfigure}{0.25\textwidth}
        \centering
        \includegraphics[width=0.99\textwidth,clip]{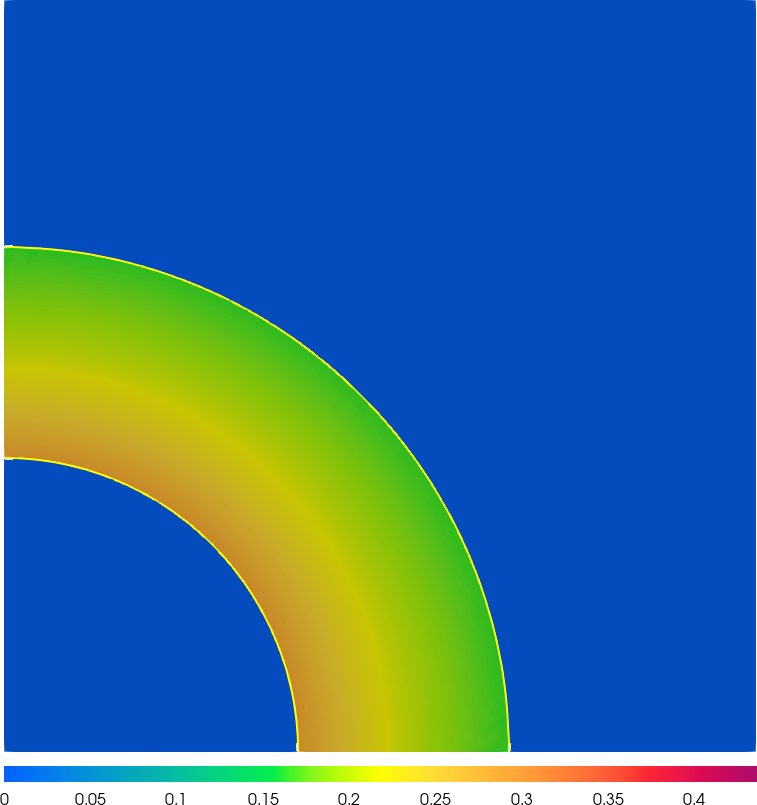}
        \caption{}%
        \label{fig:shell-collapse-1}
    \end{subfigure}%
    \begin{subfigure}{0.25\textwidth}
        \centering
        \includegraphics[width=0.99\textwidth,clip]{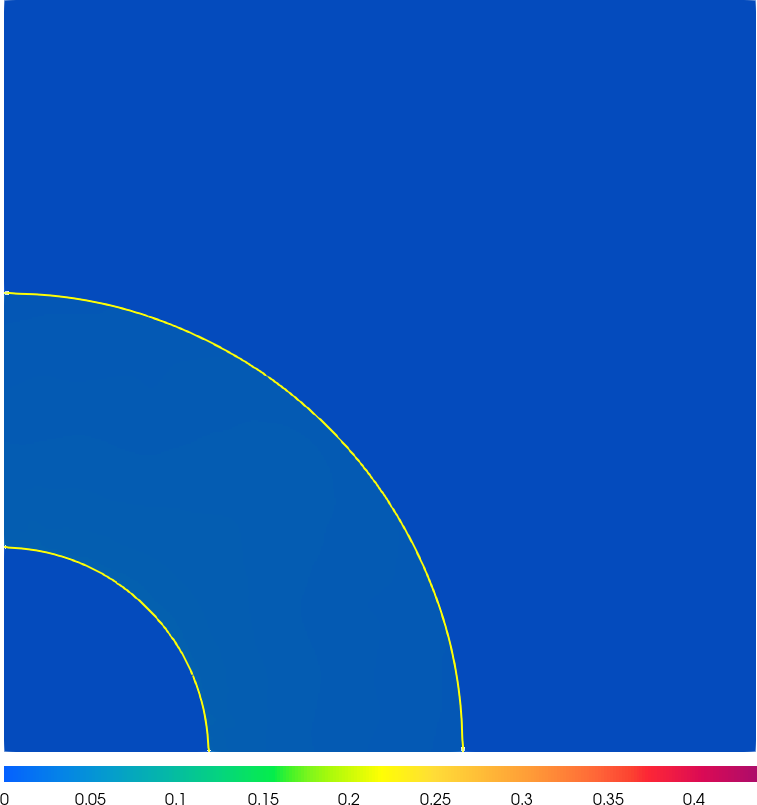}
        \caption{}%
        \label{fig:shell-collapse-2}
    \end{subfigure}%
    \begin{subfigure}{0.25\textwidth}
        \centering
        \includegraphics[width=0.99\textwidth,clip]{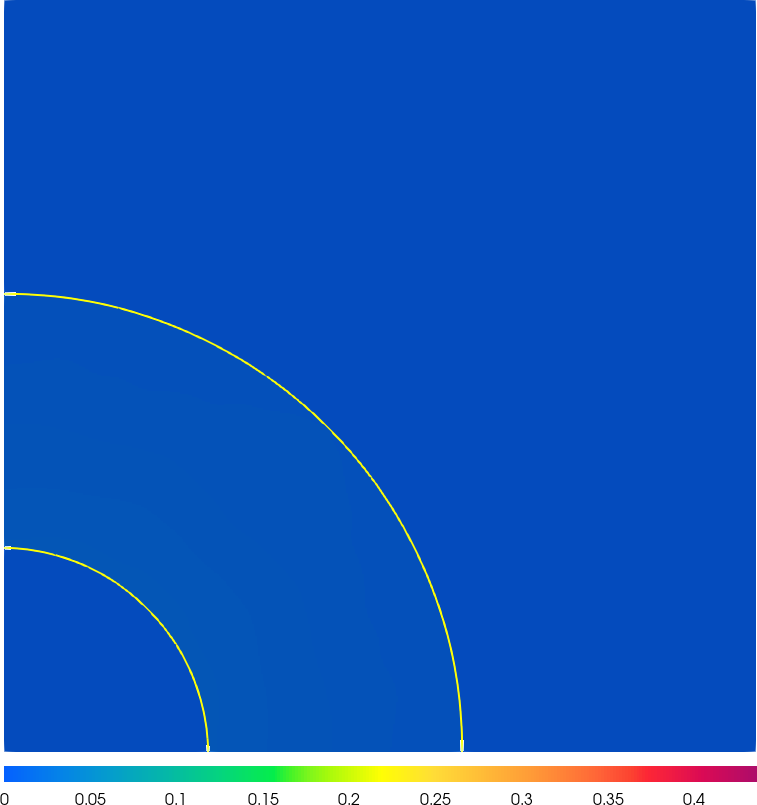}
        \caption{}%
        \label{fig:shell-collapse-3}
    \end{subfigure}

    \caption{The collapsing shell problem: magnitude of velocity within the
    shell measured in $\mathrm{km\,s}^{-1}$ (velocity of surrounding air not
    shown) and $\phi = 0.5$ contour at intervals of $10\,\mu\mathrm{s}$ from
    $t=0\,\mu\mathrm{s}$ to $t=30\,\mu\mathrm{s}$.}%

    \label{fig:shell-collapse-velocity}
\end{figure*}

A cylindrical aluminium shell is modelled with initial inner radius of $0.8 \,
\mathrm{cm}$ and initial outer radius $1.0 \, \mathrm{cm}$; the domain size is
$[0, 1.1]^2 \, \mathrm{cm}$. An initial velocity distribution is applied that
causes the shell to collapse incompressibly.  Plastic distortional effects
convert kinetic energy to internal energy, and eventually the shell stops.
Howell and Ball present formulae for the stopping radii~\cite{howell02}. In this
case we have a final inner radius of $0.3 \, \mathrm{cm}$ and a final outer
radius of approximately $0.671 \, \mathrm{cm}$.  The initial radial velocity is
given by $0.8 u_0 / r$ where $u_0 = 0.437 \, \mathrm{km\,s}^{-1}$. The equation
of state parameters for the aluminium are given in
Table~\ref{tab:dorovskii-parameters}, and ideal plasticity is used with constant
yield stress $\sigma_Y = 0.2976 \, \mathrm{GPa}$. The remainder of the domain is
filled with air governed by an ideal gas equation of state with $\gamma = 1.4$
and $\rho_0 = 1.225\cdot10^{-3} \, \mathrm{g\,cm}^{-3}$. The problem is run to
time $30 \, \mu\mathrm{s}$ with CFL number 0.6.
Figure~\ref{fig:shell-collapse-velocity} shows snapshots of velocity magnitude
within the shell, and the location of the interface, at $t=0$, $10$, $20$ and
$30\,\mu\mathrm{s}$.

\begin{figure*}
    \centering
    \begin{subfigure}{0.5\textwidth}
        \centering
        \includegraphics[width=\textwidth,clip]{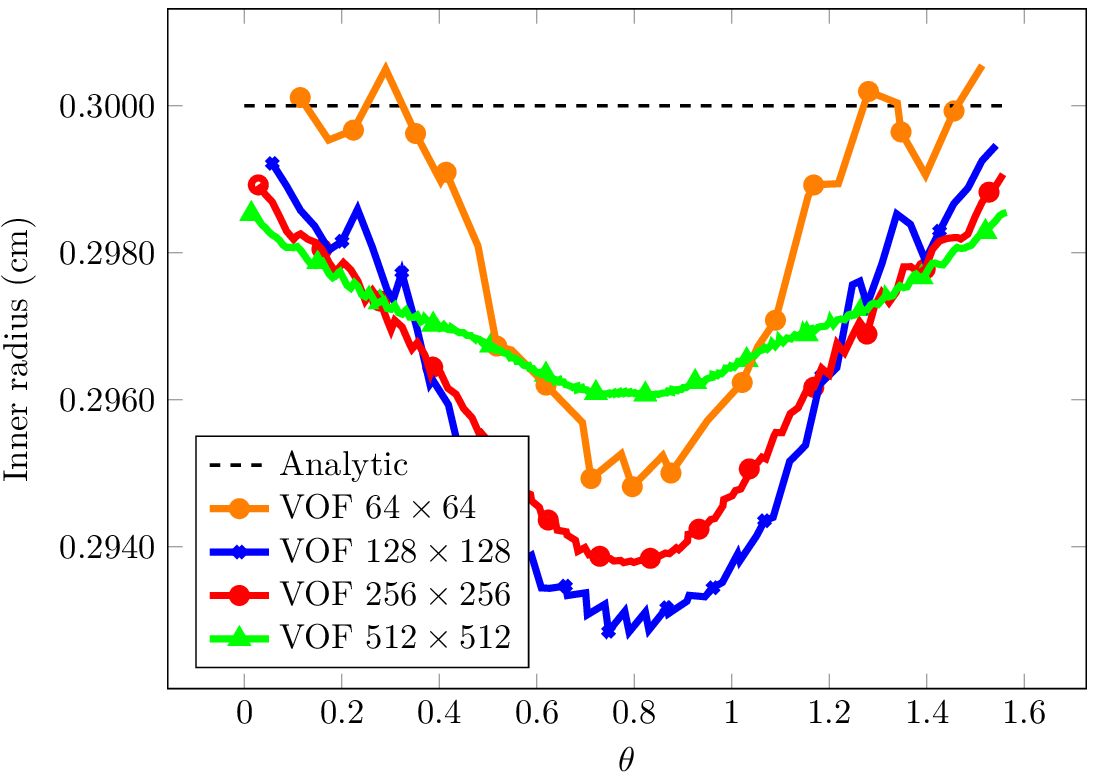}
        \caption{}%
        \label{fig:shell-collapse-inner}
    \end{subfigure}%
    \begin{subfigure}{0.5\textwidth}
        \centering
        \includegraphics[width=\textwidth,clip]{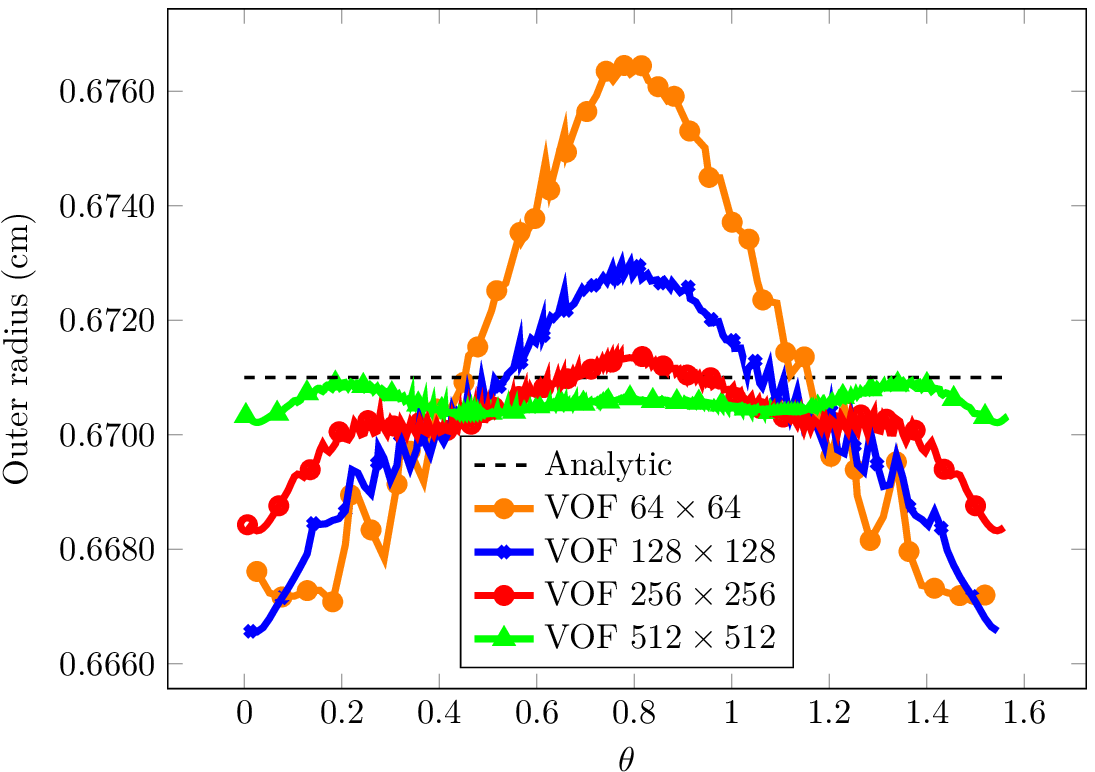}
        \caption{}%
        \label{fig:shell-collapse-outer}
    \end{subfigure}
    \begin{subfigure}{0.5\textwidth}
        \centering
        \includegraphics[width=\textwidth,clip]{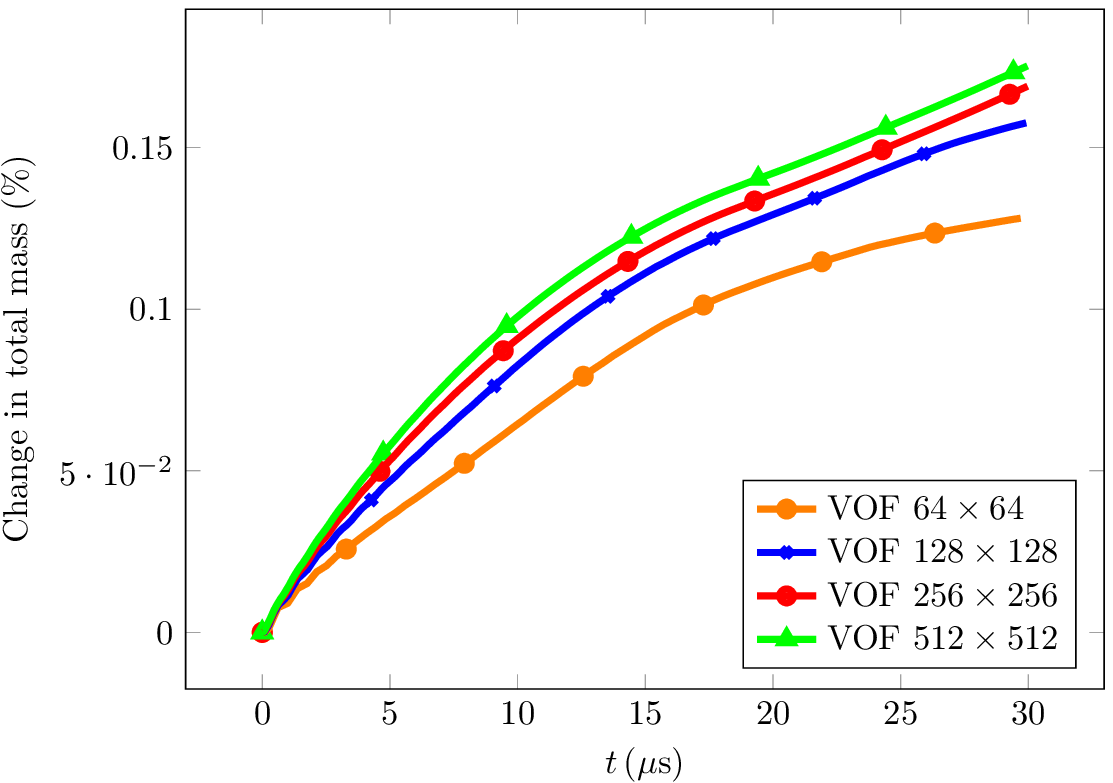}
        \caption{}%
        \label{fig:shell-collapse-mass}
    \end{subfigure}%
    \begin{subfigure}{0.5\textwidth}
        \centering
        \includegraphics[width=\textwidth,clip]{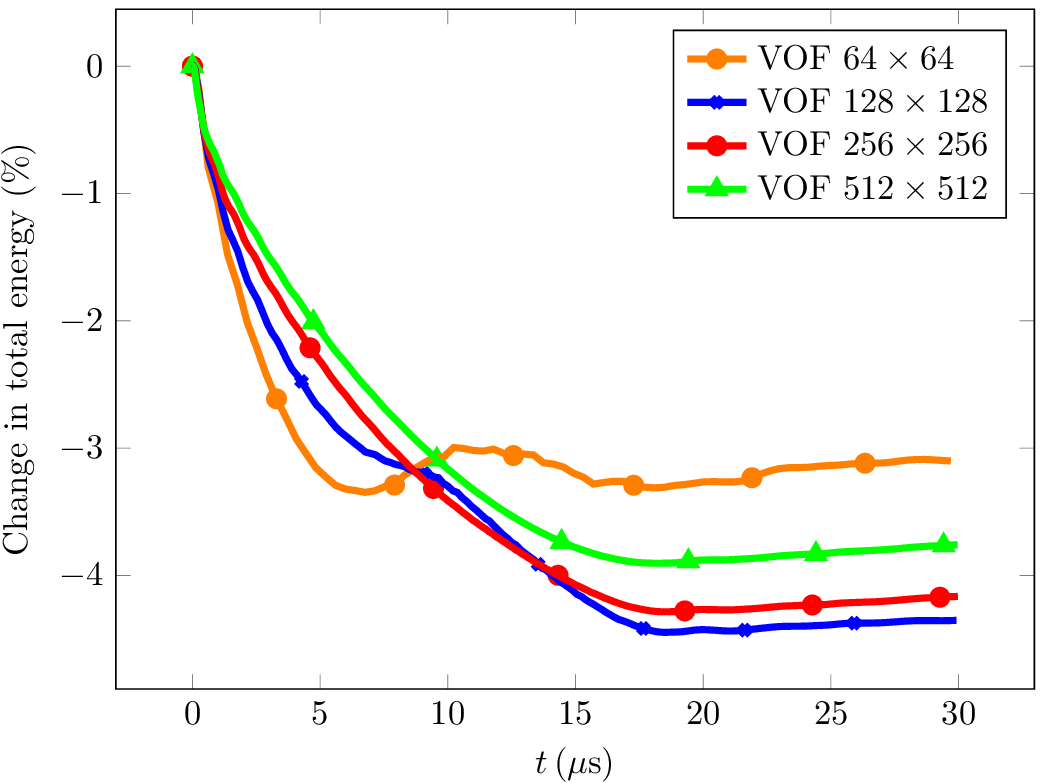}
        \caption{}%
        \label{fig:shell-collapse-energy}
    \end{subfigure}

    \caption{Final radii and energy conservation for the aluminium shell
    collapse problem at a range of mesh sizes. (a) shows the predicted inner
    radius as a function of the angle $\theta \in [0, \frac{\pi}{2}]$, (b) shows
    the predicted outer radius, (c) shows percentage conservation of total mass
    and (d) shows percentage conservation of total energy.  The dashed lines
    show the analytic radii computed using the formulae from Howell and
    Ball~\cite{howell02}. The volume-of-fluid method produces excellent results
    even at the coarsest mesh resolution, in stark contrast to previously
    published ghost-fluid method results. Radial symmetry is well preserved and
    improves as the mesh is refined, and conservation errors are low.}%

    \label{fig:shell-collapse}
\end{figure*}

The final inner and outer radii are shown as a function of angle in
Figure~\ref{fig:shell-collapse-inner} and \ref{fig:shell-collapse-outer}
respectively for a range of mesh sizes. In all cases it can be seen that the
radial error is limited to less than $2\%$, even on a very coarse mesh with only
64 cells in each coordinate direction. In contrast, both
Barton~\textit{et~al.}~\cite{barton13} and
Lopez~Ortega~\textit{et~al.}~\cite{lopezortega14} report errors in the region of
$35\%$ for similarly sized meshes using ghost-fluid methods. At higher (but
still moderate) resolutions, the deviations from radial symmetry are reduced.

Conservation histories for total mass and total energy as shown in
Figure~\ref{fig:shell-collapse-mass} and \ref{fig:shell-collapse-energy}
respectively for the same range of mesh sizes. Although the volume-of-fluid
method is not fully conservative, it can be seen that the error in total mass
for this problem is less than a fifth of a percent, and the error for total
energy is limited to less than five percent.

\subsection{Cylinder impact}

\noindent Cylinder impacts provide a strenuous test of elasto-plastic flow and
are included here to validate the rate-sensitive plasticity treatment. A
cylinder strikes a rigid anvil (here modelled using a perfectly reflecting
boundary) and deforms plastically, flowing radially outwards until all kinetic
energy has been converted to internal energy and the cylinder stops, with a
reduced length. Wilkins and Guinan provide experimental data
in~\cite{wilkins73}, and we also compare against results given by
Barton~\cite{barton19} using the underlying diffuse-interface method.

\begin{figure*}
    \centering
    \begin{subfigure}{0.25\textwidth}
        \centering
        \includegraphics[width=0.99\textwidth,clip]{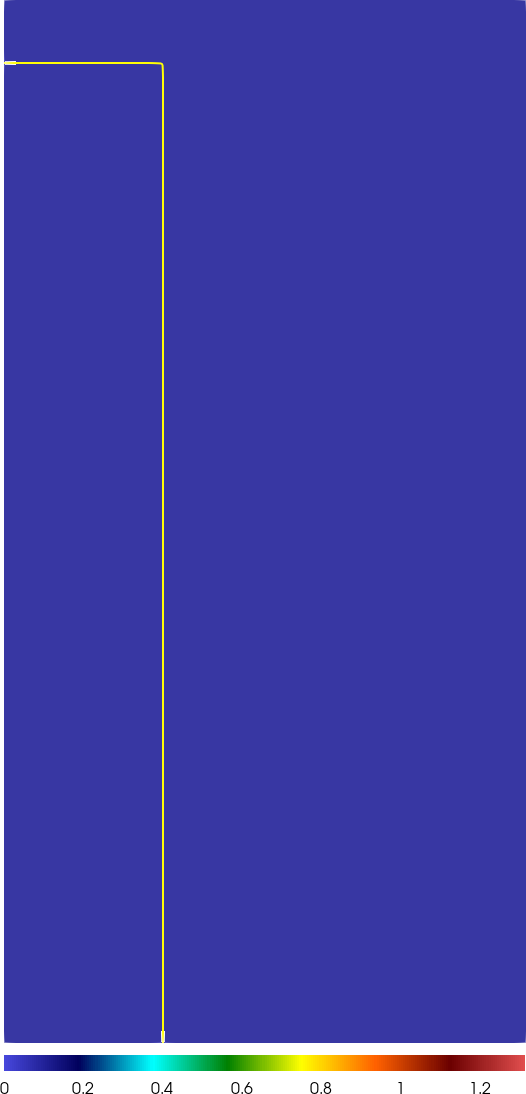}
        \caption{}%
        \label{fig:taylor-rod-0}
    \end{subfigure}%
    \begin{subfigure}{0.25\textwidth}
        \centering
        \includegraphics[width=0.99\textwidth,clip]{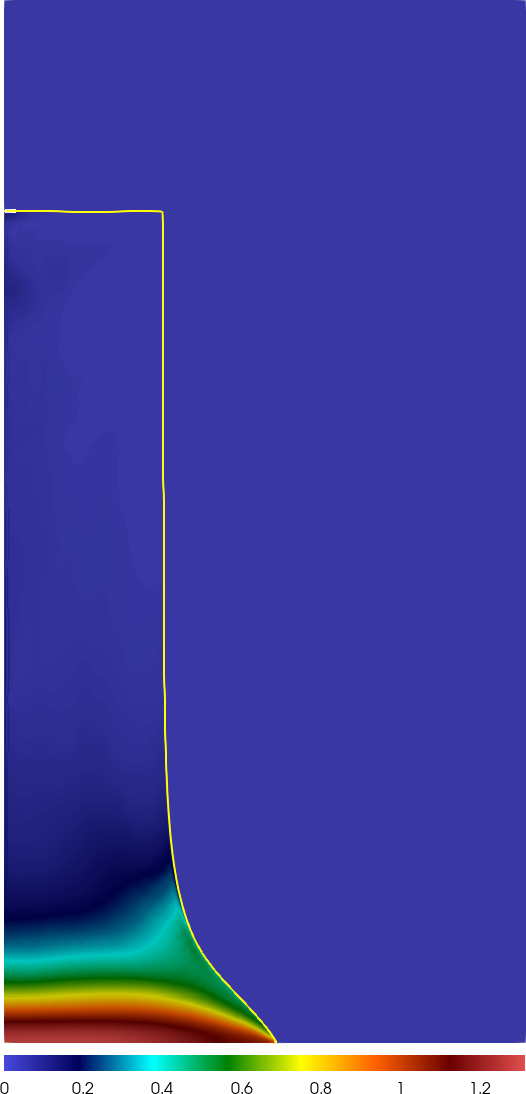}
        \caption{}%
        \label{fig:taylor-rod-1}
    \end{subfigure}%
    \begin{subfigure}{0.25\textwidth}
        \centering
        \includegraphics[width=0.99\textwidth,clip]{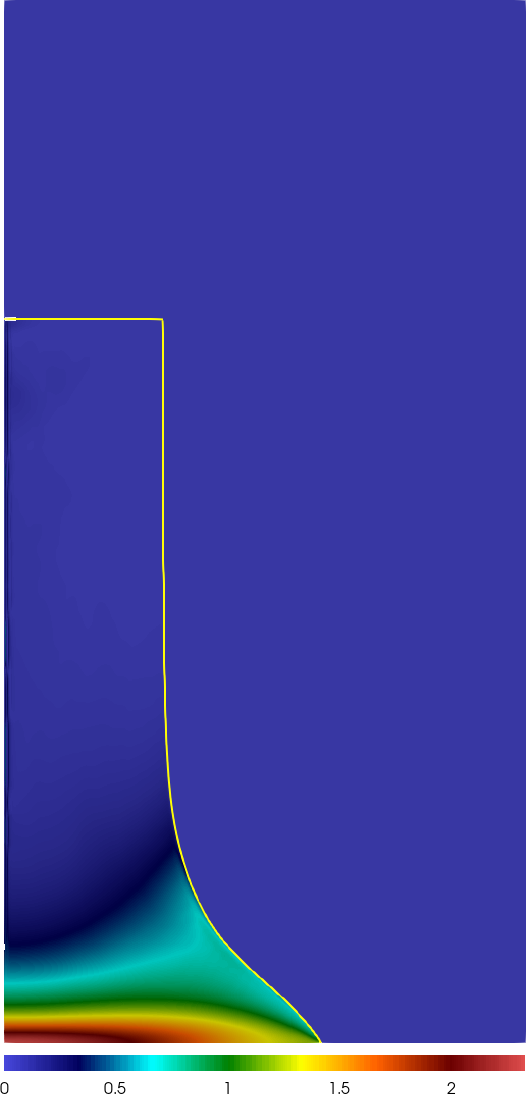}
        \caption{}%
        \label{fig:taylor-rod-2}
    \end{subfigure}%
    \begin{subfigure}{0.25\textwidth}
        \centering
        \includegraphics[width=0.99\textwidth,clip]{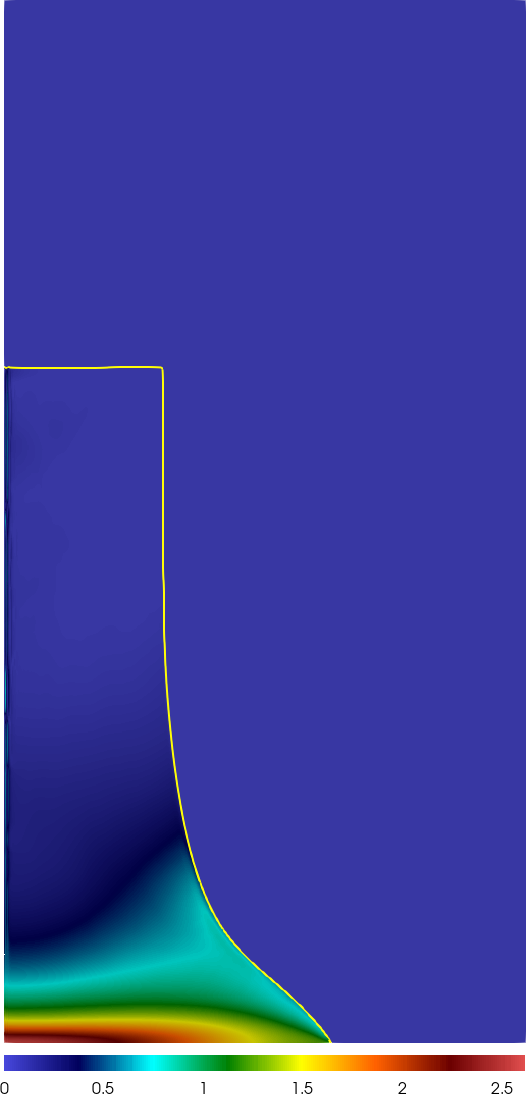}
        \caption{}%
        \label{fig:taylor-rod-3}
    \end{subfigure}

    \caption{The cylinder impact problem with an impact velocity of
    $373\,\mathrm{m\,s}^{-1}$: equivalent plastic strain is shown along with the
    $\phi = 0.5$ contour at intervals of $10\,\mu\mathrm{s}$ from
    $t=0\,\mu\mathrm{s}$ to $t=30\,\mu\mathrm{s}$.}%

    \label{fig:cylinder-impact-plot}
\end{figure*}

The initial length of the cylinder is $L_0 = 2.347\,\mathrm{cm}$ and the radius
is $0.381\,\mathrm{cm}$. We predict the final length $L_f$ for a range of impact
velocities in 2D axisymmetric geometry, using a domain of $1.25 \times 2.5
\,\mathrm{cm}$ and a mesh size of $512\times1024$. The cylinder is aluminium
6061-T6 with parameters given in Table~\ref{tab:dorovskii-parameters}. The
Johnson-Cook rate-sensitive plasticity model is used, including work-hardening
effects, with parameters given in Table~\ref{tab:jc-parameters}.  Thermal
softening is disabled here by setting $T_{\mathrm{melt}} \rightarrow \infty$.
The cylinder is immersed in quiescent air governed by an ideal gas equation of
state with $\gamma = 1.4$ and $\rho_0 = 1.225\cdot10^{-3}\,\mathrm{g\,cm}^{-3}$.
All materials are initialised to a pressure of $1\,\mathrm{atm}$ and a CFL
number of 0.6 is used.  The time evolution of the cylinder/air interface and the
equivalent plastic strain is shown for the $373 \,\mathrm{m\,s}^{-1}$ case in
Figure~\ref{fig:cylinder-impact-plot}.

\begin{table*}
    \centering
    \begin{tabular}{lrrrrrrrr}
        \toprule
        Material & $c_1$ & $c_2$ & $c_3$ & $n$ & $\chi_0$ & $m$ & $T_0\,(\mathrm{K})$ & $T_\mathrm{melt}\,(\mathrm{K})$ \\
        \midrule
        Aluminium 6061-T6 & 0.324 & 0.114 & 0.002 & 0.45 & $10^{-5}$ & - & - & - \\
        Copper & 0.09 & 0.292 & 0.025 & 0.31 & $10^{-5}$ & 1.09 & 288.15 & 1356 \\
        \bottomrule
    \end{tabular}

    \caption{Parameters for the Johnson-Cook rate-sensitive plasticity model.}%
    \label{tab:jc-parameters}
\end{table*}

\begin{figure}
    \centering
    \includegraphics[width=0.6\textwidth,clip]{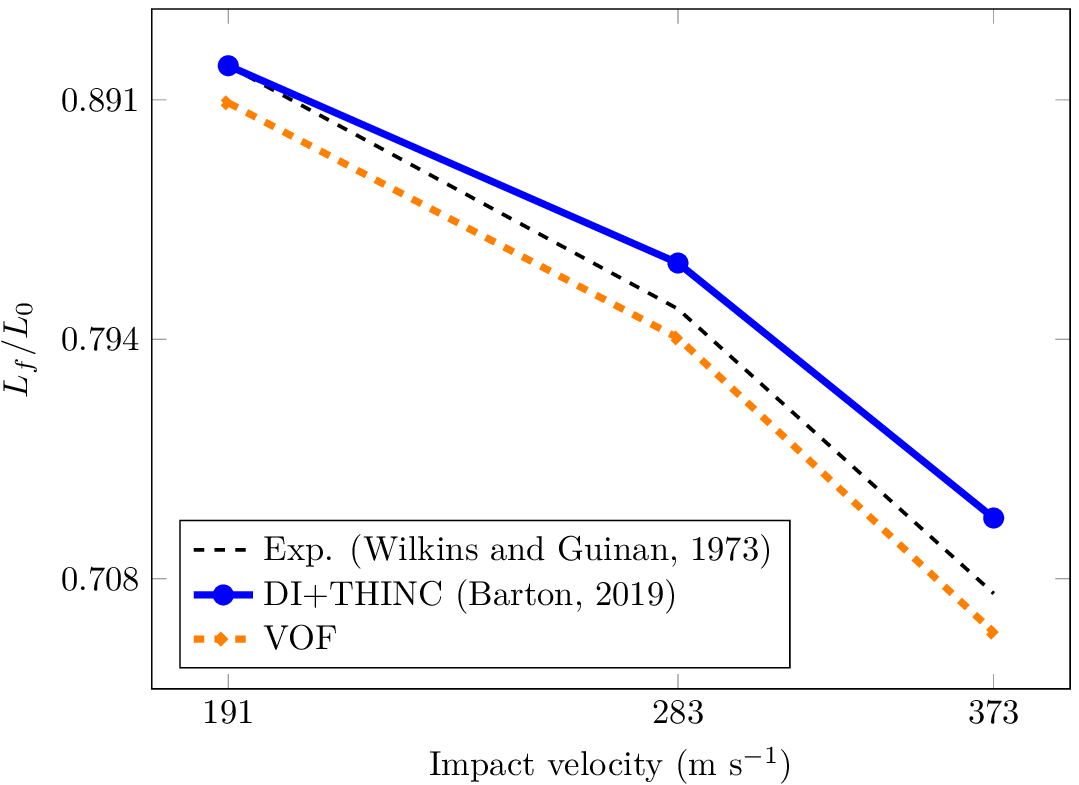}

    \caption{Results for the cylinder impact test case. The ratio of the final
    length of the cylinder to its initial length, both from experiment and
    predicted using the DI+THINC and VOF methods.}%

    \label{fig:cylinder-impact-comp}
\end{figure}

The final ratios $L_f/L_0$ for each impact velocity are compared in
Figure~\ref{fig:cylinder-impact-comp}. The volume-of-fluid method matches
experiment very well, correctly predicting the non-linear trend.

\subsection{Buried explosive}

\noindent The buried explosive problem previously studied in~\cite{barton19} is
useful to contrast the behaviour of the volume-of-fluid method against the
diffuse-interface method for problems involving large deformations and interface
breakup. A charge buried in soft clay detonates resulting in ejecta and crater
formation.

For 2D axisymmetric simulations the domain is $[0, 2] \times [-2, 4]
\,\mathrm{m}$. For full 3D simulations, one quadrant is modelled, and the domain
is $[0, 2] \times [-2, 2] \times [0, 2] \,\mathrm{m}$. The initial location of
the clay/air interface is $y=0$. The material parameters for the clay are given
in Table~\ref{tab:dorovskii-parameters}. Rather than using a fixed Gr\"uneisen
parameter, a power-law form is used

\begin{equation}
    \label{eq:power-law-gruneisen}
    \Gamma = \Gamma_0 \left(\frac{\rho}{\rho_0}\right)^q
\end{equation}

\noindent with $q = 1$. The air is modelled using an ideal gas equation of state
with $\gamma = 1.4$ and $\rho_0 = 1.225\cdot10^{-3} \,\mathrm{g\,cm}^{-3}$. The
clay and air are given an initial pressure of $1 \,\mathrm{atm}$.  The explosive
products are initialised in the cylinder centred at $(0, -13.081, 0)
\,\mathrm{cm}$ with radius $8.763 \,\mathrm{cm}$ and height $5.842
\,\mathrm{cm}$.  The phenomenological balloon model is used to model the
explosive products, sharing the equation of state with the
air~\cite{larcher2010}. The initial product density is $1.63 \,
\mathrm{g\,cm}^{-3}$, and the initial specific internal energy is $6.0568
\,\mathrm{kJ\,g}^{-1}$. Gravitational and atmospheric effects are not important
over the short timescales considered and so are not included.

\begin{figure*}
    \centering
    \begin{subfigure}{0.24\textwidth}
        \centering
        \includegraphics[width=\linewidth,clip]{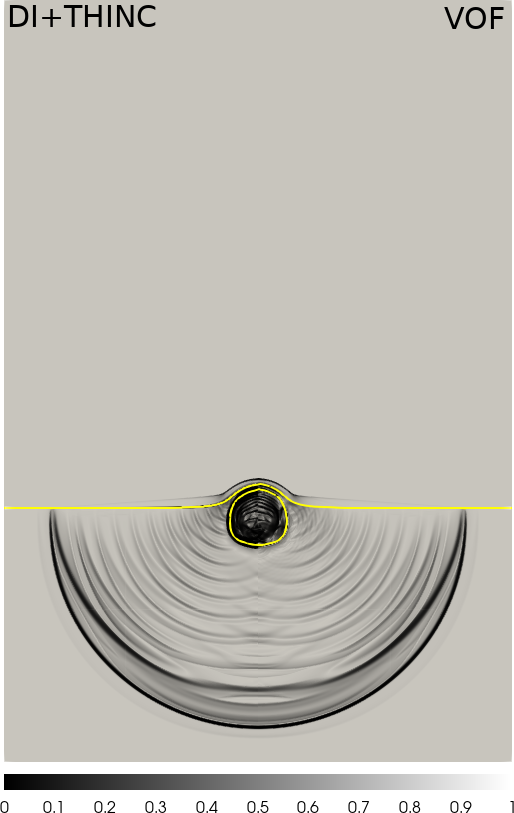}
        \caption{}%
        \label{fig:be-2d-s-30}
    \end{subfigure}\hspace*{0.01\textwidth}%
    \begin{subfigure}{0.24\textwidth}
        \centering
        \includegraphics[width=\linewidth,clip]{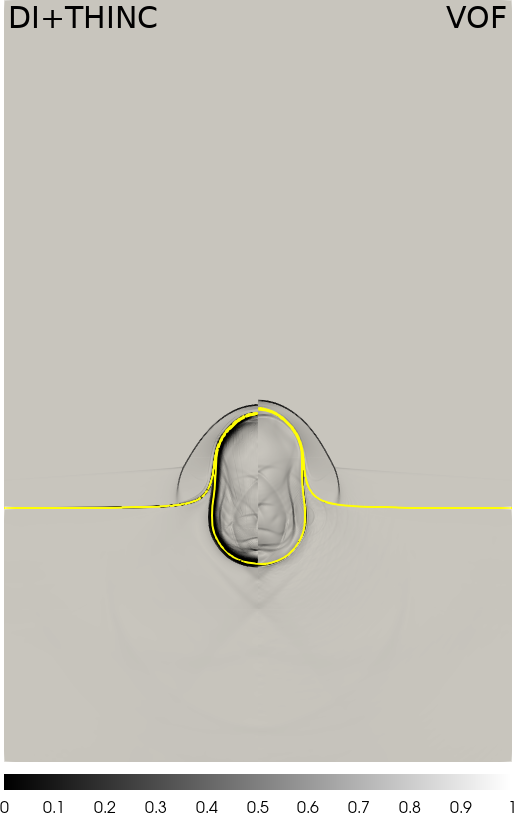}
        \caption{}%
        \label{fig:be-2d-s-100}
    \end{subfigure}\hspace*{0.01\textwidth}%
    \begin{subfigure}{0.24\textwidth}
        \centering
        \includegraphics[width=\linewidth,clip]{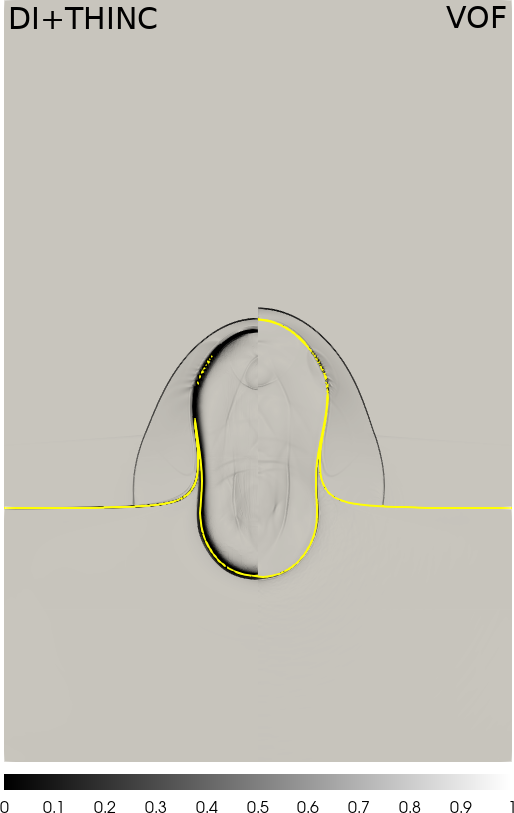}
        \caption{}%
        \label{fig:be-2d-s-180}
    \end{subfigure}\hspace*{0.01\textwidth}%
    \begin{subfigure}{0.24\textwidth}
        \centering
        \includegraphics[width=\linewidth,clip]{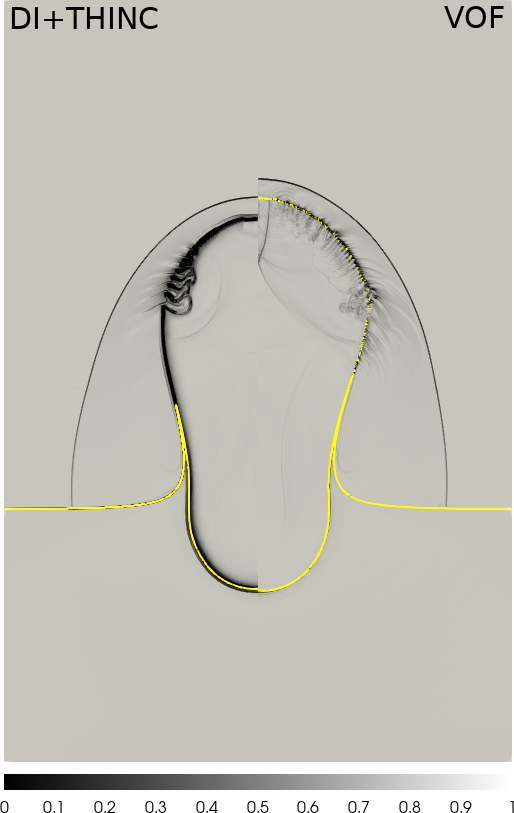}
        \caption{}%
        \label{fig:be-2d-s-300}
    \end{subfigure}
    \begin{subfigure}{0.24\textwidth}
        \centering
        \includegraphics[width=\linewidth,clip]{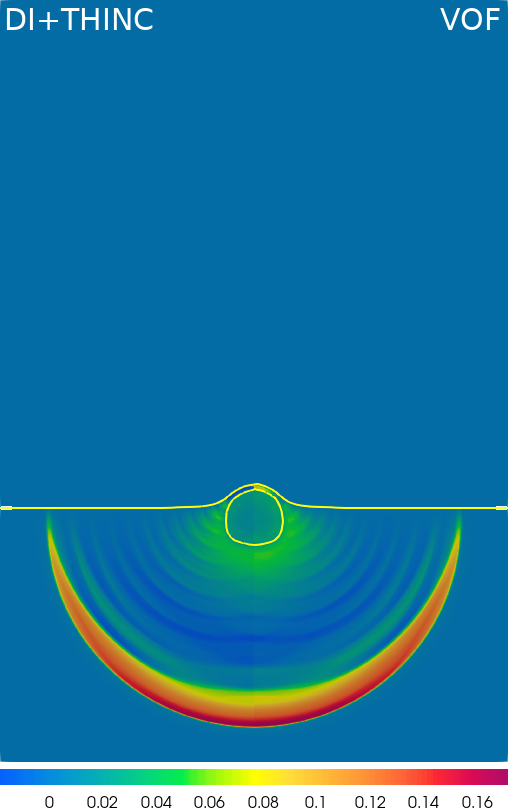}
        \caption{}%
        \label{fig:be-2d-p-30}
    \end{subfigure}\hspace*{0.01\textwidth}%
    \begin{subfigure}{0.24\textwidth}
        \centering
        \includegraphics[width=\linewidth,clip]{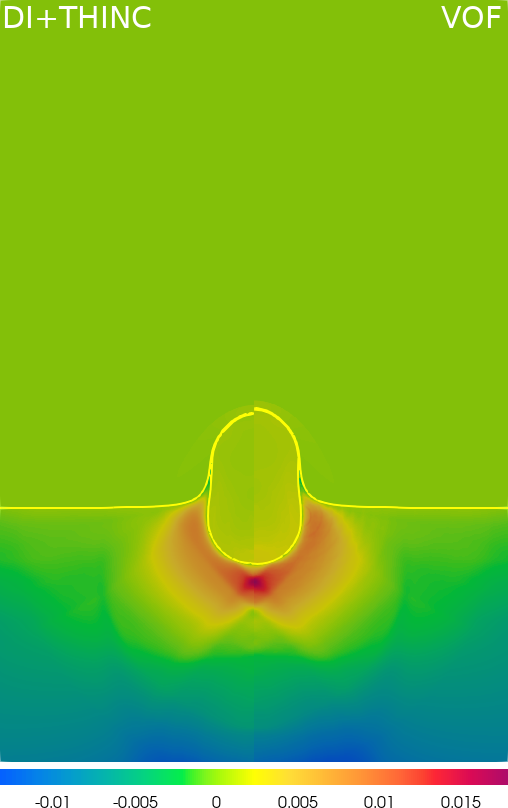}
        \caption{}%
        \label{fig:be-2d-p-100}
    \end{subfigure}\hspace*{0.01\textwidth}%
    \begin{subfigure}{0.24\textwidth}
        \centering
        \includegraphics[width=\linewidth,clip]{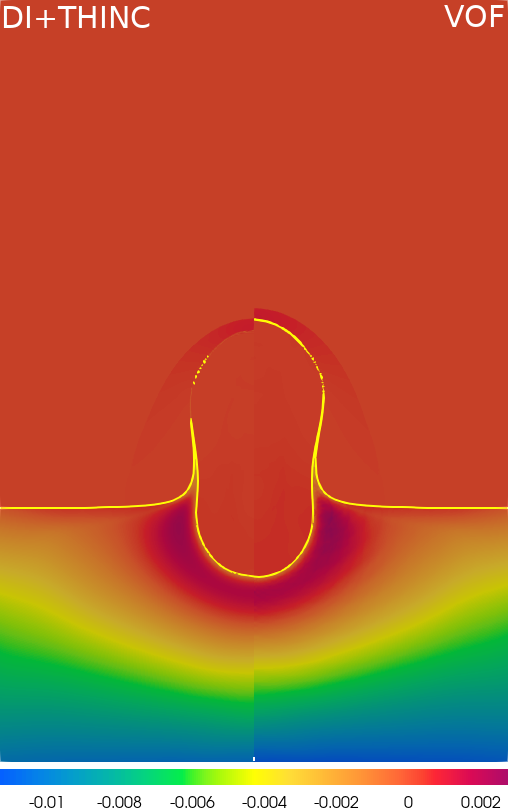}
        \caption{}%
        \label{fig:be-2d-p-180}
    \end{subfigure}\hspace*{0.01\textwidth}%
    \begin{subfigure}{0.24\textwidth}
        \centering
        \includegraphics[width=\linewidth,clip]{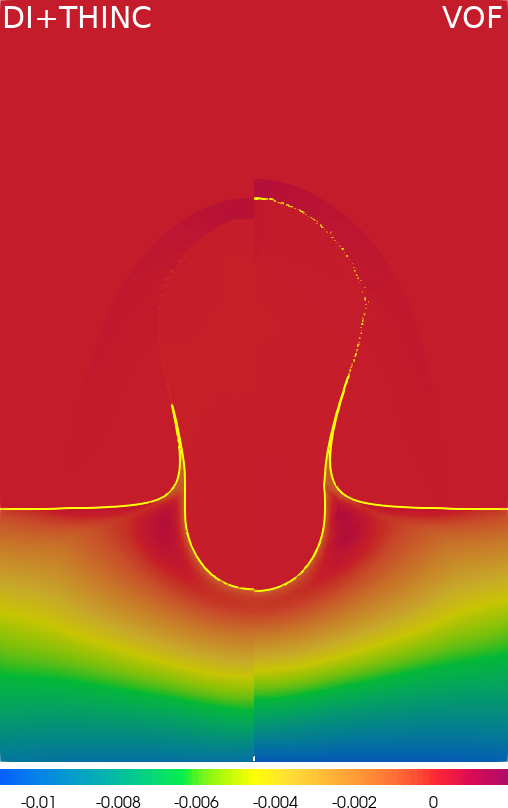}
        \caption{}%
        \label{fig:be-2d-p-300}
    \end{subfigure}

    \caption{Comparison between DI+THINC and VOF for the 2D axisymmetric buried
    explosive problem. The mesh size is $800\times2400$. Panels (a), (b), (c)
    and (d) show $t=0.3$, $1.0$, $1.8$ and $3 \,\mathrm{ms}$ respectively, and
    similarly for panels (e)--(h). In each panel the left half of the image
    shows DI+THINC and the right half shows the corresponding VOF solution. The
    first row shows a numerical Schlieren of density and the second shows
    pressure measured in $\mathrm{GPa}$.  The yellow lines denote the $\phi = 0.5$ contour.  While the two
    methods agree well on ground shock and crater dimensions, it is clear that
    the DI+THINC method has suppressed the growth of turbulent instabilities in
    the wake of lofted ejecta, which are much more pronounced in the VOF
    solution.}%

    \label{fig:be-2d}
\end{figure*}

\begin{figure*}
    \centering
    \begin{subfigure}{0.475\textwidth}
        \centering
        \includegraphics[width=\linewidth,clip]{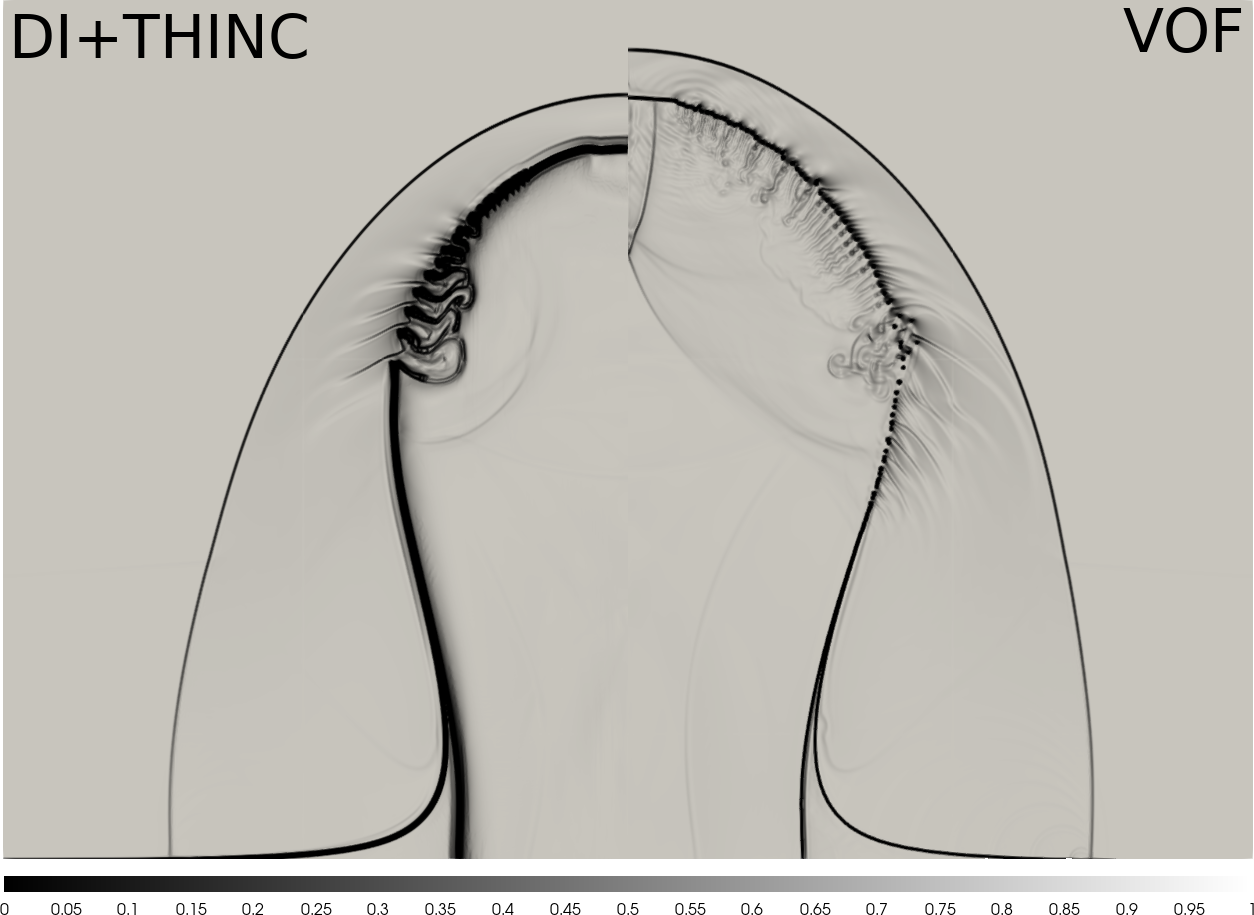}
        \caption{}%
        \label{fig:be-2d-close-schlieren}
    \end{subfigure}\hspace*{0.01\textwidth}%
    \begin{subfigure}{0.48\textwidth}
        \centering
        \includegraphics[width=\linewidth,clip]{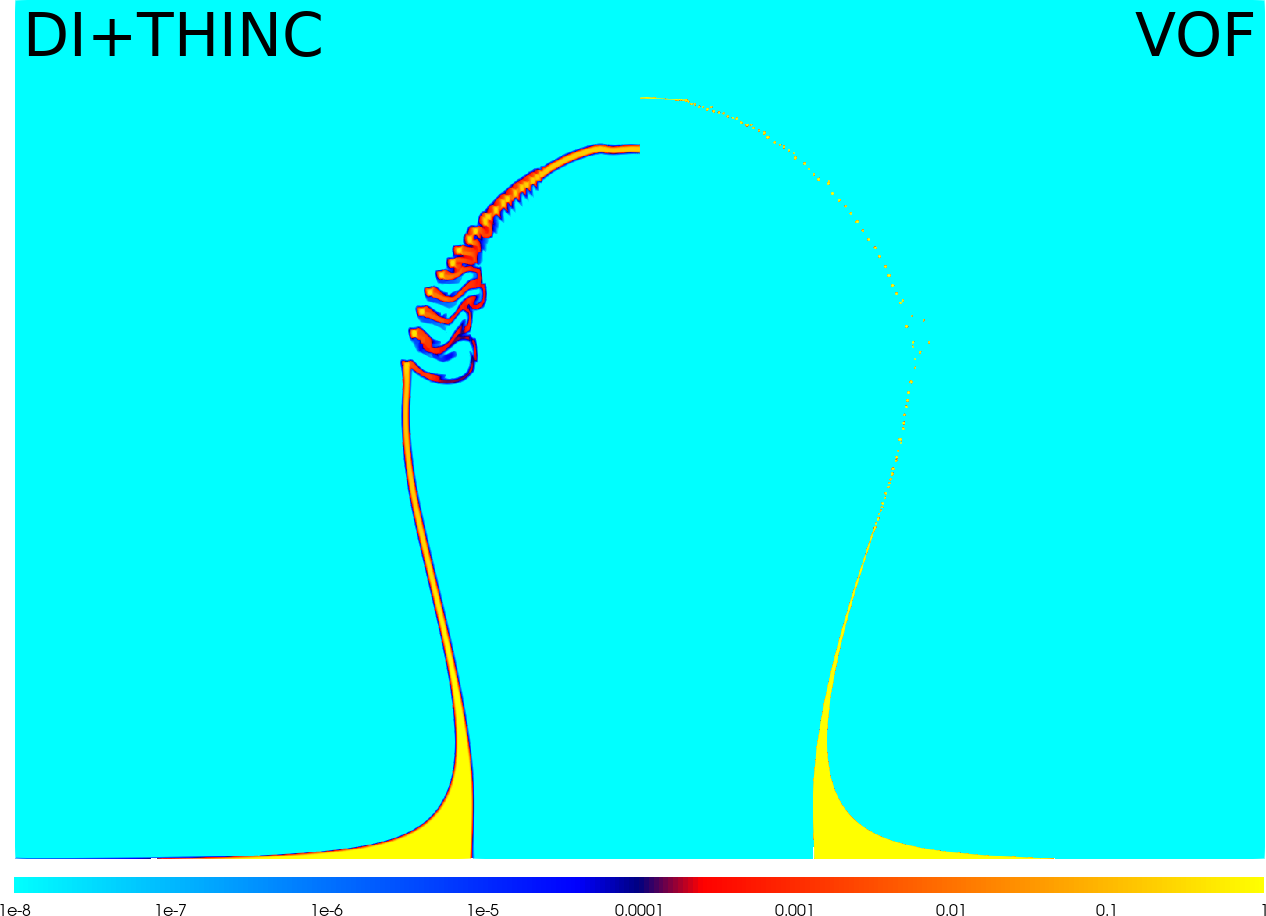}
        \caption{}%
        \label{fig:be-2d-close-vof}
    \end{subfigure}

    \caption{Close-up of $t=3\,\mathrm{ms}$ for the 2D axisymmetric buried
    explosion problem. (a) shows the numerical Schlieren of density, showing how
    the VOF method captures fine-scale turbulent flow in the wake of the lofted
    ejecta. (b) compares the log-scale clay ejecta volume fraction fields
    between DI+THINC and VOF. Unphysical ligament formation can be clearly
    observed in the DI+THINC solution, the interface never truly breaks.  This
    is absent in the VOF solution.}%

    \label{fig:be-2d-close}
\end{figure*}

Results for the 2D axisymmetric case are shown in Figure~\ref{fig:be-2d}.
Snapshots of pressure and a numerical Schlieren of density are shown at
$t=0.3$, $1$, $1.8$ and $3\,\mathrm{ms}$ for the volume-of-fluid method and
the underlying diffuse-interface method with THINC interface sharpening. The
function used to define the Schlieren plots is from~\cite{quirk96}

\begin{equation*}
    \mathrm{exp}\left(-\kappa \frac{|\nabla \rho|}{\underset{ijk}{\mathrm{max}}\,|\nabla \rho_{ijk}|}\right)
\end{equation*}

\noindent where we take $\kappa = 8000$. At the earlier times, excellent
agreement is seen between the two methods in terms of cavity dimensions and
ground shock propagation. At the later times, the crater dimensions agree well
and the volume-of-fluid method shows its strengths by resolving fine-scale
turbulent structures resulting from the break up of the lofted clay, which are
suppressed in the diffuse-interface solution. A zoomed in density Schlieren at
the final time is shown in Figure~\ref{fig:be-2d-close-schlieren} that
highlights these details.  Closer examination of the log-scale volume-fraction
field in Figure~\ref{fig:be-2d-close-vof} shows that the interface does not
fully break when using the diffuse-interface method, instead unphysical thin
ligaments with low but non-zero volume fraction develop.

\begin{figure*}
    \centering
    \begin{subfigure}{0.48\textwidth}
        \centering
        \includegraphics[width=\linewidth,clip]{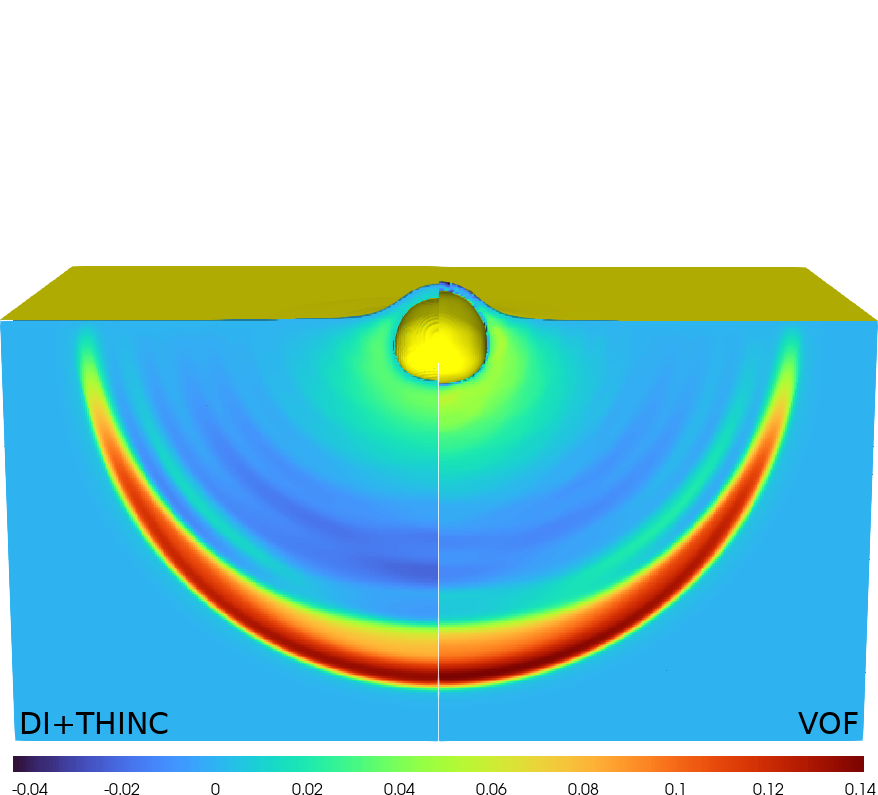}
        \caption{}%
        \label{fig:be-3d-30}
    \end{subfigure}\hspace*{0.01\textwidth}%
    \begin{subfigure}{0.48\textwidth}
        \centering
        \includegraphics[width=\linewidth,clip]{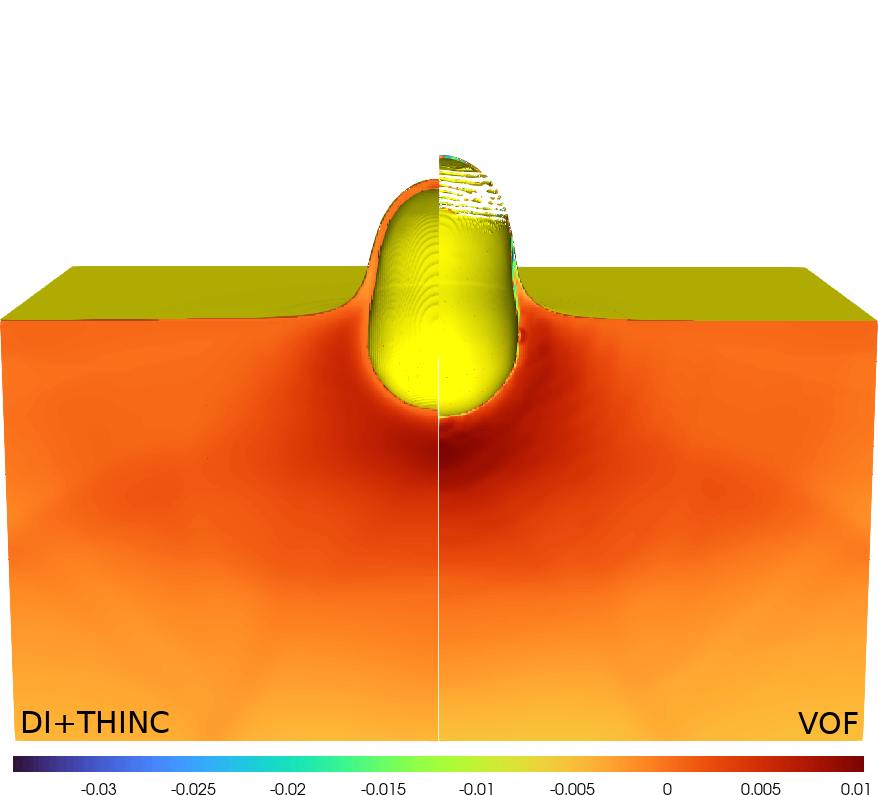}
        \caption{}%
        \label{fig:be-3d-100}
    \end{subfigure}
    \begin{subfigure}{0.48\textwidth}
        \centering
        \includegraphics[width=\linewidth,clip]{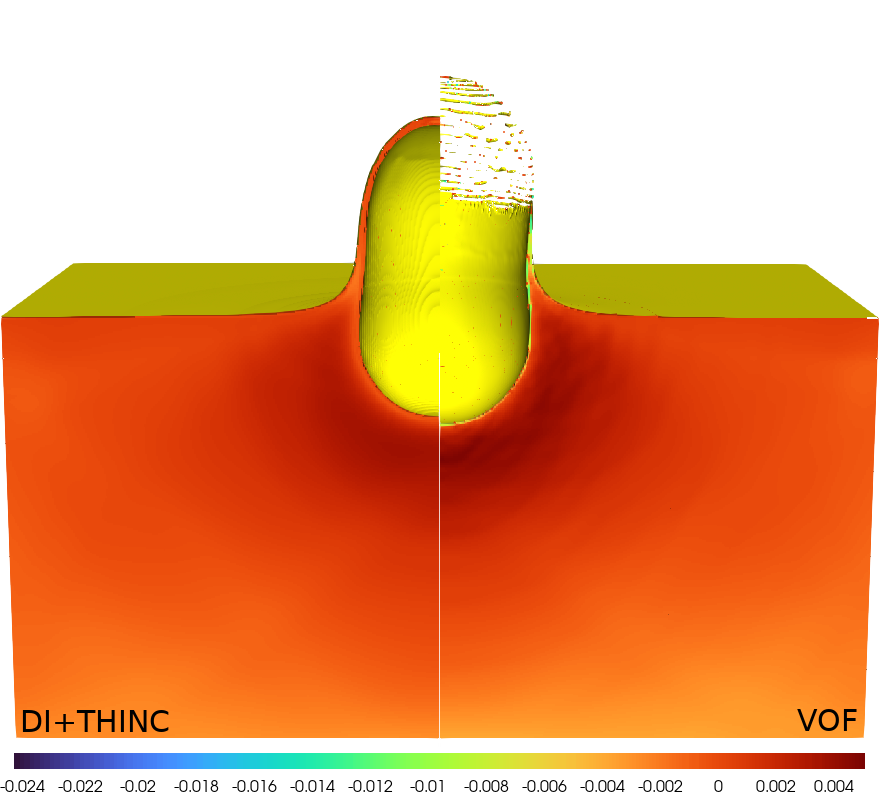}
        \caption{}%
        \label{fig:be-3d-140}
    \end{subfigure}\hspace*{0.01\textwidth}%
    \begin{subfigure}{0.48\textwidth}
        \centering
        \includegraphics[width=\linewidth,clip]{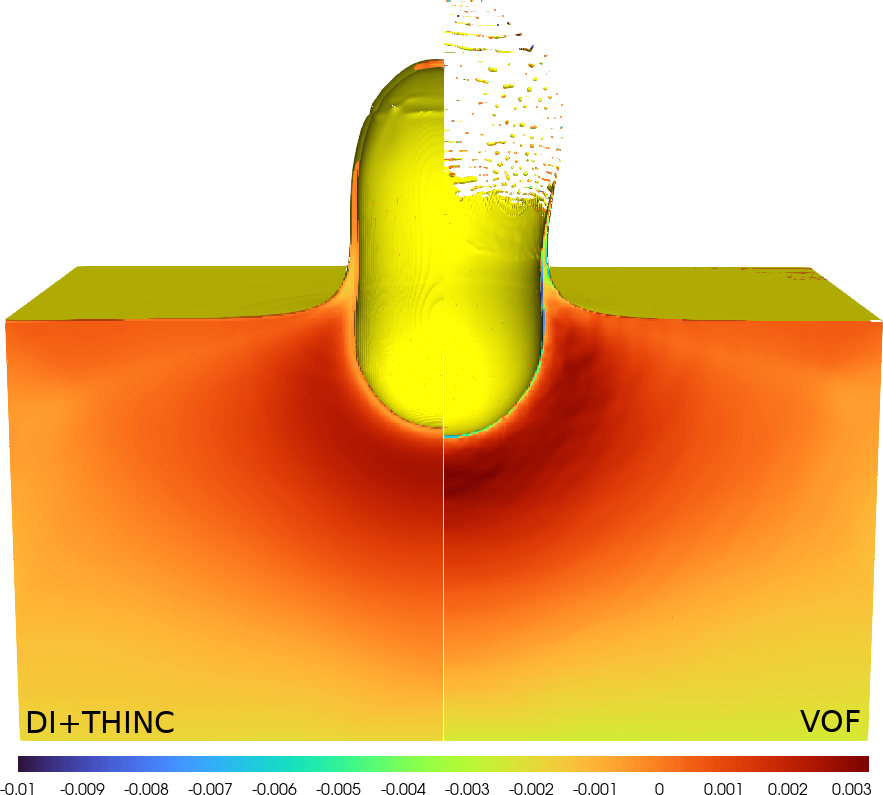}
        \caption{}%
        \label{fig:be-3d-180}
    \end{subfigure}\hspace*{0.01\textwidth}

    \caption{Comparison between DI+THINC and VOF for the 3D buried explosive
    problem. The mesh size is $240\times480\times240$. (a)--(d) show $t=0.3$,
    $1.0$, $1.4$ and $1.8 \,\mathrm{ms}$ respectively. In each image the left
    half shows DI+THINC and the right half shows the corresponding VOF solution.
    The yellow contour shows the clay/air interface and the colouring indicates
    pressure. As in 2D the two methods agree well on ground shock and crater
    dimensions, but VOF offers much improved fidelity for the lofted ejecta.}%

    \label{fig:be-3d}
\end{figure*}

Results from the 3D simulations are shown in Figure~\ref{fig:be-3d} including the
contours of the clay/air interface and the pressure in the ground. Our findings
are analogous to the 2D axisymmetric case, with good agreement between the two
methods on cavity and crater dimensions, as well as ground shock propagation.
The 3D volume-of-fluid method exhibits the same advantages as in 2D, showing
excellent fidelity in the break-up of the lofted clay.

\section{Application: modelling the BRL 105mm shaped charge}
\label{sec:application}
\noindent In this section we apply the method to modelling the BRL $105
\,\mathrm{mm}$ unconfined shaped charge as detailed in~\cite{brl105mm}.  Shaped
charges use convergent detonation waves in a high-explosive to produce a
hyper-velocity metal jet with many industrial and defence applications. They are
also stringent tests of elasto-plastic hydrodynamics methods as they typically
include multiple materials arranged in a complex configuration and subject to
extreme deformation. This particular shaped charge has been modelled by other
authors~\cite{pagosa,barlow14}, allowing us to draw useful comparisons. Note
that it is not our intention to fine tune the model to match experiment, only to
show that the volume-of-fluid method is capable of treating this extremely
strenuous scenario and producing reasonable results. It is noted that the
underlying diffuse-interface method with THINC interface sharpening does not
work well for this problem as the thin jet becomes smeared over a large area.

\begin{figure}
    \centering
    \includegraphics[width=0.6\textwidth,clip]{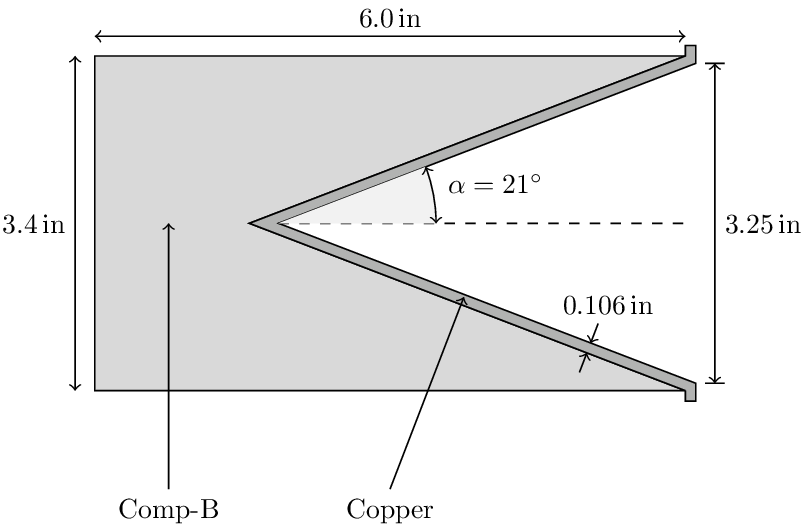}

    \caption{The BRL 105mm unconfined shaped charge, simplified from the diagram
    given in~\cite{brl105mm}.}%

    \label{fig:brl105mm}
\end{figure}

The 2D axisymmetric domain is $[0, 6] \times [-6, 40] \,\mathrm{cm}$, with the
origin chosen to be coincident with the inner liner apex. The geometry of the
shaped charge is given in~\cite{brl105mm}, but unfortunately not all pertinent
details are specified.  For simplicity we omit the tetryl booster pellet and
detonator from our model and ignore the rounded apex of the liner.  Our modified
geometry is shown in Figure~\ref{fig:brl105mm}.

The liner is copper with parameters given in
Table~\ref{tab:dorovskii-parameters}. The Johnson-Cook rate-sensitive plasticity
model is used, including both work-hardening and thermal-softening effects, with
parameters given in Table~\ref{tab:jc-parameters}.  The high-explosive (Comp-B)
is treated using the reactive burn model from~\cite{wallis21}, summarised in
Section~\ref{sec:closure-models}. The reactants and products both obey the JWL
equation of state, with parameters given in Table~\ref{tab:jwl-parameters}. The
explosive is assumed to have no strength hence $G = 0$. The parameters for the
ignition and growth model are $a = 0.01$, $b = 0.222$, $c = 0.222$, $d = 0.666$,
$e = 0.0$, $g = 0.0$, $x = 4.0$, $y = 2.0$, $z = 0.0$, $F_{ig} = 0.3$, $F_{G_1}
= 1.0$, $F_{G_2} = 1.0$, $I = 4.4\cdot{}10^2\,(10\mu\mathrm{s})^{-1}$, $G_1 =
4.14\cdot{}10^{-1} \,\mathrm{GPa}^{-y}(10\mu\mathrm{s})^{-1}$ and $G_2 = 0.0
\,\mathrm{GPa}^{-z}(10\mu\mathrm{s})^{-1}$, taken from~\cite{murphy93}. The heat
of detonation associated with the Comp-B is $6.3 \, \mathrm{kJ\,g}^{-1}$. The
shaped charge is immersed in air governed by an ideal gas equation of state with
$\gamma = 1.4$ and $\rho_0 = 1.225\cdot{}10^{-3} \,\mathrm{g\,cm}^{-3}$.

\begin{table*}
    \centering
    \begin{tabular}{lrrrrrrr}
        \toprule
        Comp-B & $\rho_0$ (g cm$^{-3}$) & $A$ (GPa) & $B$ (GPa) & $R_1$ & $R_2$ & $\Gamma_0$ & $C_V$ (kJ K$^{-1}$ g$^{-1}$) \\
        \midrule
        Reactants & $1.717$ & $7.781\cdot{}10^4$ & $-5.031$ & $11.3$ & $1.13$ & $0.8938$ & $2.487\cdot{}10^{-3}$ \\
        Products & $1.717$ & $5.242\cdot{}10^2$ & $7.678$ & $4.2$ & $1.1$ & $0.34$ & $10^{-3}$ \\
        \bottomrule
    \end{tabular}
    \caption{JWL equation of state parameters for Comp-B, taken
    from~\cite{murphy93}.}%
    \label{tab:jwl-parameters}
\end{table*}

All materials are initialised to a pressure of $1\,\mathrm{atm}$, except for a
hemispherical booster region centred at the back of the explosive with radius
$1.439 \,\mathrm{cm}$ and pressure $27\,\mathrm{GPa}$, used to initiate the
detonation.  Twenty five Lagrangian tracer particles are initially located along
the inner surface of the liner, evenly spaced along the $z$-axis at intervals of
$0.25\,\mathrm{cm}$ starting from $z=3.0\,\mathrm{cm}$. The time history of the
velocity is measured for each tracer particle.

\begin{figure*}
    \centering
    \begin{subfigure}{0.42\textwidth}
        \centering
        \includegraphics[width=\linewidth,clip]{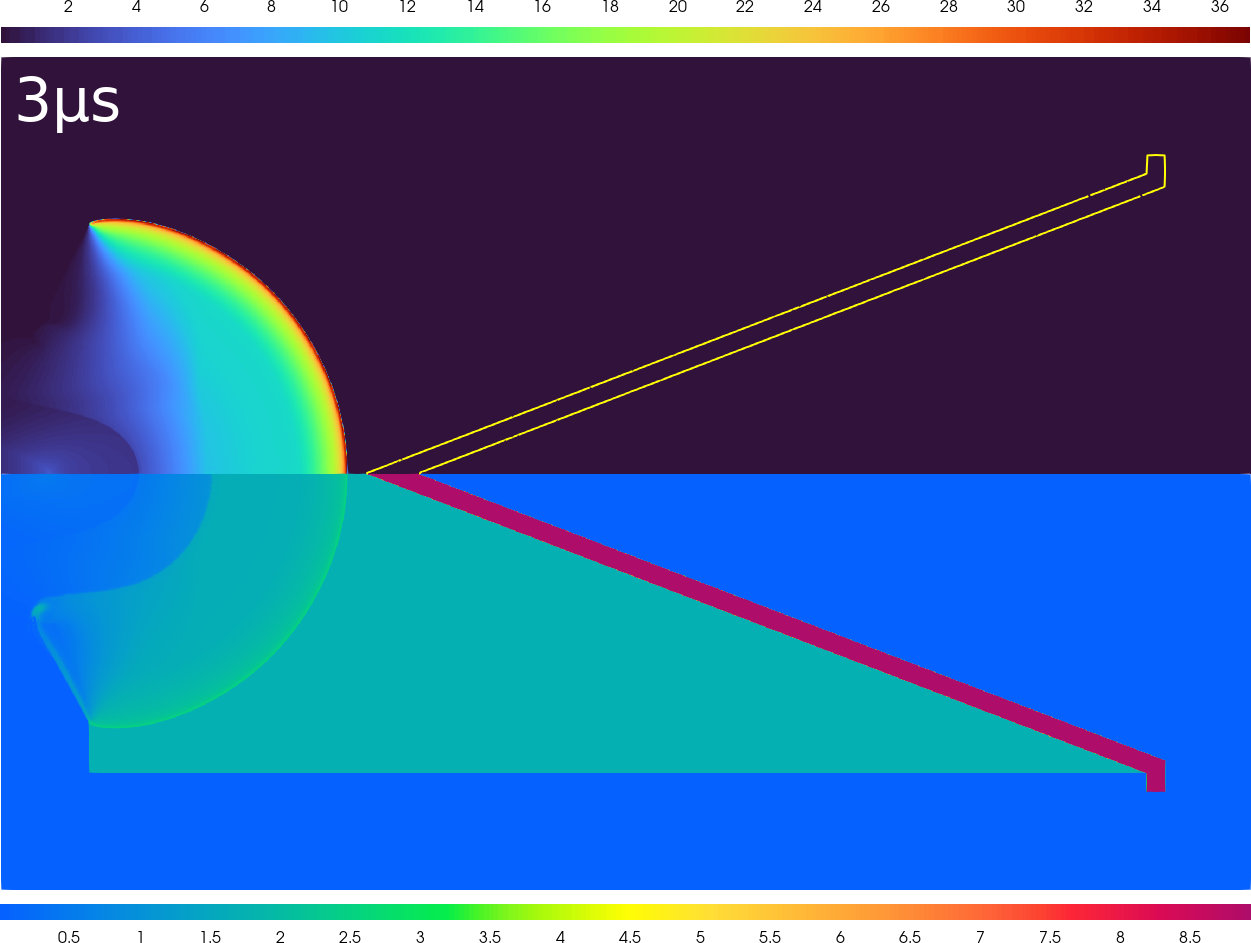}
        \caption{}%
        \label{fig:brl-3}
    \end{subfigure}\hspace*{0.05\textwidth}%
    \begin{subfigure}{0.42\textwidth}
        \centering
        \includegraphics[width=\linewidth,clip]{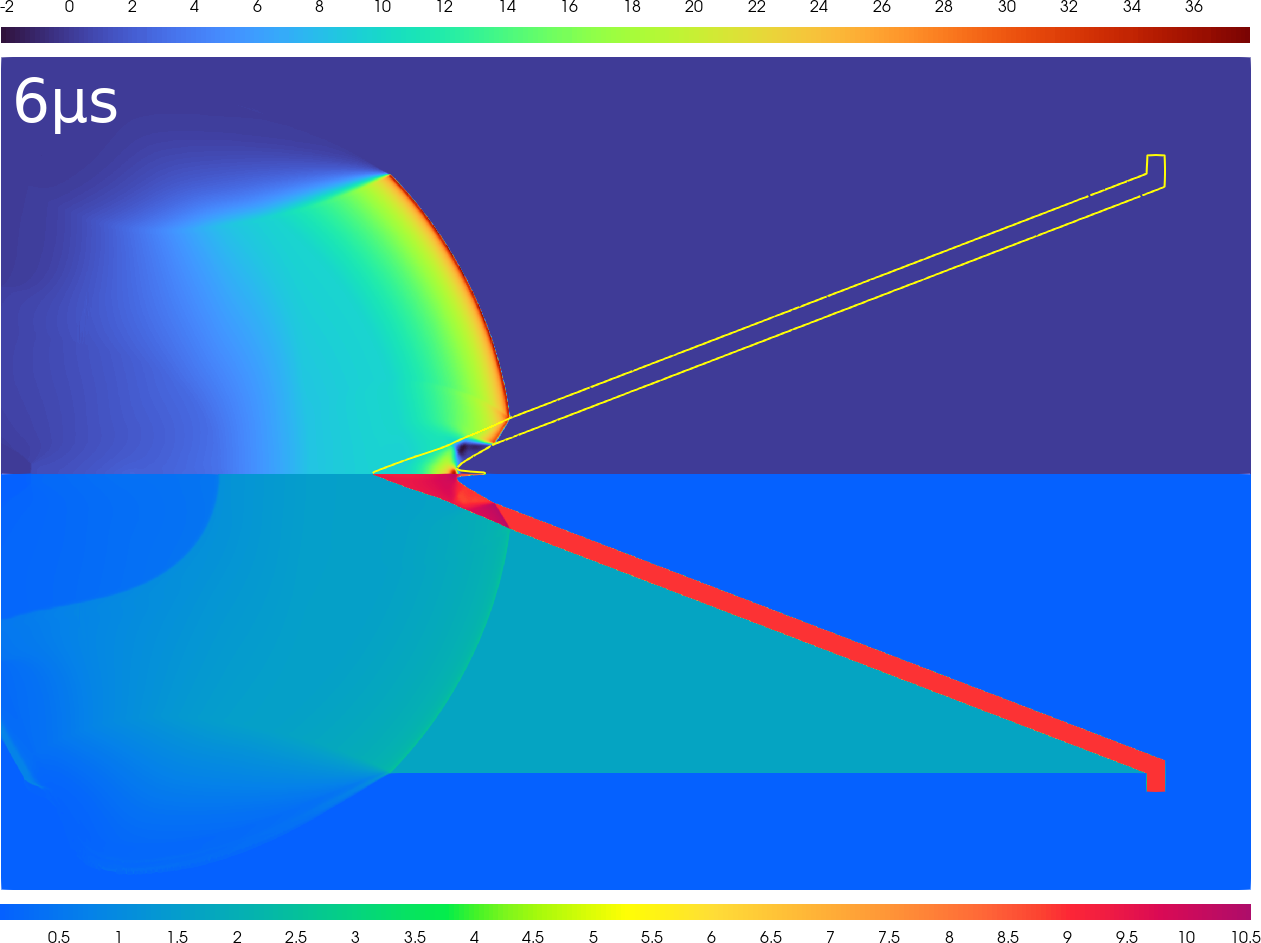}
        \caption{}%
        \label{fig:brl-6}
    \end{subfigure}
    \begin{subfigure}{0.42\textwidth}
        \centering
        \includegraphics[width=\linewidth,clip]{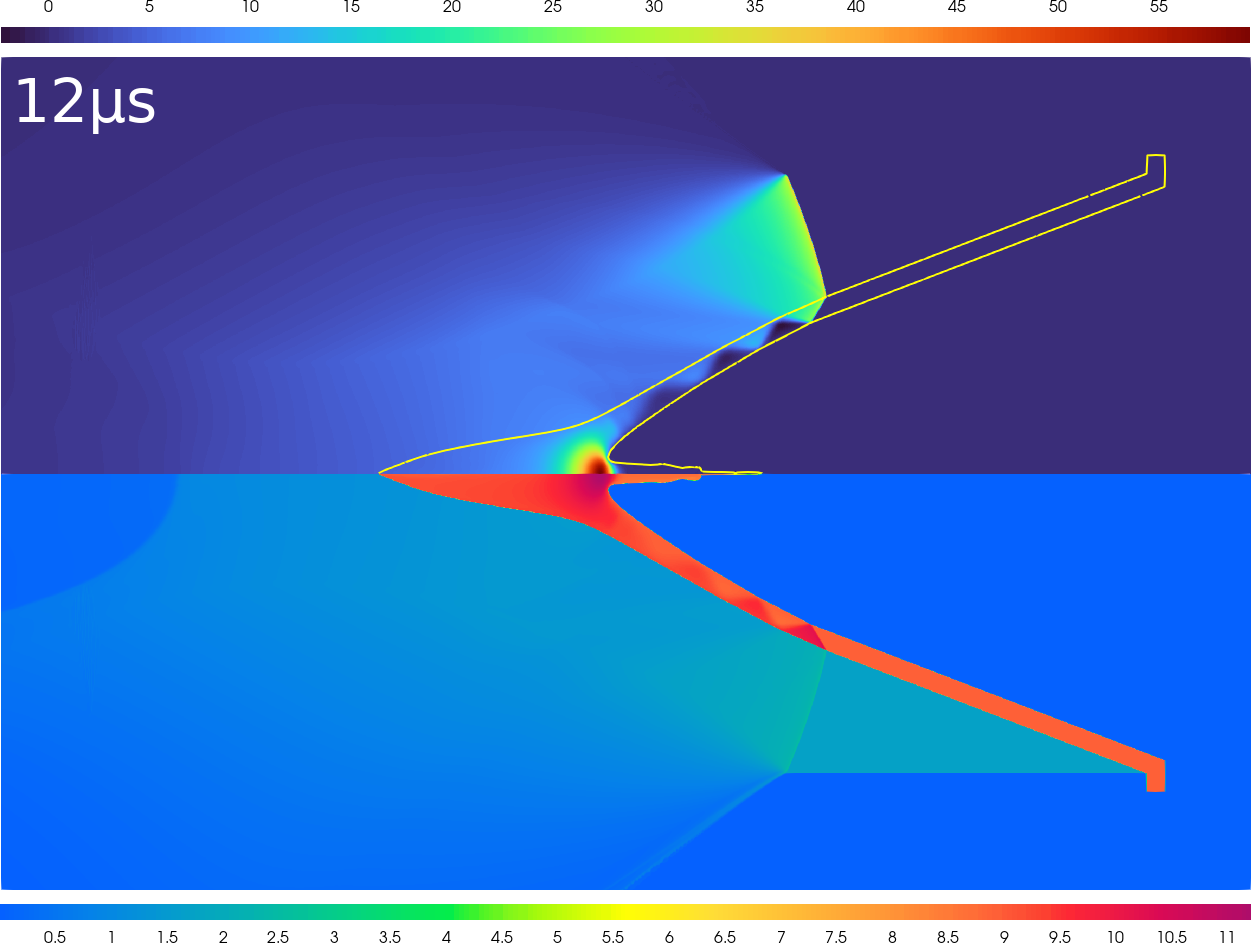}
        \caption{}%
        \label{fig:brl-12}
    \end{subfigure}\hspace*{0.05\textwidth}%
    \begin{subfigure}{0.42\textwidth}
        \centering
        \includegraphics[width=\linewidth,clip]{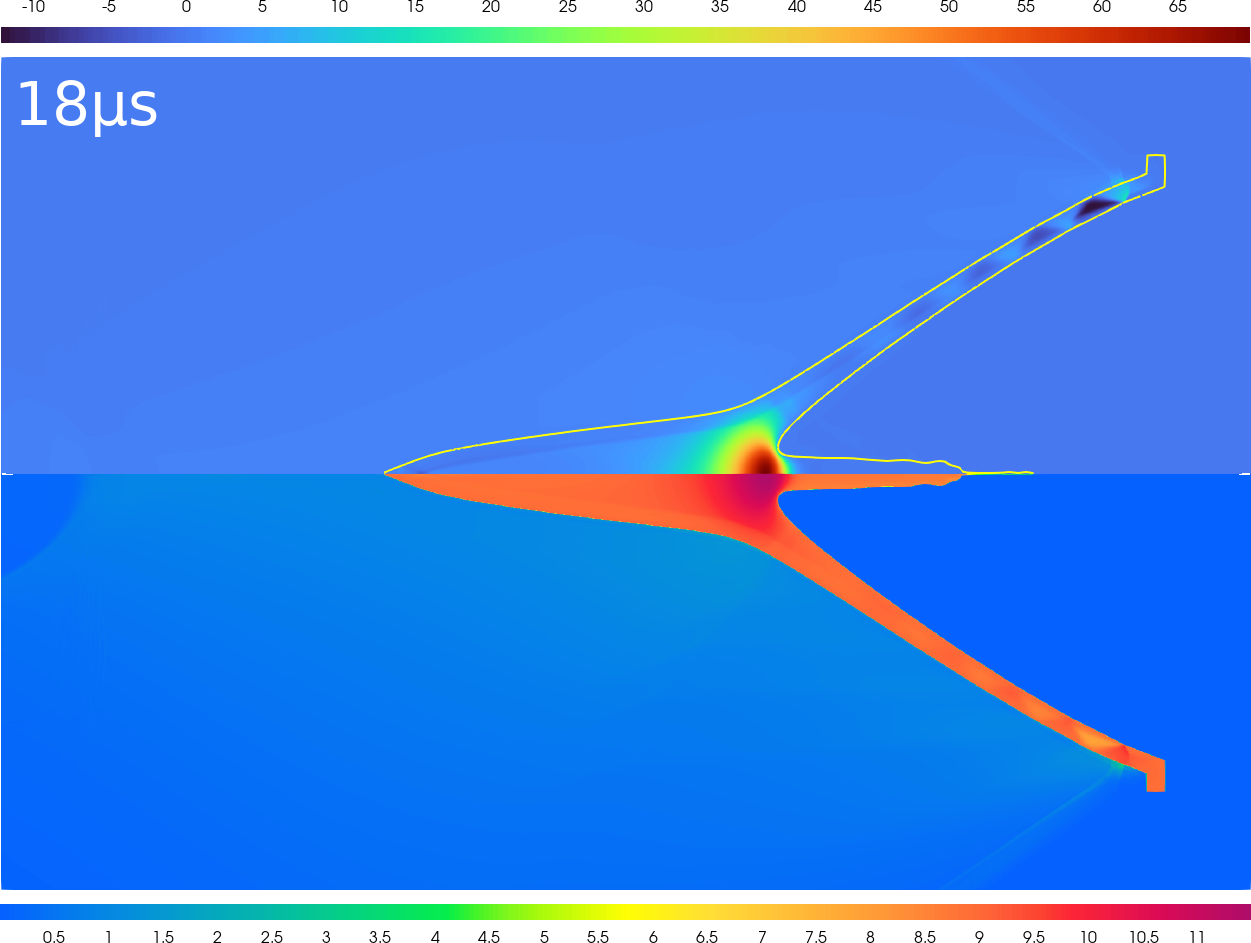}
        \caption{}%
        \label{fig:brl-18}
    \end{subfigure}
    \begin{subfigure}{0.9\textwidth}
        \centering
        \includegraphics[width=\linewidth,clip]{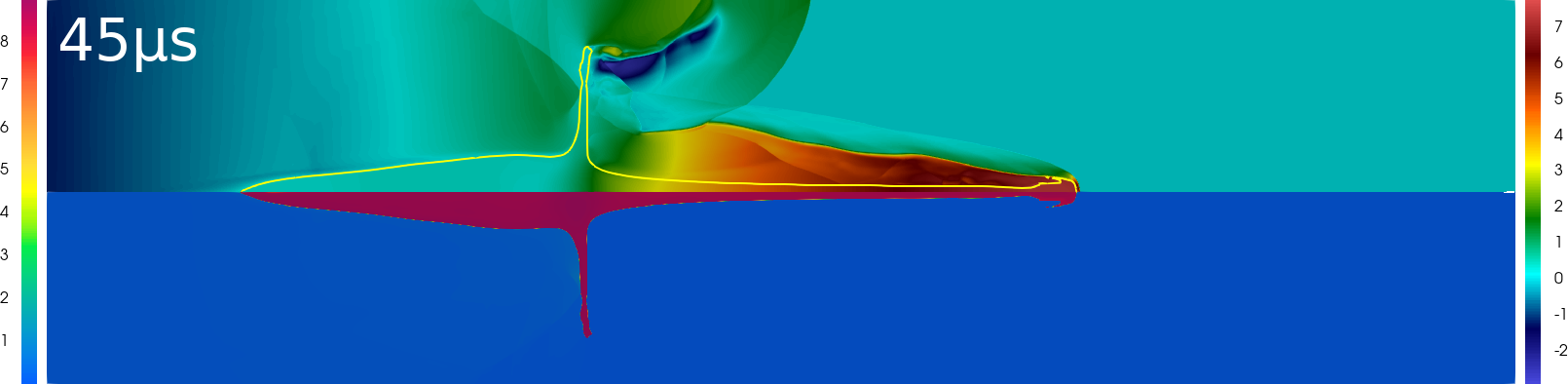}
        \caption{}%
        \label{fig:brl-45}
    \end{subfigure}
    \begin{subfigure}{0.9\textwidth}
        \centering
        \includegraphics[width=\linewidth,clip]{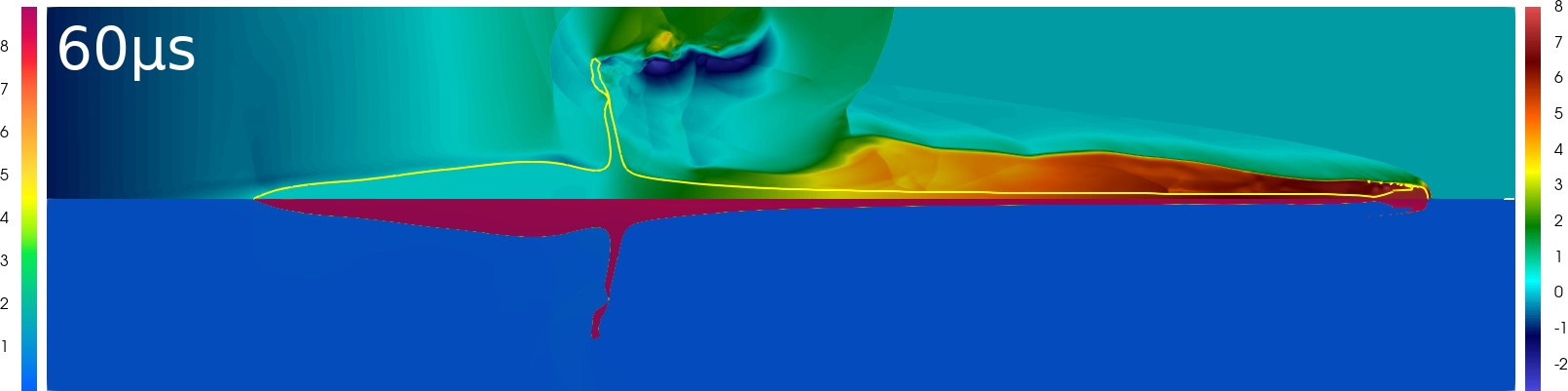}
        \caption{}%
        \label{fig:brl-60}
    \end{subfigure}

    \caption{Time evolution of the shaped charge problem. (a)--(d) shows the HE
    burn and the liner collapse during the first eighteen microseconds. In each
    panel, the top half shows pressure measured in $\mathrm{GPa}$ and the bottom
    half shows density measured in $\mathrm{g\,cm}^{-3}$.  (e) and (f) show the
    late-time jet formation at $t=45$ and $60\,\mu\mathrm{s}$. The top half of
    each panel shows the axial speed measured in
    $\mathrm{mm}\,\mu\mathrm{s}^{-1}$ and the bottom half shows the density. The
    yellow line denotes the $\phi=0.5$ contour.}%

    \label{fig:brl-stills}
\end{figure*}

Figure~\ref{fig:brl-3}--\ref{fig:brl-18} shows the high-explosive burn and
subsequent collapse of the liner that occurs during the first eighteen
microseconds.  The detonation wave is initiated by the high pressure booster
region. Just after $t=3\,\mu\mathrm{s}$ the wave strikes the apex of the liner
and pressure begins to build up in that region.  At $t=12\,\mu\mathrm{s}$ the
jet is beginning to form as the liner turns in and is rapidly accelerated.  By
$t=18\,\mu\mathrm{s}$ the high-explosive is fully burnt and the high pressure
stagnant region can be clearly seen. Figure~\ref{fig:brl-45} and
\ref{fig:brl-60} show the late-time jet formation at $t=45$ and
$60\,\mu\mathrm{s}$ respectively. The jet reaches a stable speed and begins to
thin out due to the velocity gradient along its length. The jet also has a blunt
tip, which is in good agreement with the Arbitrary Lagrangian-Eulerian results
presented by Barlow~\textit{et~al.}~\cite{barlow14}, and as they note matches
what is typically seen in experiment. Close inspection of the density field
reveals fragments of copper breaking off the tip.

\begin{figure*}
    \centering
    \begin{subfigure}{0.5\textwidth}
        \centering
        \includegraphics[width=\textwidth,clip]{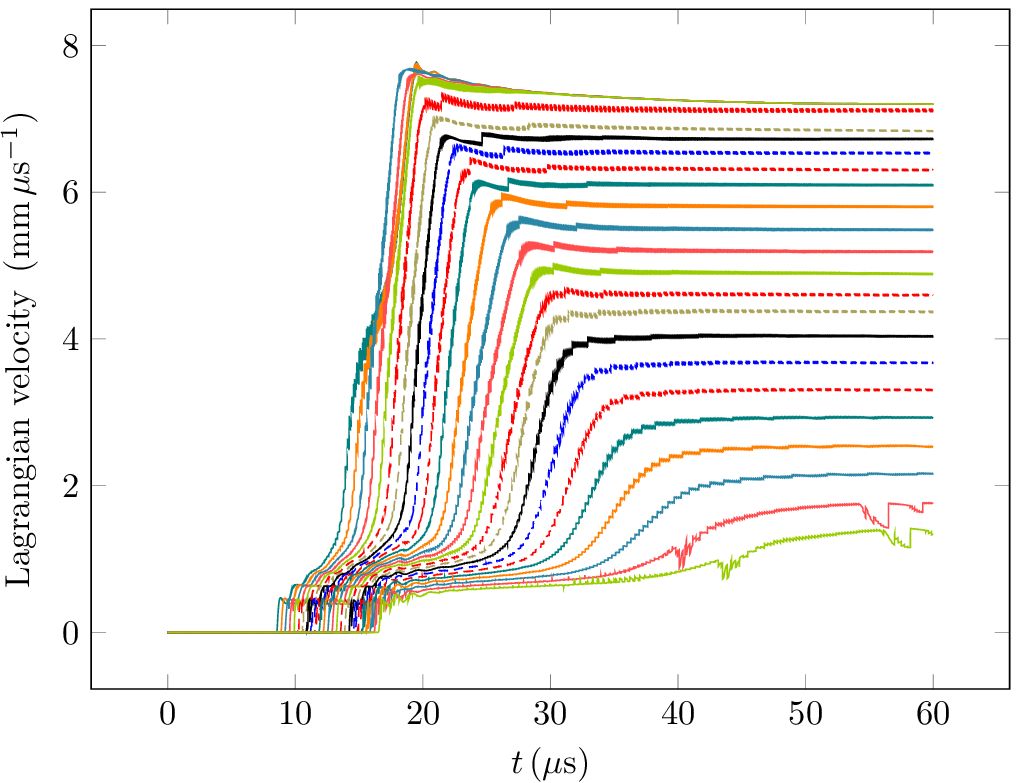}
        \caption{}%
        \label{fig:brl-2d-tracers}
    \end{subfigure}%
    \begin{subfigure}{0.5\textwidth}
        \centering
        \includegraphics[width=\textwidth,clip]{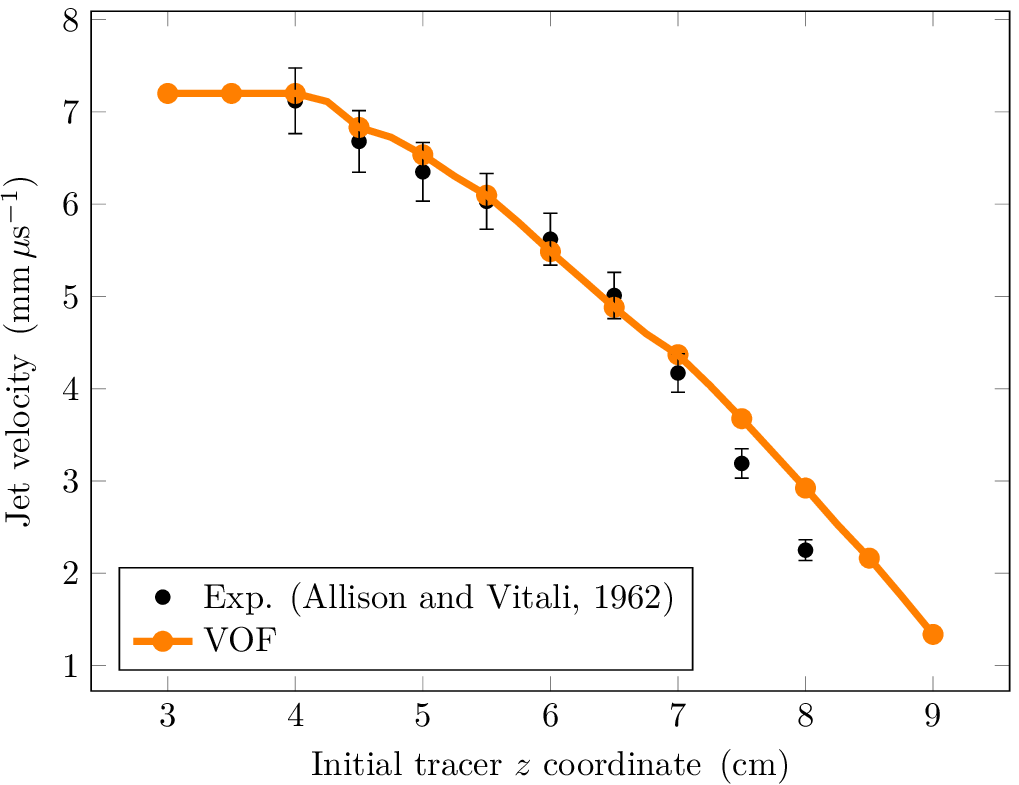}
        \caption{}%
        \label{fig:brl-2d-jet}
    \end{subfigure}

    \caption{(a) shows the time history of the velocity of each Lagrangian
    tracer particle, initially distributed along the inner liner. (b) shows the
    stabilised jet velocity at $t=60\,\mu\mathrm{s}$ as a function of the
    initial position of each Lagrangian particle, alongside the tabulated
    experimental results from~\cite{brl105mm}, where the error bars correspond
    to a $5\%$ uncertainty.}%

    \label{fig:brl-jet}
\end{figure*}

Figure~\ref{fig:brl-2d-tracers} shows the time history of the velocity of each
Lagrangian tracer particle over the course of the simulation. The inner apex of
the liner is accelerated close to the detonation velocity of the Comp-B
($7.98\,\mathrm{mm}\,\mu\mathrm{s}^{-1}$). By the end of the simulation the
particle velocities have stabilised. Figure~\ref{fig:brl-2d-jet} shows the slice
corresponding to $t=60\,\mu\mathrm{s}$ as a function of the initial
$z$-coordinate of each tracer particle. The experimental results from
\cite{brl105mm} are overlaid. The jet velocity is particularly sensitive to the
parameters used in the reactive burn model which have not been tuned, but
nevertheless a fairly good match is obtained.

\section{Conclusions}
\label{sec:conclusions}
\noindent We have presented a novel Godunov-type volume-of-fluid method for the
simulation of large deformation multi-material problems featuring elasto-plastic
solids and fluids. The method can be viewed as an interface-sharpening extension
to the diffuse-interface method of Barton~\cite{barton19}. The method is
practical, robust and extensible, and has been shown capable of treating a range
of challenging problems with excellent fidelity.



\bibliographystyle{plain}
\bibliography{main}

\end{document}